\documentclass{aa}  

\usepackage{graphicx}
\usepackage{float}
\usepackage{txfonts}
\usepackage{color}
\usepackage{ulem}

\newcommand\gaia{\textit{Gaia}\xspace}
\newcommand\gdrtwo{\gaia~DR2\xspace}

\newcommand{\gbp}{{G$_\mathrm{BP}$}\xspace}
\newcommand{\grp}{{G$_\mathrm{RP}$}\xspace}
\usepackage{lscape}
\usepackage{threeparttable}
\begin{document}

   \title{Kinematics and dynamics of Gaia red clump stars}

  \subtitle{Revisiting north-south asymmetries and dark matter density at large heights}

   \author{Jean-Baptiste Salomon
          \inst{1}
          \and
          Olivier Bienaym\'e\inst{2}
          \and
          C\'eline Reyl\'e\inst{1}
          \and
          Annie C. Robin\inst{1}
          \and
          Benoit Famaey\inst{2}
          }

   \institute{Institut UTINAM, CNRS UMR6213, Universit\'e Bourgogne Franche-Comt\'e, OSU THETA Franche-Comt\'e-Bourgogne,\\
   Observatoire de Besançon, 41 bis avenue de l'Observatoire, BP 1615, 25010 Besan\c{c}on C\'edex, France\\
              \email{jean-baptiste.salomon@utinam.cnrs.fr}
         \and
             Observatoire astronomique de Strasbourg, Université de Strasbourg, CNRS, 11 rue de l’Université, 67000 Strasbourg, France\\
             \email{olivier.bienayme@unistra.fr}
             }

   \date{Received May 29, 2020; accepted September 3, 2020}

\abstract{
In this study, we analyse the kinematics and dynamics of a homogeneous sample of red clump stars, selected from the second Gaia data release catalogue in the direction of the Galactic poles, at five different positions in the plane. The level of completeness of the sample at heights between 0.6 and 3.5 kpc was asserted through a comparison with the 2 Micron All Sky Survey catalogue. We show that both the density distribution and velocity dispersion are significantly more perturbed in the north than in the south in all analysed regions of our Galactic neighbourhoods. We provide a detailed assessment of these north-south asymmetries at large heights, which can provide useful constraints for models of the interaction of the Galactic disc with external perturbers. We proceeded to evaluate how such asymmetries could affect determinations of the dynamical matter density under equilibrium assumptions. We find that a Jeans analysis delivers relatively similar vertical forces and integrated dynamical surface densities at large heights above the plane in both hemispheres. At these heights, the densities of stars and gas are very low and the surface density is largely dominated by dark matter (DM), which allows us to estimate, separately in the north and in the south, the local dark matter density derived under equilibrium assumptions. In the presence of vertical perturbations, such values  should be considered as an upper limit. This Jeans analysis yields values of the local dark matter density above 2~kpc, namely, $\rho_{\rm DM} \sim 0.013 \, {\rm M}_\odot/{\rm pc}^3$ ($ \sim 0.509 \, {\rm GeV/cm}^3$) in the perturbed northern hemisphere and $\rho_{\rm DM} \sim 0.010 \, {\rm M}_\odot/{\rm pc}^3$ ($ \sim 0.374 \, {\rm GeV/cm}^3$) in the much less perturbed south. As a comparison, we determine the local dark matter density by fitting a global phase-space distribution to the data. We end up with a value in the range of $\rho_{\rm DM} \sim 0.011 - 0.014 \, {\rm M}_\odot/{\rm pc}^3$ , which is in global agreement with the Jeans analysis. These results call for the further development of non-equilibrium methods with the aim of obtaining a more precise estimate for the dynamical matter density in the Galactic disc.}

   \keywords{Galaxy: evolution -- Galaxy: kinematics and dynamics -- Galaxy: disc -- Galaxy: structure
               }
   \maketitle
%

\section{Introduction}\label{sectionIntro}

The structure and dynamics of our Galaxy have long been studied, revealing, in the process, an ever more complex picture. A great deal can be learned about these properties from the Galaxy's vertical structure. This knowledge can, in principle, allow us to recover the vertical restoring force above the Galactic plane and teach us about the external perturbations that have affected the disc.

The first studies of the vertical dynamics of Milky Way stars date back to the early works of \citet{Kapteyn20}, \citet{Kapteyn22}, \citet{Oort26}, and \citet{Oort32}, who noted that the motions of nearby stars perpendicular to the Galactic plane were mostly uncorrelated with horizontal motions, at least for low vertical amplitude. For a stationary stellar population, the collisionless Boltzmann equation or the Jeans equations make it possible to link variations perpendicular to the Galactic plane of velocities and density to the vertical variations of the gravitational potential.  Close to 25 works were published up through the 1980s on the measurement of
the Galactic potential in the solar neighbourhood. Due to the small size and inhomogeneity of the stellar samples used at the time, it was not possible to obtain an accurate measurement of this potential. The first large homogeneous sample of a few hundreds K dwarfs used to measure both their vertical density variation and kinematics allowed for one of the first accurate determinations of the vertical potential \citep{Kuijken89}. Moreover, \cite{Bienayme87}, in comparing a Galactic model of stellar populations \citep{Robin86} and Galactic star counts, as well as the age-velocity dispersion relation of stellar populations, was also able to accurately measure the vertical variation of the potential at the Galactic position of the Sun. The first accurate measurements of the Oort's limit, that is, the dynamical determination of the total local mass density, were obtained using Hipparcos observations \citep{Creze98,Holmberg00}, which provided the first complete samples of thousands of stars with truly accurate distances within 120 pc around the Sun. The 2000s saw the progressive use of samples of stars more distant from the Galactic plane ranging from a few hundred \citep{Siebert03} to a few thousands of stars \citep{Bienayme14}; see also \citet{Read14} and references therein. The advent of the \gaia\ space mission \citep{Prusti16} with its successive data releases, has led to a few new analyses \citep[e.g.][]{Hagen18, WidmarkMonari, Widmark19, Buch19, Nitschai20}. For these new samples, the main sources of error were no longer statistical limitations or measurement accuracy, but systematic biases and errors related to the approximate modelling and to the underlying assumptions. Here, we carefully re-examine the problem and, especially, how the north-south asymmetries systematically affect the results.

The Galaxy is not in a stationary state and at present, this appears  to be  the main limitation in accurately determining the gravity field outside the Galactic plane. Non-stationarity is directly visible, for example, by examining the differences in star counts as well as velocity fields towards the northern and southern galactic poles \citep{Widrow12,yanny13,williams13, xu15}. These north-south asymmetries can be classified into bending (odd parity in density, even in vertical velocity) and breathing (even parity in density, odd in vertical velocity) modes \citep{Widrow14}, and the Milky Way disc is experiencing both. \citet{Faure14}, \citet{Debattista14}, and \citet{Monari16} showed that breathing modes can be caused by internal disturbances, such as spiral arms. Bending modes, on the other hand, are likely caused by external perturbations \citep{Gomez13, Widrow14, Laporte18, Chequers18}, even though, to a certain extent, an internal buckling bar instability can also cause such modes \citep[e.g.,][]{Khoperskov19}. More recently, the extremely precise measurements of the second Gaia data release (\gdrtwo) \citep{Brown18} have made it possible to highlight strong correlations between horizontal and vertical velocities up to $z=\pm 1$kpc \citep{Antoja18, Binney18, Laporte19, bennett19}.

Therefore, the equilibrium of the Galactic disc is no longer a fully credible assumption. \citet{Banik17} and \citet{Haines19} have investigated in simulations how departures from equilibrium could affect the determination of the surface density of the disc and of the local dark matter (DM) density. They both concluded that non-stationarity of the disc causes a systematic overestimate of both the vertical force (by a factor that can be as large as 1.5 close to the plane in under-dense regions) and local density (up to 20\%) when wrongly assuming equilibrium. They concluded that these biases, especially in under-dense regions, call for the development of non-equilibrium methods to estimate the dynamical matter density.

In this work, we do not propose a non-equilibrium method for the estimate of the dark matter density but, rather, we follow-up on the above-cited works by investigating, for the first time and based on actual data, how making an equilibrium assumption separately for data in the northern and southern Galactic hemispheres affect the estimates. To do so, we carefully selected red clump stars from the \gdrtwo catalogue and, incidentally, revisited in detail the north-south asymmetries with regard to their density and kinematics. We show, in particular, how the southern kinematics is significantly less perturbed than the northern one.

The article is organised as follows. In Section~\ref{sampleSelection}, we present the sample used in our study and describe the different cuts applied to target red clump stars from the \gdrtwo catalogue. We also check the completeness of the sample. In Section~\ref{sectionDensProf}, we specify how we partition the space to obtain both density profiles and, later, velocity dispersion gradients. We then derive the density profiles separately for the north and south and we discuss in detail their north-south asymmetries at large heights. We then analytically approximate them by way of several functions to derive scale heights and scale length. In a similar approach, velocity dispersion profiles and gradients are derived in Section~\ref{sectionVelocity}. Radial and vertical forces are derived in the north and south under the assumption of equilibrium and used to estimate different surface mass density and dark matter density in Section~\ref{sectionForces}. An alternative estimate of the dark matter density via the direct modelling of the stellar phase-space distribution is given for comparison in Section~\ref{sectionAltern}. Discussions and conclusions about the degree of north-south asymmetry of our Galactic suburbs as well as the local dark matter density are given in Section~\ref{sectionCCL}.


\section{Sample selection}\label{sampleSelection}

\subsection{Selection of clump stars}

The \gdrtwo catalogue  provides accurate astrometry and photometry for more than 1 billion stars. Detailed information on the data processing, validation, and catalogue content is given in \cite{Lindegren18} for the astrometry, and \cite{Riello18} and \cite{Evans18} for the photometry. The global validation of the catalogue can be found in \cite{Arenou18} and the description of the Gaia archive\footnote{https://gea.esac.esa.int/archive/} in \cite{Salgado17}.

Our stellar sample was extracted from \gdrtwo catalogue in the aim of use stars with the best precision in the six-dimensional space (positions and velocities) to maximise the volume probed and the number of stars and, finally, to achieve the best possible completeness. In order to avoid extinction problems within the plane and contamination by dwarf stars, which are dominant at lower galactic latitudes, we first selected all \gdrtwo star members with an absolute galactic latitude larger than 45 degrees, which was then reduced to include only the red clump stars. The sample was further restricted to stars with a magnitude in the G-band (G) smaller than 15. The final sample is made up of 2 539 093 stars.
 
\begin{figure}
  \centering
   \includegraphics[width=\hsize]{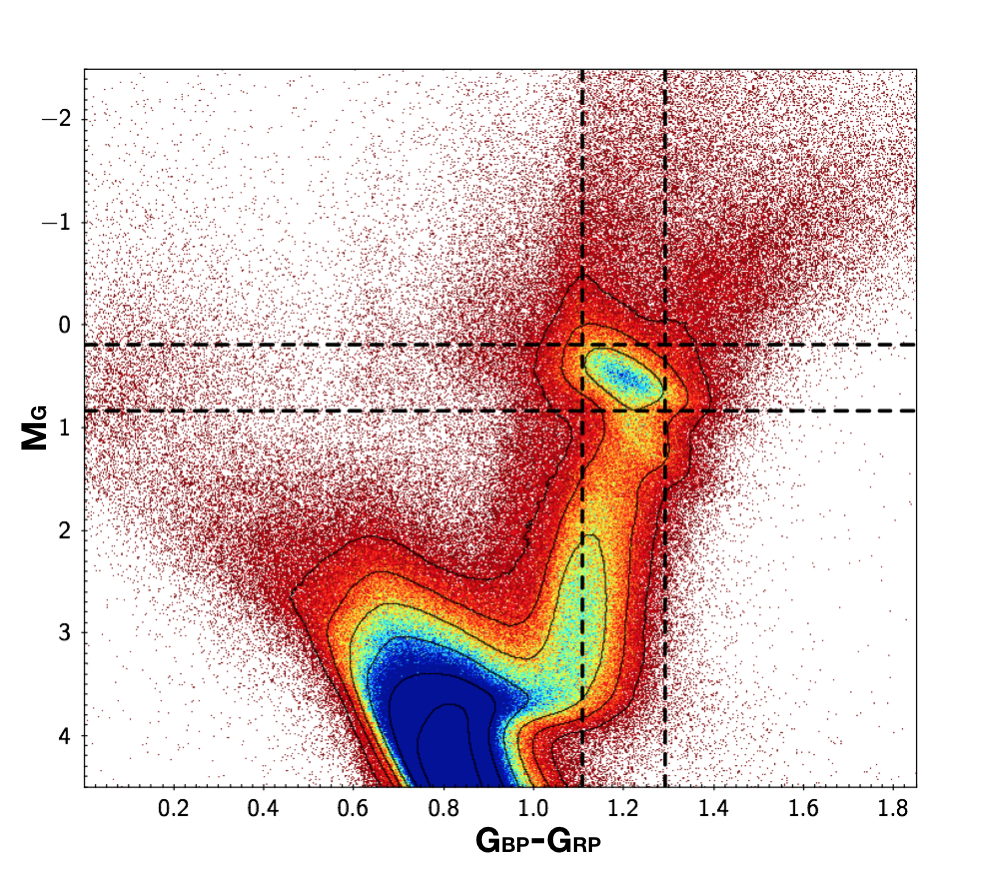}
   \caption{Part of the colour-magnitude diagram of stars from the \gdrtwo catalogue with $|b| > 45$ degrees and $G < 15$. The selection of our sample of red clump stars from these data is delimited with the dashed lines. }
              \label{selectionRC}%
    \end{figure}

In this sample, the red clump was identified in a colour magnitude diagram as the over-density near the red giant branch (see Figure~\ref{selectionRC}). The proxy for the absolute \gaia magnitude in the $G$ band was computed for individual stars, using $M_G = G-5 \log_{10}(1000/\varpi)+5$, where $\varpi$ is the parallax in milli-arcsec. For the completeness motivation (see Section~\ref{completnessSection}), only stars with $G < 15$ were considered.
 
The vertical cuts in colour \gbp-\grp are 1.107 and 1.291. The superior limit is enforced because at larger \gbp-\grp, the sample has incomplete velocity information. We  need radial velocities in our study to later derive the different velocity gradients in every direction as well as velocity dispersions. The horizontal cuts in absolute magnitude, M$_G,$ are 0.185 and 0.83 (see Figure~\ref{selectionRC}). These strict limits leave us with 43 589 stars. For the measurement of velocity dispersions (Section~\ref{sectionVelocity}), further restrictions based on stars belonging to the {\it Gaia} radial velocity spectrometer catalogue (RVS) are set.
 
To compute the spatial positions of stars, we used simple distance determinations from $\varpi$. However, this approximation is only valid when $\sigma_\varpi / \varpi \lesssim$ 20\% \citep{Luri18}. We check that more than 95\% of the selected sample (41 457 stars) comply with this criterion, which justifies our choice. Nevertheless, a distance derived from a parallax with 20\% errors can still be biased at the level of a few percents -- a caveat to consider. An alternative could have been to work directly in parallax space for the dynamical density estimates, while a joint spectroscopic analysis of the red clump stars could allow for a Bayesian refinement of the distances based on their atmospheric parameters and distance modulus estimates in combination with their parallax. Here, we take the most direct route by simply inverting the parallax, but the above caveats should be kept in mind for potential future improvements.
 
We then used the classical Galactic cylindrical coordinates (R, $\phi$, z), where $\phi$ equal 0 corresponds to the line connecting the Sun and the Galactic centre and positive towards Galactic rotation. We adopted the distance of the Sun to the Galactic of R$_0$ = 8.34 kpc \citep{Reid14}. The vertical distance of the Sun to the Galactic plane is 14 pc according to \cite{Binney97}. After correcting the vertical distances of the sample by the height of the Sun, there are 21 065 stars remaining in the north (z > 0) and 22 524 in the south (z < 0).
 
 \subsection{Completeness}\label{completnessSection}

The sample has been built in order to be complete up to at least 3 kpc from the Galactic plane. At this distance, the stars at our limit magnitude, $G = 15,$ have an absolute magnitude,  $M_G = 2.3$, which is 1.5 magnitude fainter than the clump stars. This ensures that all the clump stars are in our sample, assuming that Gaia is complete within our colour and apparent magnitude selection.

\begin{figure}[!ht]
  \centering
   \includegraphics[width=\hsize]{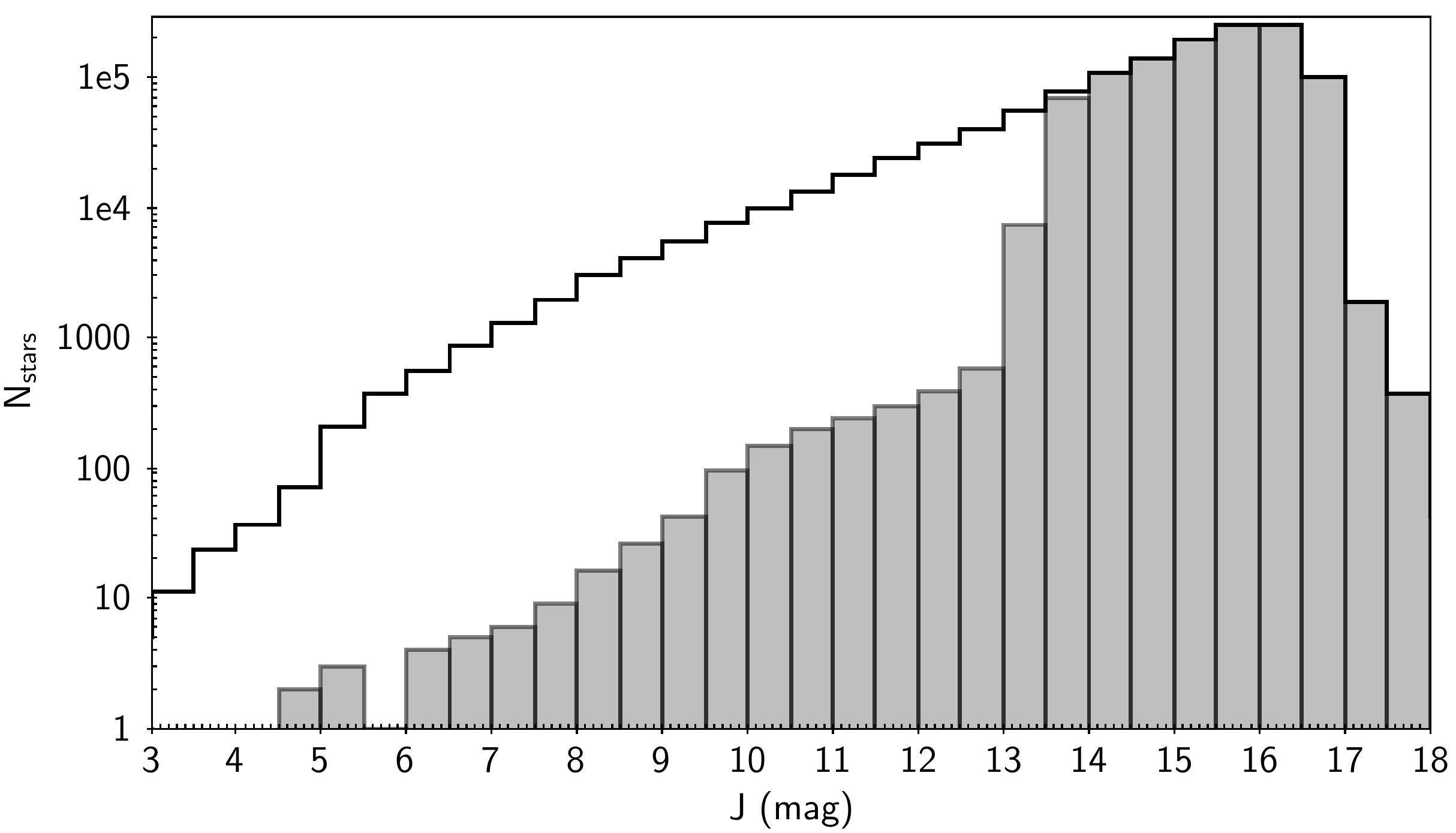}
   \caption{Number of stars with respect to their $J$ magnitude. Black line shows stars selected in 2MASS with $0.55 < J-K_s < 0.7$. Grey steps show stars from the previous selection that are not found in \gdrtwo,\ with a magnitude cut at $G=15$.}
              \label{2MASS}%
    \end{figure}

To check this assertion, we made a comparison with the 2 Micron All Sky Survey (2MASS) Point Source Catalogue, which is complete to $J=15.9$ and $K_s=14.3$ in unconfused regions \citep{Skrutskie06}. We first used the cross-match of \gdrtwo\ with 2MASS provided in the \gaia\ archive
\citep{Marrese19} to locate the colour of our clump stars sample in the 2MASS photometric system. The colour of the sample mainly ranges from $J-K_s = 0.55$ to $0.7$, with $J<13.3$. Next we selected all stars in 2MASS within this colour range and tried to retrieve them in \gdrtwo. Figure~\ref{2MASS} shows the $J$ magnitude distribution of the stars in 2MASS, superimposed with those that are not found in \gdrtwo. It is quite clear from this plot that many stars with $J>13$ were not retrieved in \gdrtwo\ due to our magnitude cut at $G=15$. If we further restrict our sample to the stars brighter than $J=13.3$, that is, the faintest stars in our clump sample, we retrieve 98\% of the stars.

\section{Density profiles}\label{sectionDensProf}
%
%
\subsection{Sub-samples}\label{subSampleSection}

\begin{figure}[!ht]  
   \centering

            {\includegraphics[trim = {0cm 0.8cm 0cm 0cm}, clip, width=\hsize]{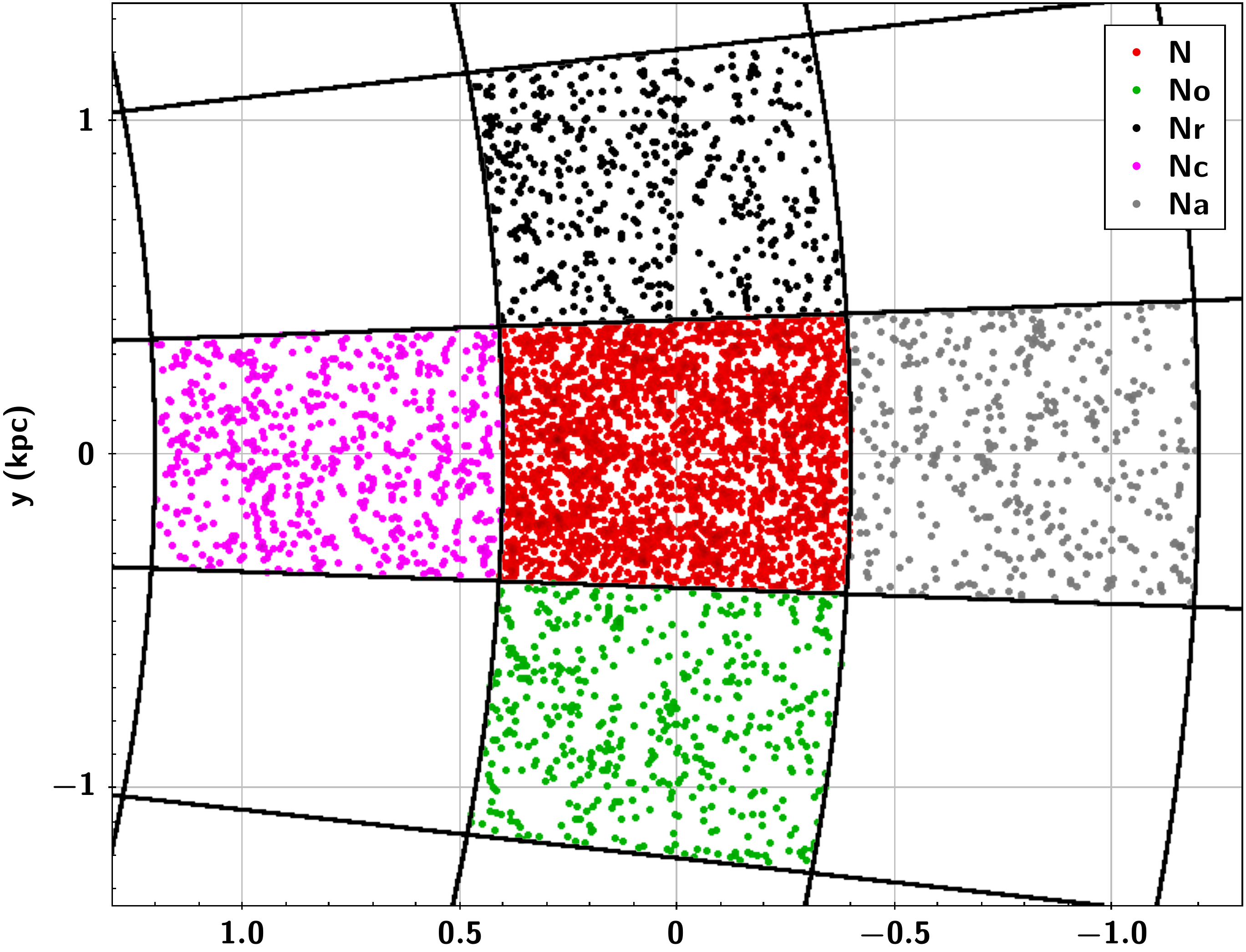}
            \includegraphics[width=\hsize]{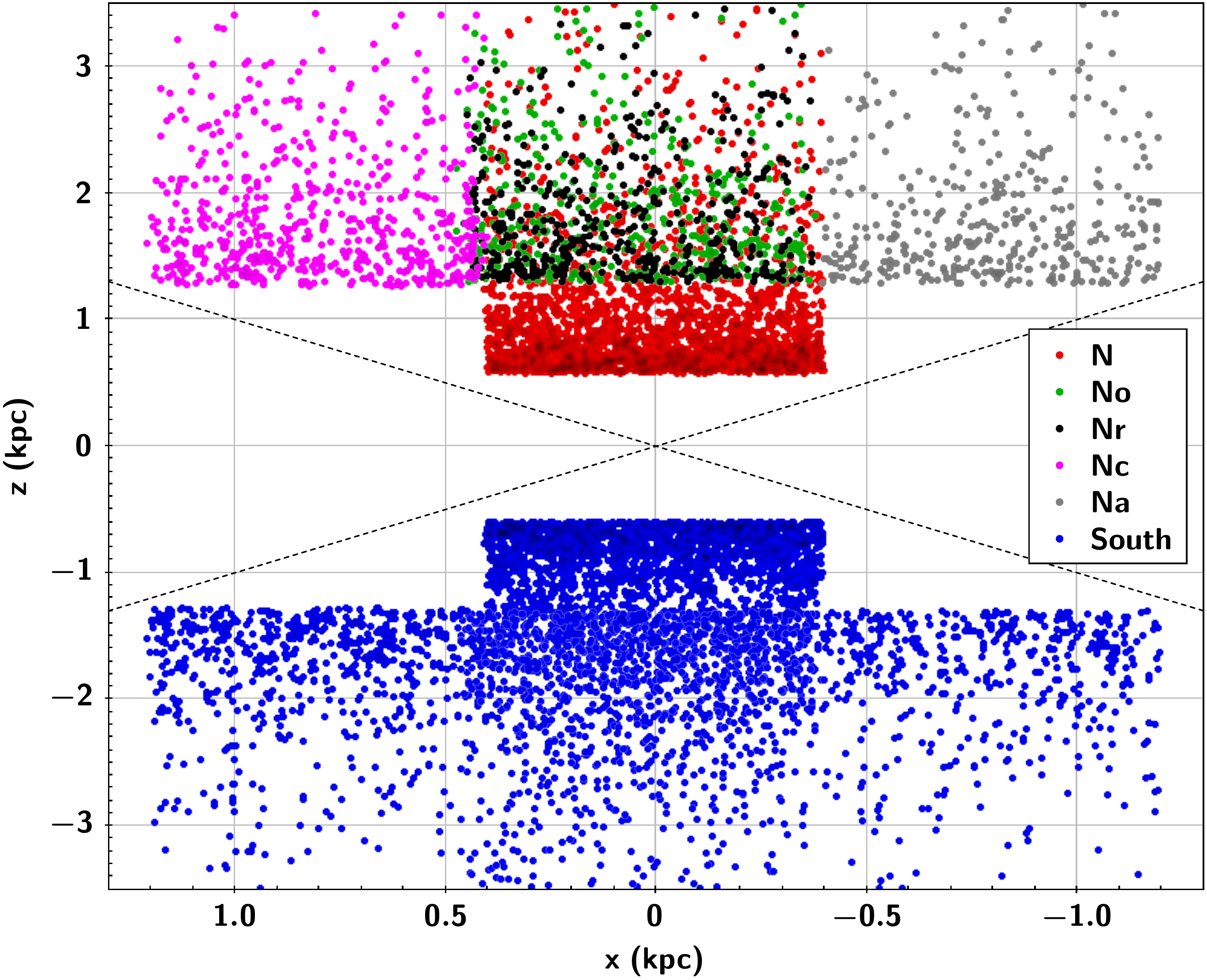}
            }
      \caption{Sub-samples of red clump stars selected from the initial conical selection described in Section~\ref{sampleSelection}, in Galactic Cartesian coordinates, centred at the solar radius, $(x,y)$ plane (top panel), and $(x,z)$ plane (bottom panel). The five sub-samples delimited in the northern hemisphere ($z>0$) are colour-coded. On the top panel, the cylindrical grid is superimposed in black solid lines. The colours given in the legends are those taking the same notation as described in the text (Subsection~\ref{subSampleSection}).
      The bottom panel illustrates the distribution of all sub-samples used in this study, 10 volumes, (colour-coded in the north, blue in the south) with respect to the Galactic plane ($z = 0$). The dashed lines illustrate the intersection of the conical selection and the plane $y = 0$.}
         \label{subSample}
   \end{figure}

From the sample of red clump stars  defined above, several sub-samples were compiled in order to make it possible to simultaneously characterise the vertical density, the scale height, and the scale length. This requires the initial volume to be described both in the radial and vertical directions. Hence, the volume of the two cones is divided in sectors (or partial hollow cylinders) in order to respect the cylindrical geometry (to the first order) of the Galaxy.
As a summary, Figure~\ref{subSample} represents selected stars that are colour-coded with respect to their inclusion among the different sub-samples.
For simple explanations of how the conical volume is split, we focus on the northern sector centred on the solar position. We  briefly note the other sub-samples later in the paper.

The central sector (called N in the north and S in the south) is characterised by the Galactocentric cylindrical coordinates ($R,\phi,z$). It is defined in $R$, in kpc, such that $R_{min} \leq R \leq R_{max}$ with $R_{min}=R_{0}-0.4$ and $R_{max}=R_{0}+0.4$.
It spans the range in $\phi$ in radians such that $\phi_{min} = -0.4/R_{0}$ and $\phi_{max} = 0.4/R_{0}$. The surface of a slice of this sector, perpendicular to the $z$-axis, is given by $\phi_{max}(R_{max}^2-R_{min}^2) = 0.64$ kpc$^{2}$.

However, this partial hollow cylinder has to be fully contained within our selection, which is done within galactic cones.
Since the cones have an angle of 45 degrees, this involves shifting the sector vertically by the largest distance to the Sun reached in the ($x,y$) plane.
It is reached when $R = R_{max}$ and $\phi = \phi_{max}$. This distance is given by $d = \sqrt{2R_{0}(1-\cos\phi_{max})(R_{0}+0.4)+0.4^2} = 0.572$ kpc.
Since our cone summit is not in the plane but at the Sun position, an additional offset of $z_\odot$ must be applied to ensure that the north-central sector is within the cone selection. Hence, our central sector is limited in $z$ with $z_{min}  = z_{\odot} + d = 0.586$ kpc. In order to keep our north and south sub-samples symmetric to the plane, this offset is also applied in the southern-central sector.

To compute the density gradient along the Galactic radius, we defined two sub-samples bordering the central one shifted toward the Galactic centre (Nc and Sc), and one shifted toward the Galactic anti-centre (Na and Sa)

The last parts of our space-cuts are those neighbouring our central sector at the same Galactic radius. We defined two sectors alongside the central sector, one in the same direction of the Galaxy rotation (Nr and Sr), and one in the opposite direction of the Galaxy rotation (No and So).

The geometrical ranges of the sectors and the resulting number of stars in each sub-sample are given in Table \ref{sectors}.
\begin{table*}[!ht]
\caption{\label{sectors} Properties of the sectors.}
\centering
\begin{tabular}{lccccc}
\hline\hline
Sector &N/S &Nc/Sc &Na/Sa &Nr/Sr &No/So\\
\hline
$R_{min}$ (kpc) &$R_0$-0.4 &$R_{0}$-1.2 &$R_{0}$+0.4 &$R_{0}$-0.4 &$R_{0}$-0.4 \\
$R_{max}$ (kpc) &$R_{0}$+0.4 &$R_{0}$-0.4 &$R_0$+1.2 &$R_0$+0.4 &$R_{0}$+0.4 \\
$\phi_{min}$ (rad) &-0.4/$R_0$ &-0.4/$R_0$ &-0.4/$R_0$ &0.4/$R_0$ &-0.4/$R_0$ \\
$\phi_{max}$ (rad) &0.4/$R_0$ &0.4/$R_0$ &-0.4/$R_0$ &1.2/$R_0$ &-1.2/$R_0$ \\
$d$ (kpc) &0.572 &1.255 &1.274 &1.291 &1.291 \\
$|z_{min}|$ (kpc) &0.586 &1.269 &1.288 &1.305 &1.305 \\
$N_{stars}$ &2397/2316 &622/663 &396/478 &483/502 &455/544 \\
\hline
\end{tabular}
\tablefoot{ Geometrical limits of the sectors, as defined in the text. The resulting number of stars in each sector is also given.}
\end{table*}
We point out that the size of the sectors was designed to conserve a minimum of about 400 stars in order to be able to perform reliable statistics up to high Galactic altitudes at a later stage. 

\subsection{Vertical density profiles}\label{zrProfilesSection}

To construct vertical profiles, each sub-sample is binned along the $z$ axis. These bins always contain a fixed number of 50 stars. The number of objects was settled after several tests in order to be sensitive enough to the actual density variations but not to statistical fluctuations. A consequence of keeping a fixed star number is that the volume of the bins is not constant since the individual height of slices $h_i$ has to be adapted.

\begin{figure}[!ht]
    \centering
        \begin{tabular}{c}
    \includegraphics[trim = {0cm 1.95cm 0cm 0cm}, clip, width=8.78cm]{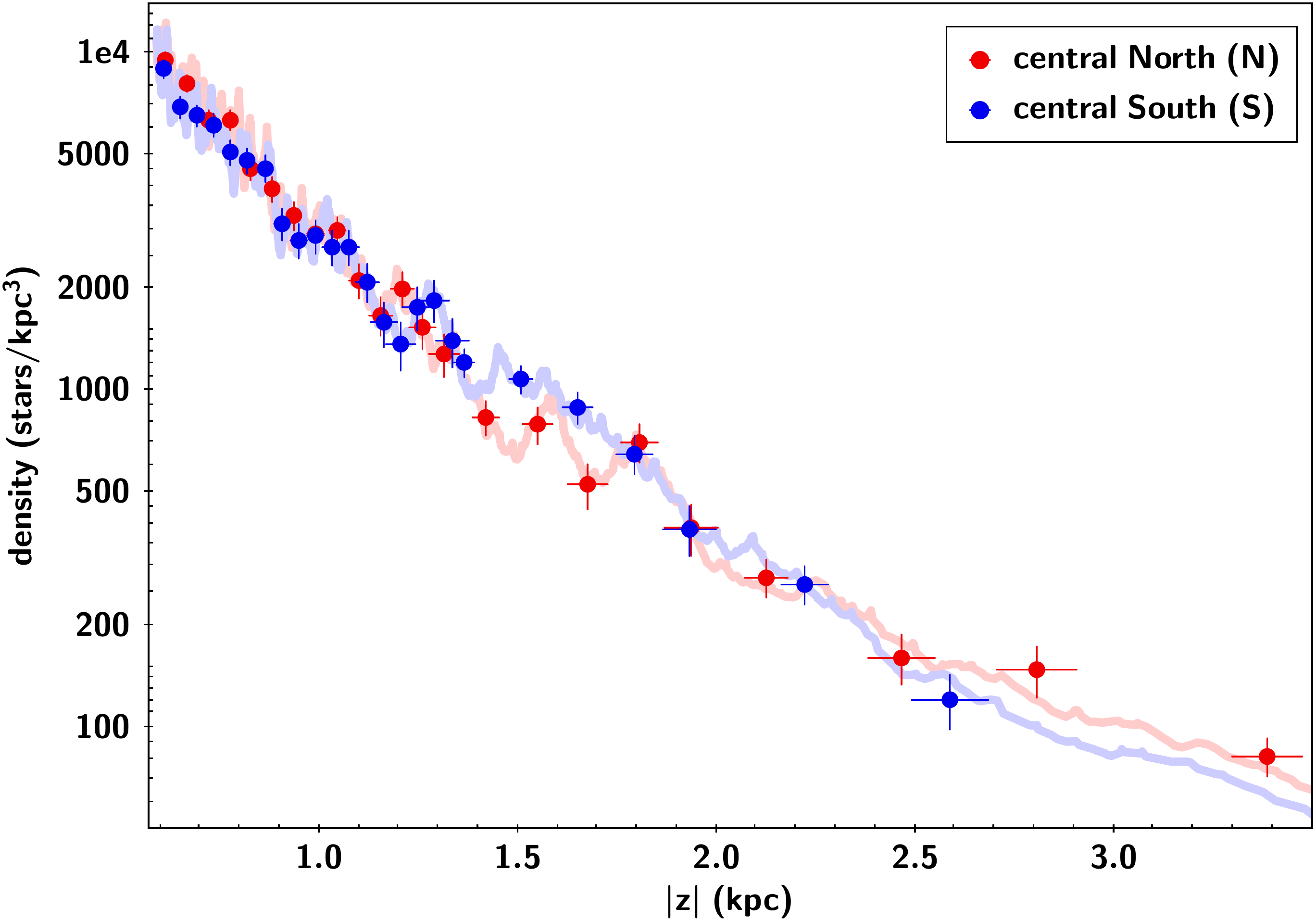} \\
    \includegraphics[width=8.78cm]{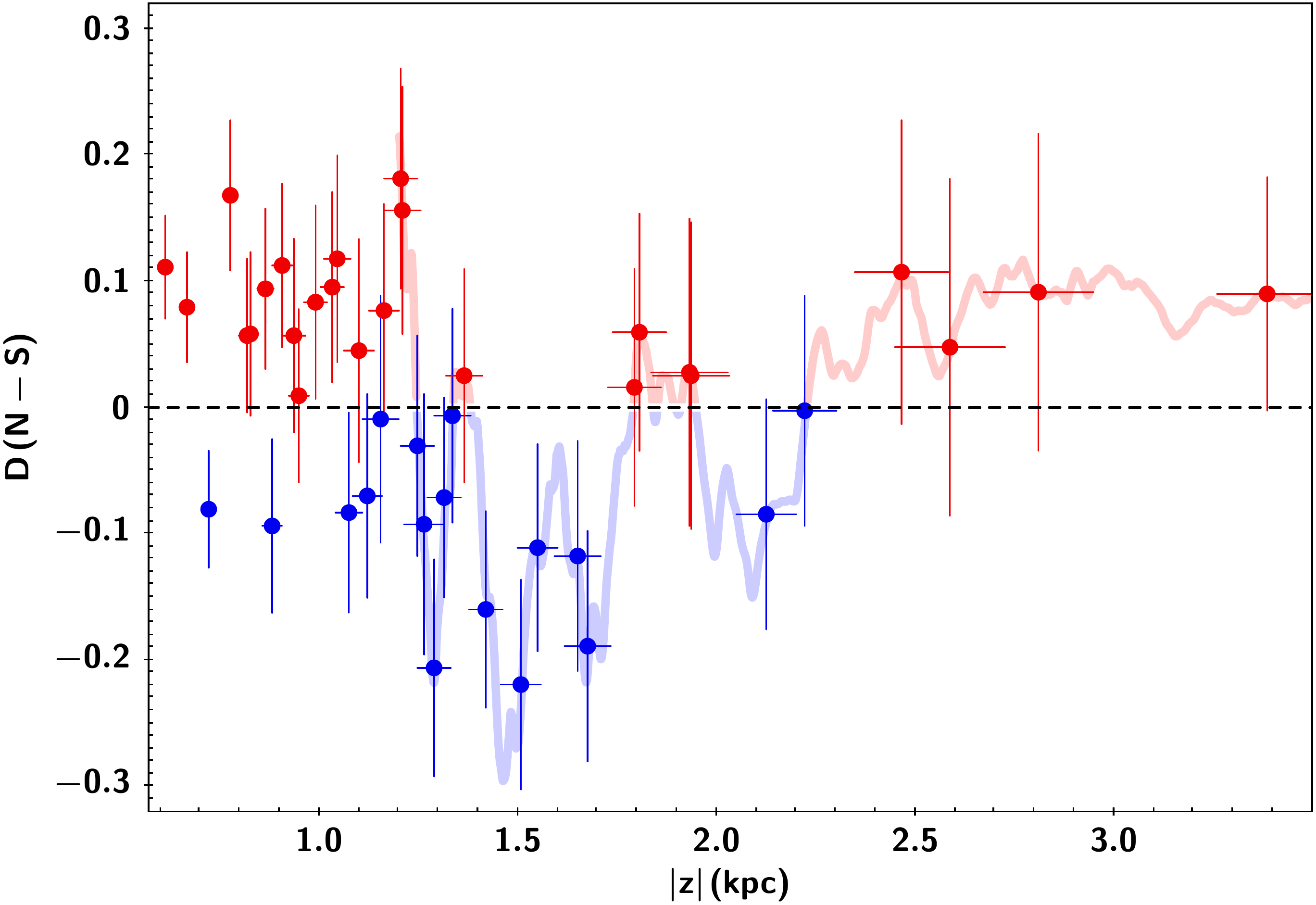}\\
   \end{tabular}
\caption{ Central sectors vertical density profiles. Top panel: Density profiles for the northern and southern central sectors (in stars per kpc$^3$) along the $|z|$ axis (in kpc). Light-coloured solid lines follow sliding bins of a constant number of 50 stars while dots are the values of the density in independent bins of fixed size (see text for details). Bottom panel: Normalised density differences between the north and south density profiles as a function of $|z|$ (see Equation~\ref{eqDiff}), red when the north is denser, and blue otherwise. Vertical error bars illustrate the normalised uncertainties added in quadrature.}

         \label{fig_dffdens_NS}
   \end{figure}

\begin{figure*}[!ht]
    \centering
        \begin{tabular}{cc}
    \includegraphics[trim = {0cm 1.95cm 0cm 0cm}, clip, width=8.78cm]{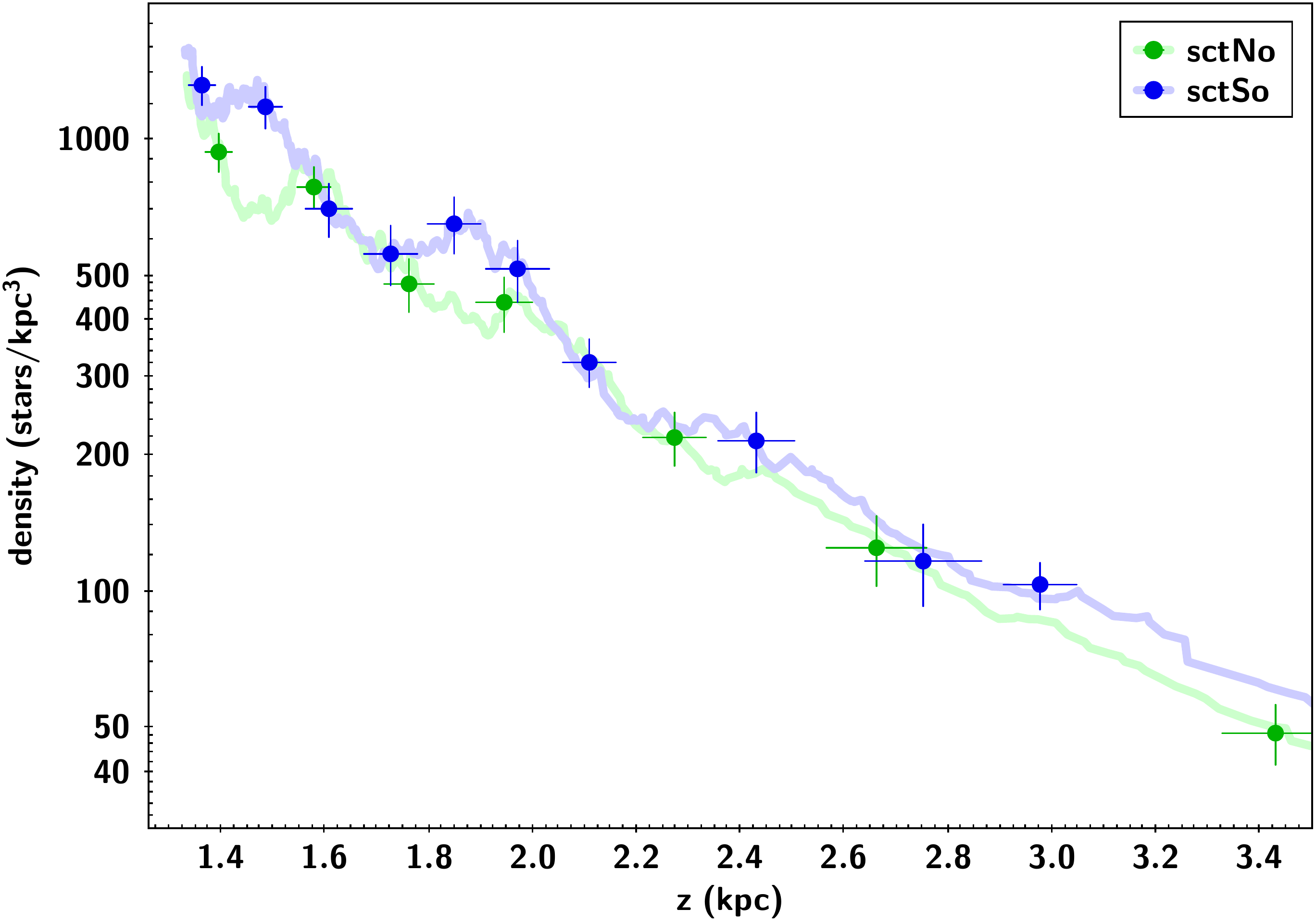} & \includegraphics[trim = {0cm 1.95cm 0cm 0cm}, clip, width=8.78cm]{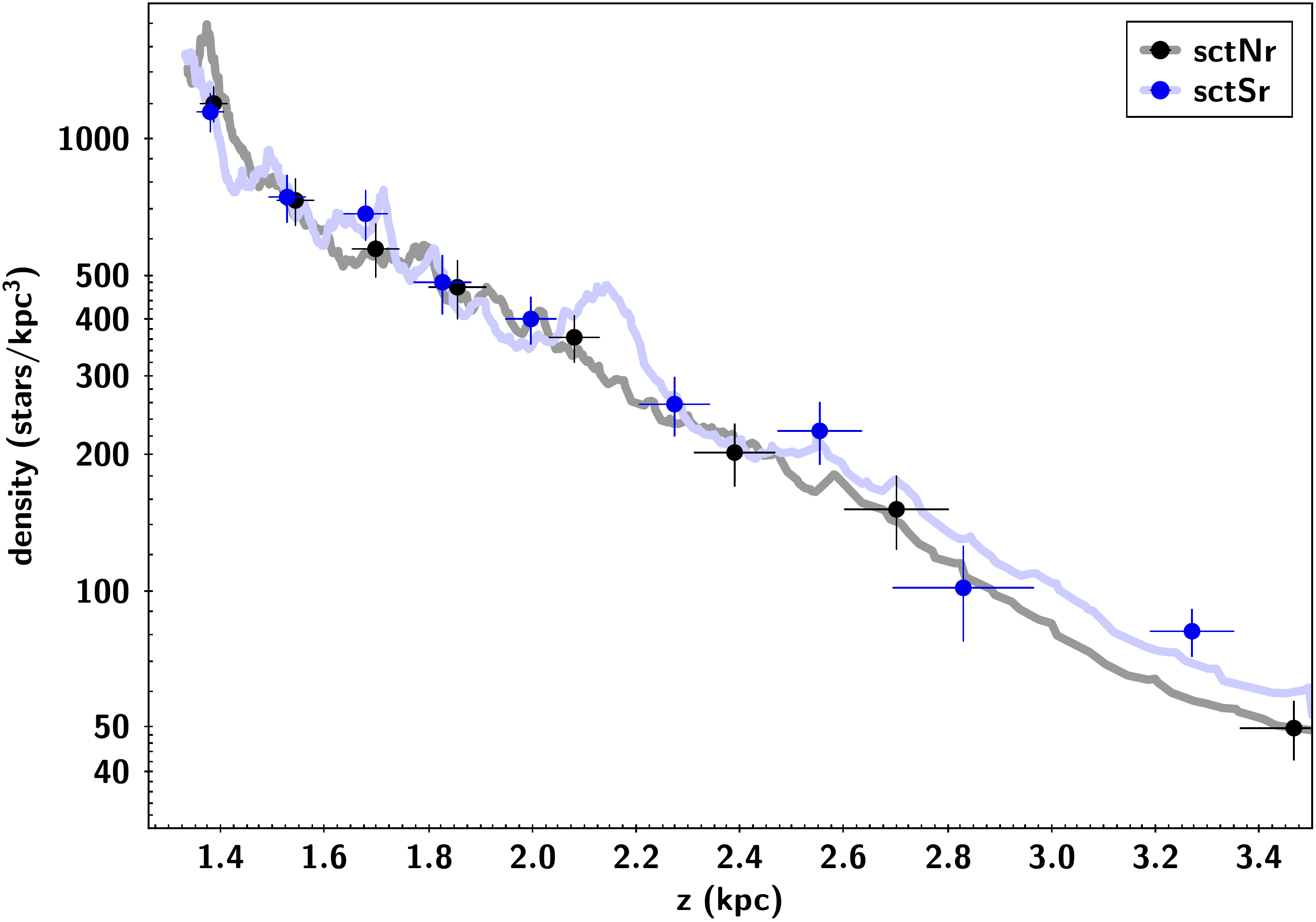} \\
    \includegraphics[width=8.78cm]{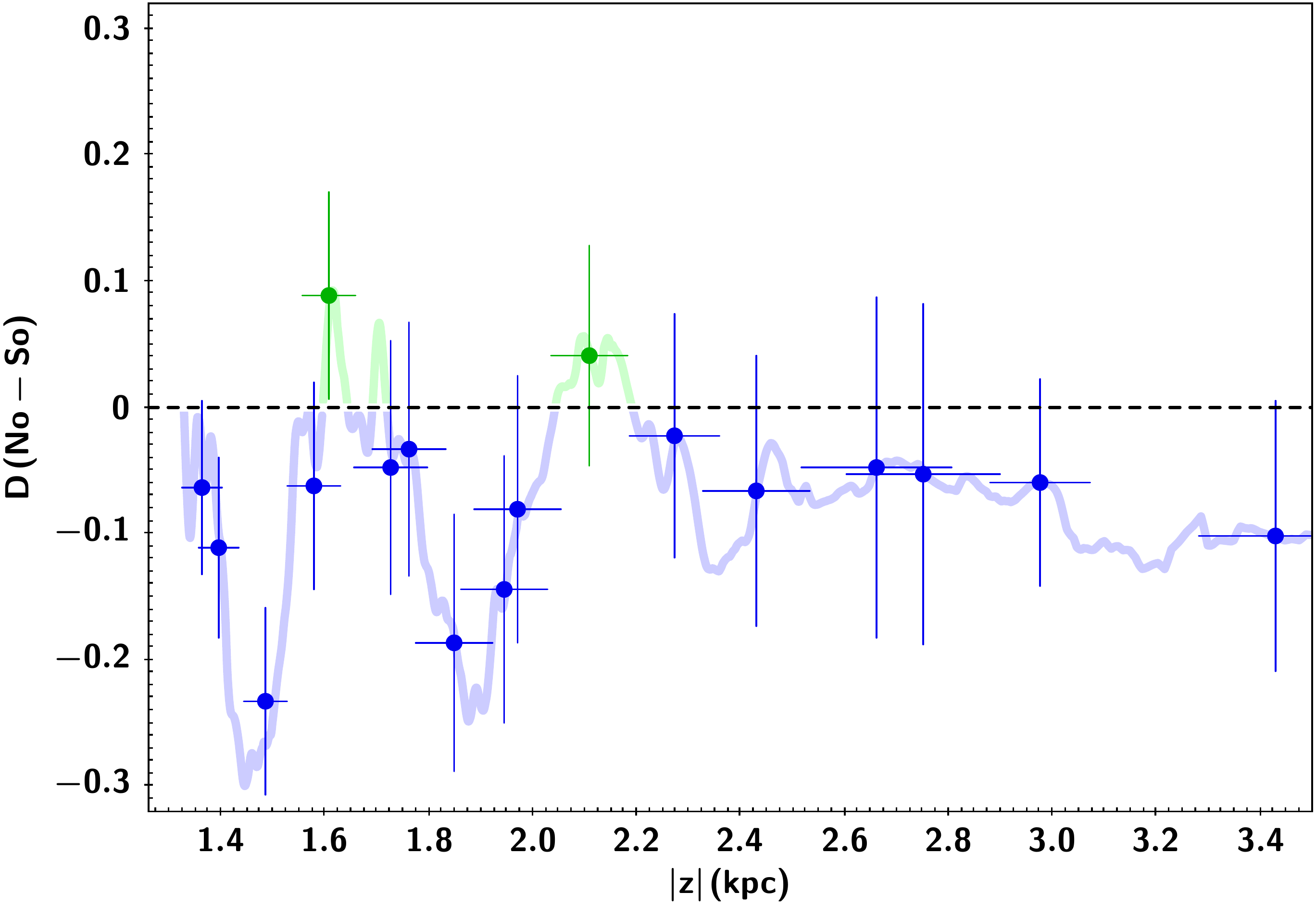} & \includegraphics[width=8.78cm]{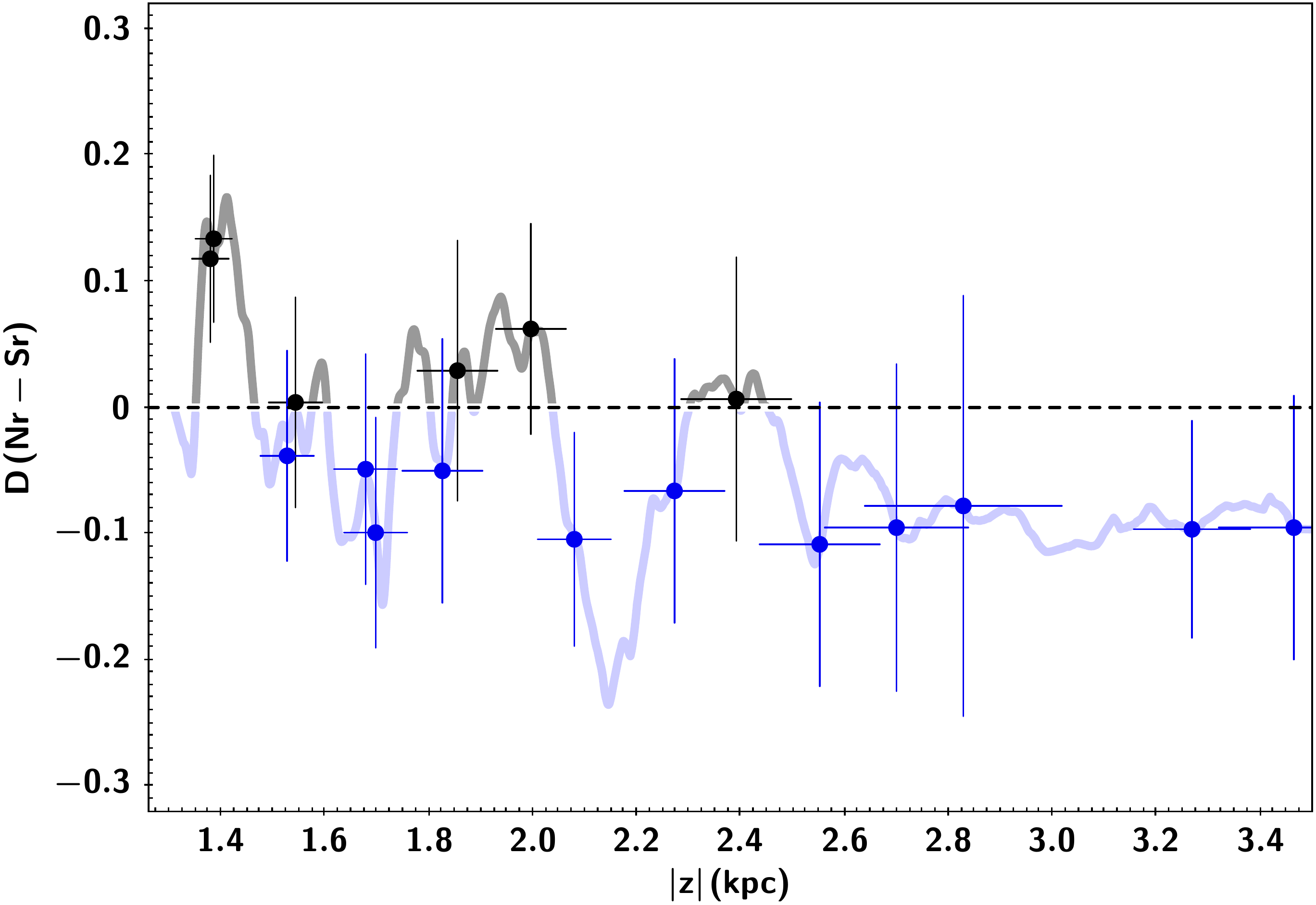} \\
   \end{tabular}
\caption{ Same as Figure~\ref{fig_dffdens_NS} but for the sectors shifted in the counter rotation (No and So) on the left and for sectors shifted in the direction of the Galaxy rotation (Nr and Sr) on the right.}

         \label{fig_diffdens_NoSo_NrSr}
   \end{figure*}

\begin{figure*}[!ht]
    \centering
        \begin{tabular}{cc}
    \includegraphics[trim = {0cm 1.95cm 0cm 0cm}, clip, width=8.78cm]{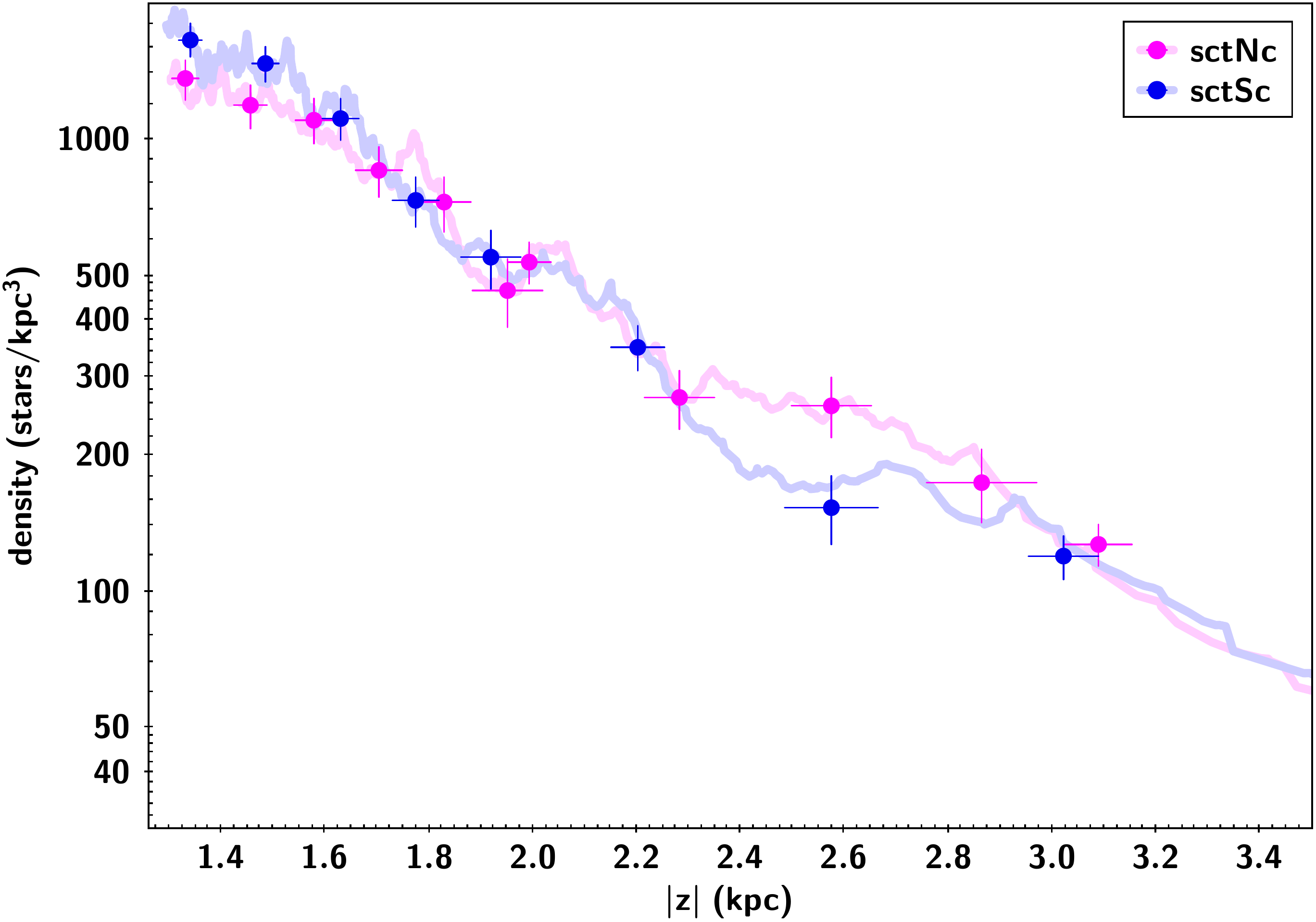} & \includegraphics[trim = {0cm 1.95cm 0cm 0cm}, clip, width=8.78cm]{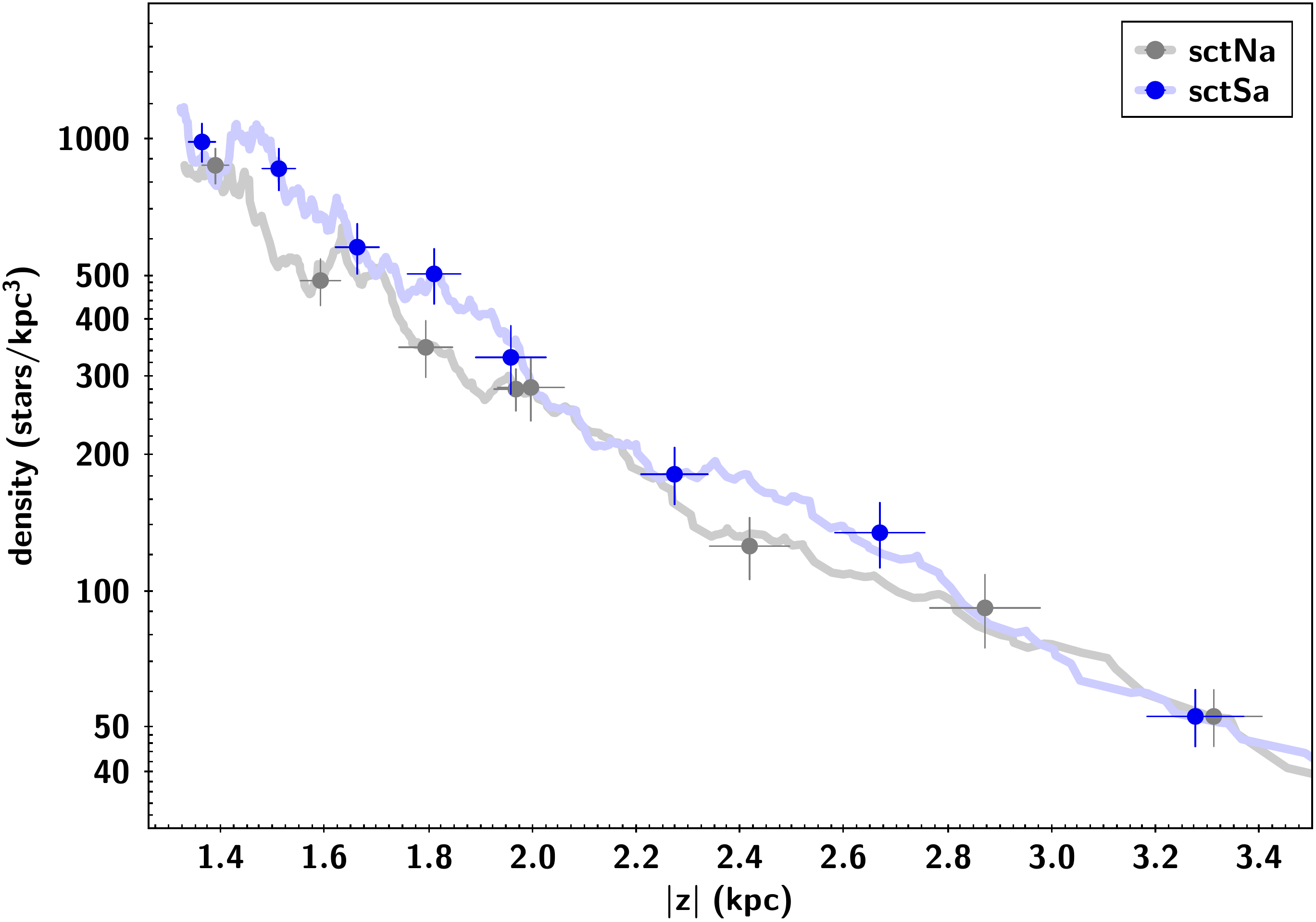} \\
    \includegraphics[width=8.78cm]{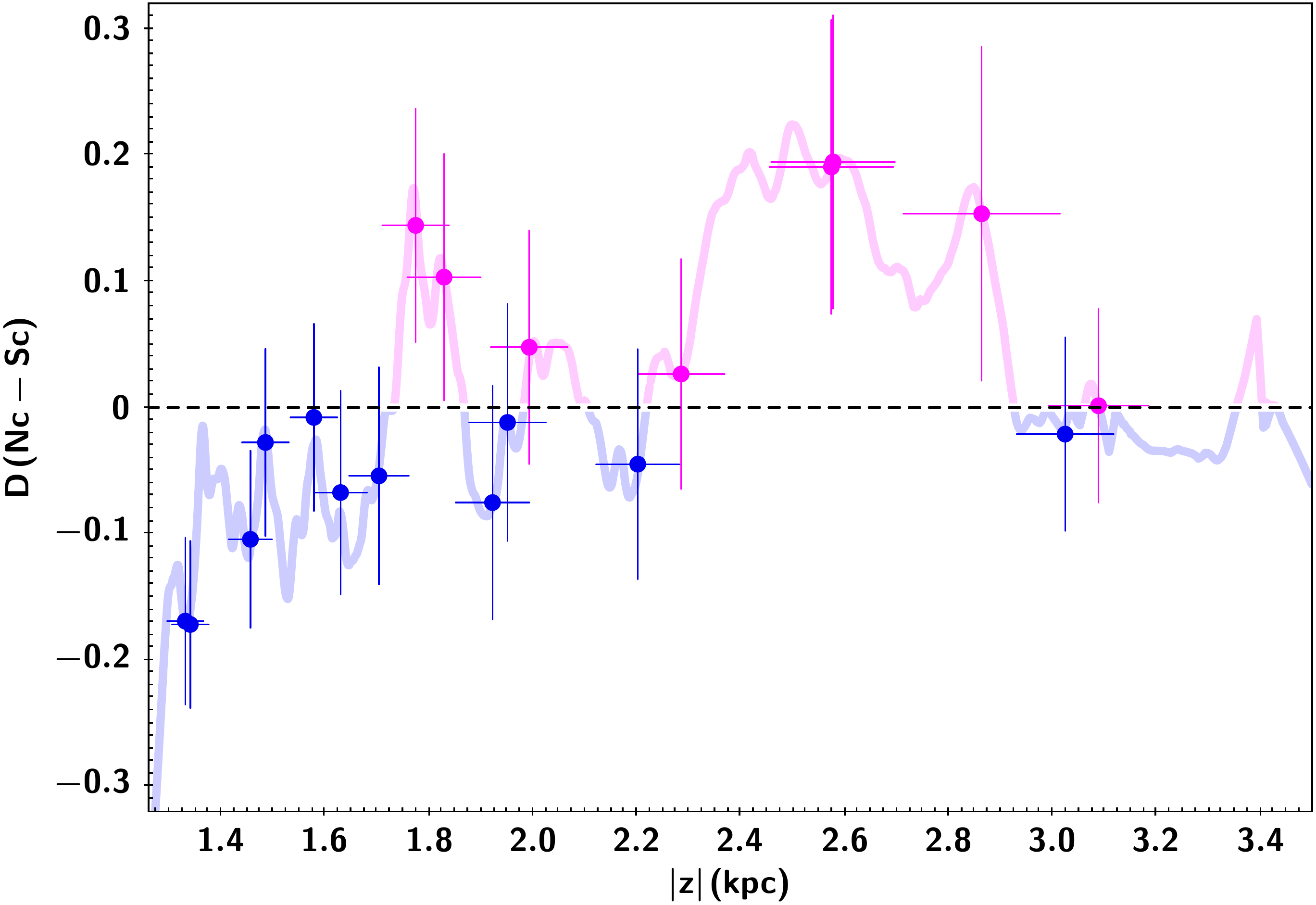} & \includegraphics[width=8.78cm]{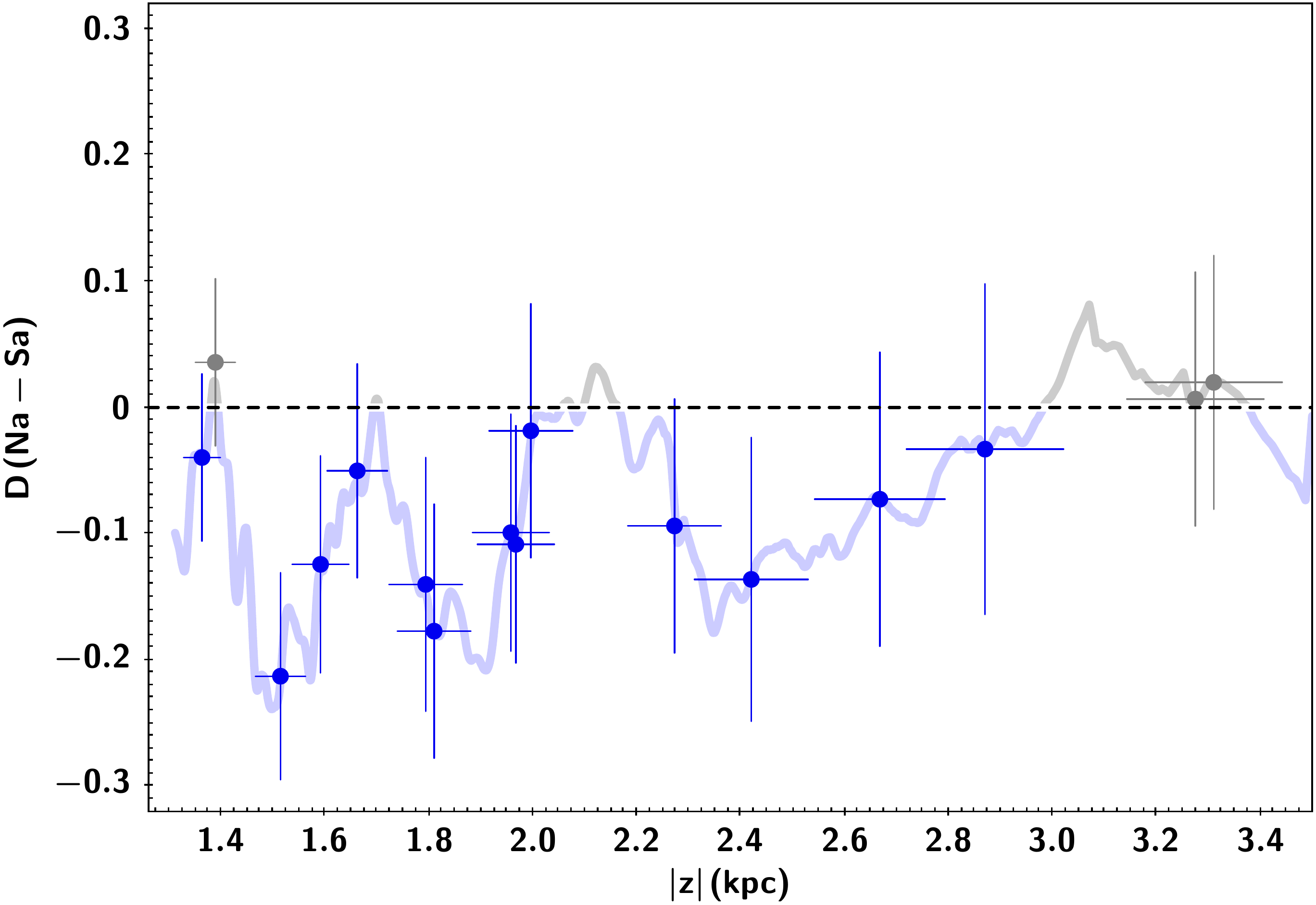} \\
   \end{tabular}
\caption{ Same as Figure~\ref{fig_dffdens_NS} but for the sectors shifted in the direction of the Galactic centre (Nc and Sc) on the left and for sectors shifted toward the anti-centre (Na and Sa) on the right.}

         \label{fig_diffdens_NcSc_NaSa}
   \end{figure*}

If stars are sorted in $|z|$ increasing, from $j = 1$ to $j = N_{stars}$, in each bin $i$, $h_i$ is calculated as the absolute distance between the $j^{th}$ star and the $j +49^{th}$ star.
This provides us with an adequate assessment of density variations (see shallow solid lines in Figures~\ref{fig_dffdens_NS},~\ref{fig_diffdens_NoSo_NrSr}, and~\ref{fig_diffdens_NcSc_NaSa}). The volume, $V_i$ , of a bin is defined by the minimum and maximum radii $R_{min}$ and $R_{max}$, the minimum and maximum angles $\phi_{min}$ and $\phi_{max}$ of the sector to which it belongs, and $h_i$ , such that $V_i = \frac{1}{2}  (\phi_{max}-\phi_{min})\times (R_{max}^2-R_{min}^2) \times h_i$.

In a second stage (also to get reliable Poisson-like statistics), we derive the densities again, but this time keeping a constant width for bins (denoted $\langle b \rangle$). This fixed width is given by the first method as the average width on a considered range. Given the large range covered by the density, we have to define four regions to derive $\langle b \rangle$ in order to get a number of stars ($n_*$) per fixed bin that is neither too high nor too low (actually between $\sim$20 and $\sim$100 with a median at 50 stars): $|$z$|$ $\le$ 1.3 kpc; 1.3$<$ $|$z$|$ $\le$1.9 kpc; 1.9$\le$ $|$z$|$ $<$2.5 kpc; 2.5 kpc $\le$  $|$z$|$. Densities derived with this second method are represented as dots in Figures~\ref{fig_dffdens_NS},~\ref{fig_diffdens_NoSo_NrSr}, and~\ref{fig_diffdens_NcSc_NaSa}.

Top panels in Figures~\ref{fig_dffdens_NS},~\ref{fig_diffdens_NoSo_NrSr}, and~\ref{fig_diffdens_NcSc_NaSa} present the vertical density profiles, in stars per kpc$^3$, up to $z=3.5$~kpc in each of the sectors, that is, sub-samples in the north and south: (i) for the central sectors; (ii) for the sectors shifted in the opposite direction of the Galaxy rotation and in the direction of the rotation; and (iii) for the sectors shifted towards the Galactic centre and anti-centre. The lower limits on each of these plots are related to the geometry of the selection described in Section~\ref{subSampleSection} and the upper limits are due to the poor statistics at overly large heights above the plane. Errors bars are illustrating one standard deviation of Poisson noise uncertainties, and our typical tolerance of $\sim$20\% for the distance determinations (see Section~\ref{sampleSelection}), $\epsilon_{z}=0.2|z|/\sqrt{n_*}$, and $\epsilon_{\nu}=\nu/\sqrt{n_*}$. 

A rough general view of the vertical profiles in the three figures exposes an identical exponentially decreasing trend for all volumes. Sub-samples at the same Galactic radius tend to overlap. As expected, there exists a radial gradient in density, visible in Figures~\ref{fig_diffdens_NcSc_NaSa} and~\ref{fit2ExpSp}, since volumes at a smaller radius (Nc and Sc) are denser than their counterparts at a larger radius (Na and Sa). The associated scale heights and scale length are studied in Section~\ref{subSectScalehl}.

Nevertheless, the density distributions in the north and south are not really identical. It has already been shown, for instance, by \citet{bennett19} that density differences are significant at $z<1$~kpc (under-density of 40 \% in the north compared to the south at $z \sim 500$~pc, which is not analysed here given our selection function). These differences are directly related to the spiral shape visible in phase space ($z,V_z$) as identified by~\cite{Antoja18}. This clearly highlights the non-stationarity near the galactic plane. 

When closely inspecting the north and south volumes, several characteristic areas can be identified with deviations from stationarity and symmetry at larger heights. To better visualise the north-south asymmetries, the bottom panels of Figures~\ref{fig_dffdens_NS},~\ref{fig_diffdens_NoSo_NrSr}, and~\ref{fig_diffdens_NcSc_NaSa} present the normalised density differences  in opposite hemispheres \citep{Widrow12, bennett19}:
\begin{equation}\label{eqDiff}
    \rm{ D(N-S) = \frac{\nu_N - \nu_S}{\nu_N+\nu_S} }.
\end{equation}

Firstly, a global mild northern over-density, already hinted at by \citet{bennett19}, ranging from $z \sim 0.6$ to $\sim$1 kpc, can be seen in the top panel of Figure~\ref{fig_dffdens_NS}, corresponding to the central sector. Given our selection function, this northern over-density cannot be followed in the other sectors. Furthermore, up to 3.5 kpc, every profile is perturbed in every sector, but generally more so in the northern hemisphere than in the south. Southern profiles tend to generally be smoother, with the notable exception of an interesting bump in the south at $z \simeq -2.1$~kpc, in the sector corresponding to the direction of Galactic rotation. Northern profiles are rather bumpy in all sectors. At large heights in the central sectors, the north ends up dominating. Interestingly, in the direction of Galactic rotation and the direction opposite to Galactic rotation, it is rather the southern density that tends to dominate at large heights, indicating a global azimuthal wobble, with a northern peak at the Solar azimuth, preceded and followed by southern peaks in the directions of rotation and counter-rotation. A similar wobble can be noted in the radial direction, with a global northern domination at large heights in the inner and central sectors, followed by a southern domination in the anti-centre sector. These indicate clearly the wave-like nature of the general north-south asymmetries, with a slightly shorter wavelength in the azimuthal direction.

The more localised wobbles and bumps in the vertical density can be classified into three categories:
In the first, some bumps and wiggles are jointly occurring in the north and south, and, therefore, they are not visible on the bottom panels (representing the asymmetry). A noticeable example is a correlated bump around $|z| \simeq 1$ kpc in the central (N, S) sector. One can again find such small correlated bumps at $|z| \simeq 1.8$~kpc in the direction of rotation (Nr, Sr), at $|z| \simeq 1.95$~kpc in (No, So), at $|z| \simeq 1.65$~kpc in (Na, Sa), or at $|z| \simeq 2.05$~kpc and 2.2~kpc in (Nc, Sc).

The second kind of disturbances are revealed by some bumps and wiggles which are clearly north-south antisymmetric, meaning that they represent anti-correlated oscillation of the density. This is clearly visible in the central sectors (N, S) at $|z| \sim$ [1.15, 1.25], [1.25, 1.35], and [2, 2.2]~kpc. These anti-correlated features are creating large asymmetries in the bottom panels of the figures. In the anti-centre (Na, Sa), we find them at [1.4, 1.7], [1.7, 2], and [2.25, 2.95]~kpc. In the sectors toward the Galactic centre (Nc, Sc), we find one large anti-correlated feature at [2.3, 2.9]~kpc. In the sectors opposite to the Galactic rotation (No, So), we find them at [1.35, 1.55], [1.75, 2.05], and [2.2, 2.45]~kpc. Finally in the direction of rotation, (Nr, Sr), we find small anti-correlated features at 1.65~kpc and $2.55$~kpc. Obviously, such  features are present in all sectors between 2 and 3 kpc, but vastly more so in the radial directions (centre and anti-centre), with only a tiny asymmetry in the three sectors at $R=R0$.

The last noticeable features are some bumps which occur only in one hemisphere. For instance, we can note in the central sectors (N, S) that the southern density has a plateau which globally dominates over the north at [1.4-1.8]~kpc, while the north goes through a bump at 1.6~kpc, and then another one at 1.8~kpc to catch up with the south. These bumps have no counterpart in the south. The most striking of these bumps in only one hemisphere is the one at 2.1~kpc in Sr. A similar, less prominent, bump is present in the north at 1.7~kpc in Nc.

\subsection{Scale heights and scale length}\label{subSectScalehl}

After having obtained the detailed density profiles in different sectors, we now want to describe their global variations. The goal here is to describe the global density scale height and scale length, which will be later used in Jeans equations. Given the north-south asymmetries noted previously, the two Galactic hemispheres will be treated separately. We want to insist that the purpose here is not to give physically fully reliable parameters for those scales but, rather, to derive approximate analytical expressions both vertically and radially, which will be useful for a Jeans analysis. In other words, we aim to parametrise, in the best possible way, the general trend of the density evolution of red clump stars with |z| between 0.6 and 3.5 kpc. At the same time, the radial variation will be also parametrized but only with three bins, namely, the central value of each radial sector ($R_0-0.8$kpc, $R_0$, $R_0+0.8$kpc).

Several analytical expressions based on exponential or secant square functions are tested to simultaneously fit radial and vertical profiles (using from three up to nine free parameters, taken from multiple components with exponential and secant square profiles, with or without scale-length). We use the Markov Chains Monte Carlo (MCMC) technique to adjust our data with analytical expressions. The best likelihood ($\cal{L}$) stemming from MCMC chains and the number of free parameters (k) in each analytical expression make the application of the Akaike information criterion (AIC, \citealt{Akaike74}) possible: $AIC = 2k -2 \ln(\cal{L})$. The following expression using an exponential law in radius and height plus one in height has been retained:
\begin{equation}\label{eq2expSp}
        \nu(R,z) = \alpha_{\nu} \exp{ \left( - \frac{|z|}{h_{za}} \right)} + \beta_{\nu} \exp{ \left( - \frac{|z|}{h_{zb}} -\frac{R}{h_{R}} \right)}.
\end{equation}
The MCMC technique is employed with the five free parameters with flat priors: two normalisation factors $\alpha_{\nu}$, $\beta_{\nu}$, two scale heights $h_{za}$ and $h_{zb}$, and the scale length $h_{R}$. 

When maximising the likelihood, we assume the data follow Poisson statistics. Accuracies on the height and length scales are constrained by the amplitude of the density deviations from the adjusted model. Accuracy is not related to the model, which only partially adjusts the counts, but is dominated by variations between the different sub-samples. The measurement of this accuracy is important since it is the dominant source of uncertainty for measuring the vertical and horizontal force fields. A key advantage of using a MCMC method is the production of a probability density function for each output parameter, providing access to realistic uncertainties (for the posterior distribution correlation matrix, see Appendix~\ref{correldensityProfileAnnex}).

\begin{figure}[!ht]
   \centering
            {\includegraphics[trim = {0cm 1.7cm 0cm 0cm}, clip, width=\hsize]{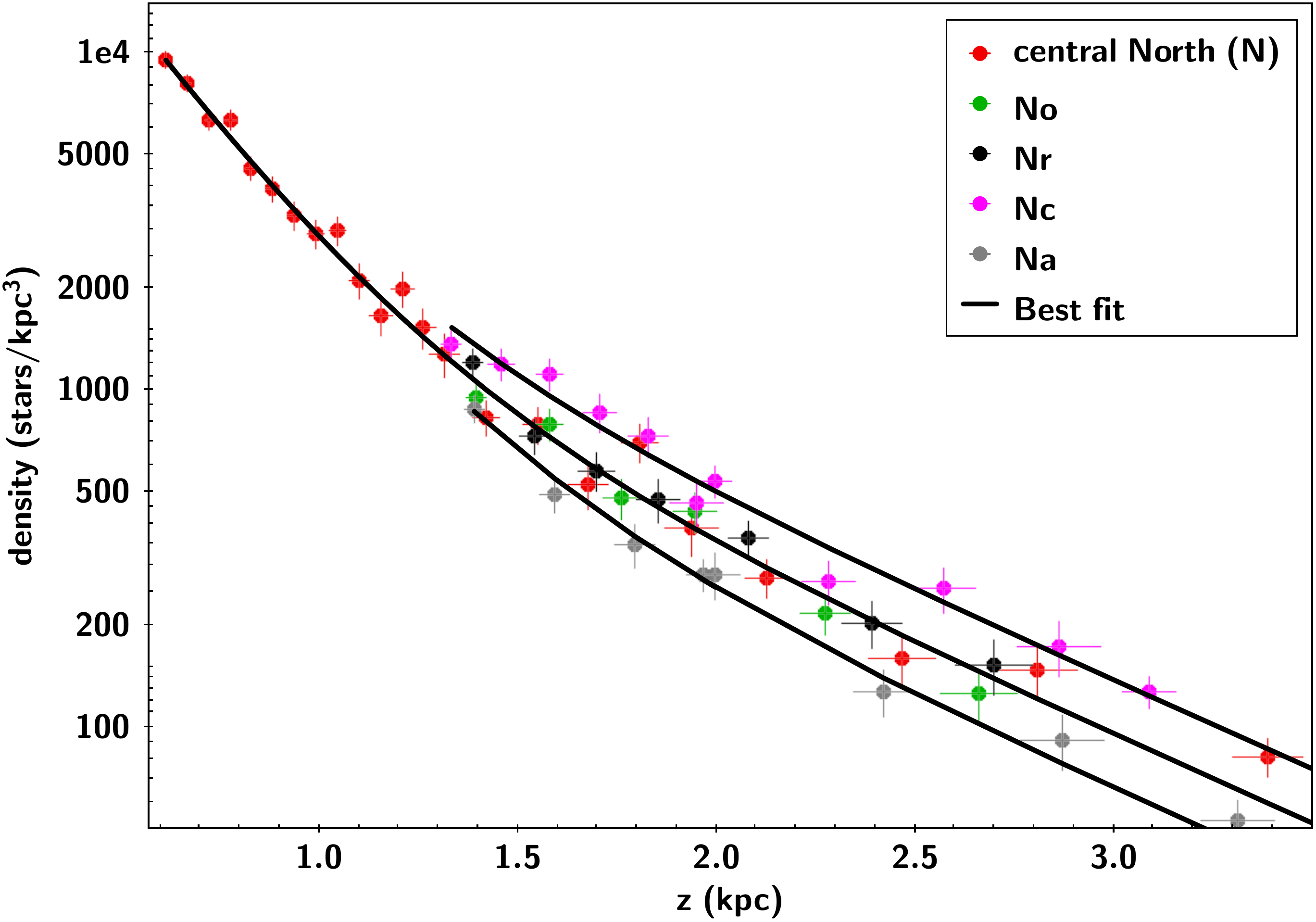}
            \includegraphics[width=\hsize]{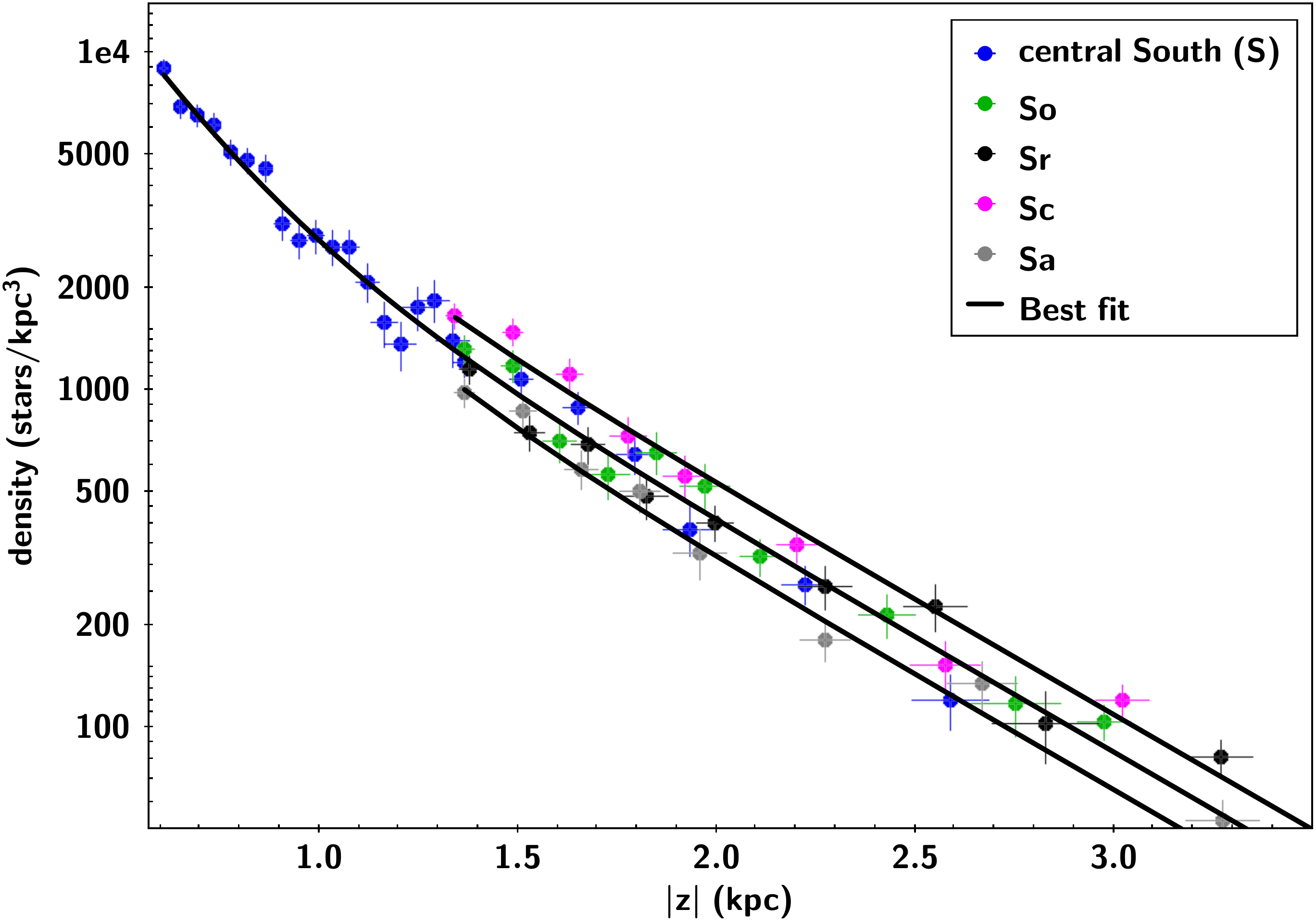}
            }
      \caption{Vertical density profiles binned in $|z|$ for the northern (top panel) and southern (bottom panel) sub-samples, superimposed with the best-fit profile at the three $R$ values (upper line: $R_0-0.8$ kpc, middle line: $R_0$, lower line: $R_0+0.8$ kpc). The colour code is the same as in Figure~\ref{subSample}}
         \label{fit2ExpSp}
   \end{figure}

Profiles are simultaneously adjusted from $|z|$ equal 0.6 kpc up to 3.5 kpc, and for $R$ = [7.54,8.34,9.14] kpc for each hemisphere. Maximum likelihood parameters and their respective uncertainties are given in Table~\ref{table2expSp} for the north and the south. The vertical density profiles are shown in Figure~\ref{fit2ExpSp} for the  north (top panel) and south (bottom panel) sub-samples. The best-fit density profile is superimposed with black lines for the three values of $R$.

\begin{table*}[!ht]
\caption{\label{table2expSp} Best-fitting parameters of the density variations. }
\centering
\begin{tabular}{lccccc}
\hline\hline
            & $\alpha_{\nu}$ $(\times10^4)$ & $h_{za}$                      & $\beta_{\nu}$ $(\times10^3)$ &   $h_{zb}$                    &  $h_R$         \\
            &   (star/kpc$^3$)              & (kpc)                         & (star/kpc$^3$)               & (kpc)                         & (kpc)          \\
\hline
North       &  8.096$^{+ 0.270}_{- 0.207}$  &  0.262$^{+ 0.003}_{- 0.004}$  &  3.556$^{+0.485}_{-0.337}$   &  0.827$^{+ 0.017}_{- 0.016}$  &  2.162$^{+0.068}_{-0.074}$  \\
South       &  8.108$^{+ 0.779}_{-0.351}$   &  0.219$^{+ 0.004}_{- 0.008}$  &  9.291$^{+0.994}_{-0.856}$   &  0.638$^{+ 0.007}_{- 0.012}$  &  3.113$^{+0.140}_{-0.091}$  \\
\hline
\end{tabular}
    \tablefoot{
{ Parameters from Equation~\ref{eq2expSp} derived from the MCMC method for two sets of volumes: the five in the north and the five in the south. The parameters $\alpha_\nu$ and $\beta_\nu$ are the normalisation factors such that the density of stars at zero height and solar radius is $\nu(R_0,0) = \alpha_\nu + \beta_\nu$. The parameters $h_{za}$ and $h_{zb}$ are the scale heights and $h_R$ is the scale length.}
}
\end{table*}

As noted in Section~\ref{zrProfilesSection}, density distributions differ by Galactic hemisphere. The scale length in the north (2.162 kpc) is shorter than in the south (3.113 kpc). Two well-defined scale heights have been found, namely, 0.219 and 0.638 kpc for the south and 0.262 and 0.827 kpc for the north. The difference between these two scale heights, $h_{za}$ and $h_{zb}$ , is larger for the north (565 pc) than for the south (419 pc), which may indicate that the density transition in the north between the two vertical density components is sharper or more pronounced than in the south. Although our goal is not to make a detailed thin-thick disc decomposition, which would not be useful without additional chemical information, these two components may, nonetheless, be broadly seen as the geometrical thin and thick discs. It is, thus, reassuring that the values are in broad agreement with the literature both for the scale heights and the scale length (see e.g. \citealt{Juric08,Robin14}). Because of the way we built the sample, we do not have information about the radial density variation below $\sim$1.3 kpc in |z| (see Table~\ref{sectors}): therefore, we cannot constraint the scale length of the thin disc and the scale-length we obtained is more representative of the thick (or thicker) disc component. Extrapolating our fits towards the plane, we have a local `thick-to-thin' disc density normalisation ($\beta_{\nu}/\alpha_{\nu}$) of about 4 \% in the north and 11 \% in the south.

In this section, we derive density distributions and their gradients by adjusting simple functions over extended intervals and defining the most likely values for the scale heights and scale length. It should be noted that deviations from the mean values are greater than expected from random statistical fluctuations and serve as indicators of non-stationarity and of the differences between the northern and southern sectors. We also note that due to this state of affairs, the formal errors on the parameters in Table~2 should only be taken as lower limits, since the reduced $\chi
^2$ values are larger than 1 (1.12 and 1.90 in the north and south).
This may also partly explain the differences in measured galactic vertical forces obtained in works prior to the publication of the \gdrtwo\ catalogue. We go on to explore how velocity dispersions differ between the north and south.

\section{Velocity dispersion gradients}\label{sectionVelocity}
\begin{figure}[!ht]   
   \centering
            \includegraphics[width=\hsize]{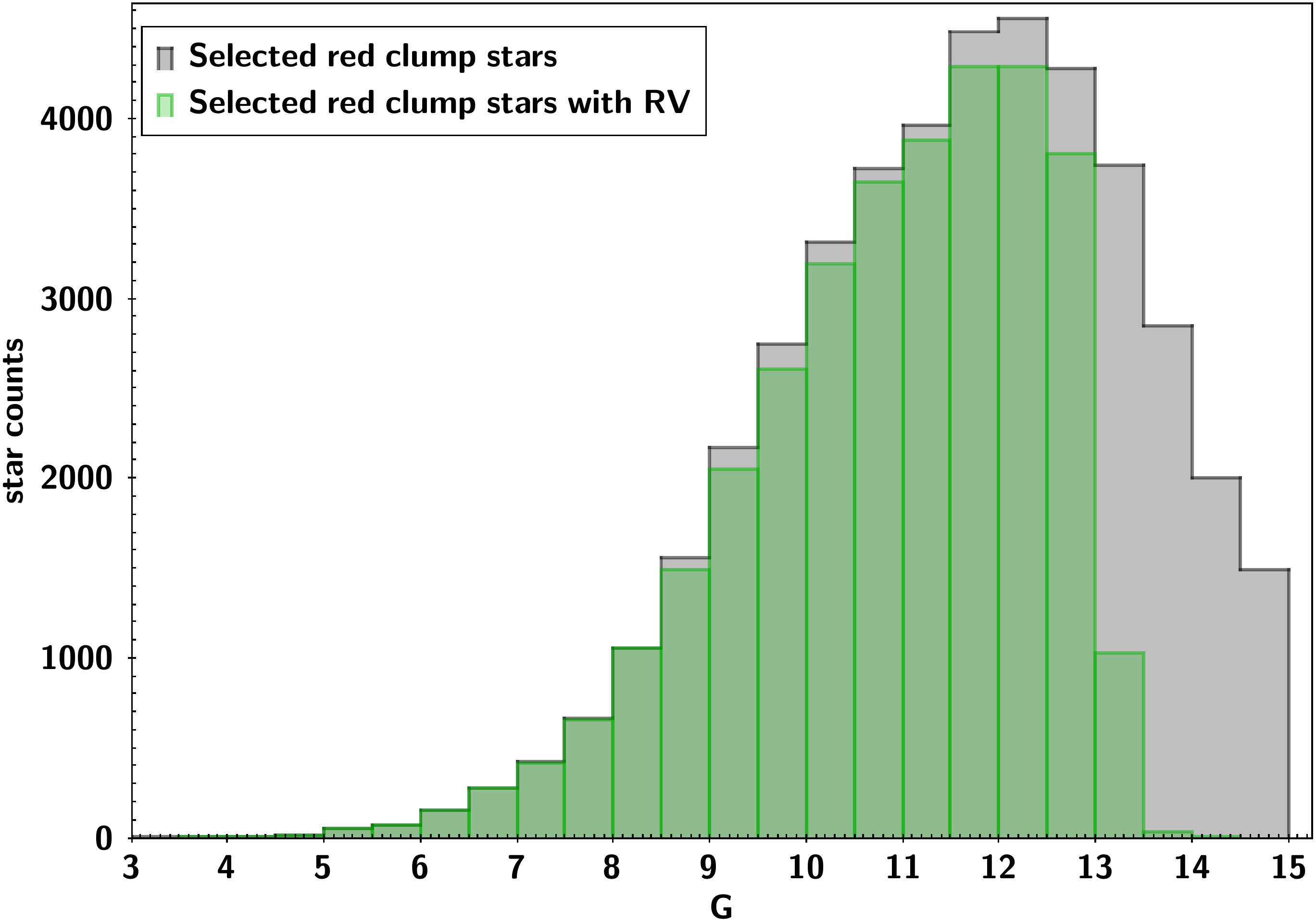}
      \caption{Distribution of red clump samples designed for this study with respect to their \gaia $G$ magnitude. The grey distribution is the main sample used to derive the density profiles (Section~\ref{sectionDensProf}). Distribution in green is a selection from the main sample where only stars with radial velocity measurements are considered in order to obtain velocity dispersion profiles (Section~\ref{sectionVelocity}).}
         \label{figmagRV}
   \end{figure}

To be able to derive gravitational forces under equilibrium assumptions from Jeans equations, velocity dispersion variations are needed. Of course, only stars with line-of-sight velocity data can be used for this. The selection will thus contain only objects with six dimensional phase space information from the {\it Gaia} RVS. About 76\% of the sample that we previously used now remains, consisting of 33 005 red clump giants (15702 in the north, 17 303 in the south). Actually, the radial velocity condition to build the new sample results in an additional cut in magnitude. This effect is clearly shown in Figure~\ref{figmagRV}, which represents the number of stars per bin of magnitude. The first sample of this study used to derive densities in grey is limited to $G=15$, as imposed in Section~\ref{sampleSelection}. The second sample in green, where only stars with radial velocity information are kept, is limited to a magnitude of 14 and strongly reduced up to a magnitude of 13. It is obvious that the sample used in this step is not complete (whereas the selection built for the density analysis is complete). Nevertheless, this is not of primary importance as the assumption that our partial sample follows the same dynamics as the complete sample is reasonable. Moreover, the number of stars is still large enough to get good statistics across the studied range.

\begin{figure}[!ht]
   \centering
            {\includegraphics[trim = {0cm 1.7cm 0cm 0cm}, clip, width=\hsize]{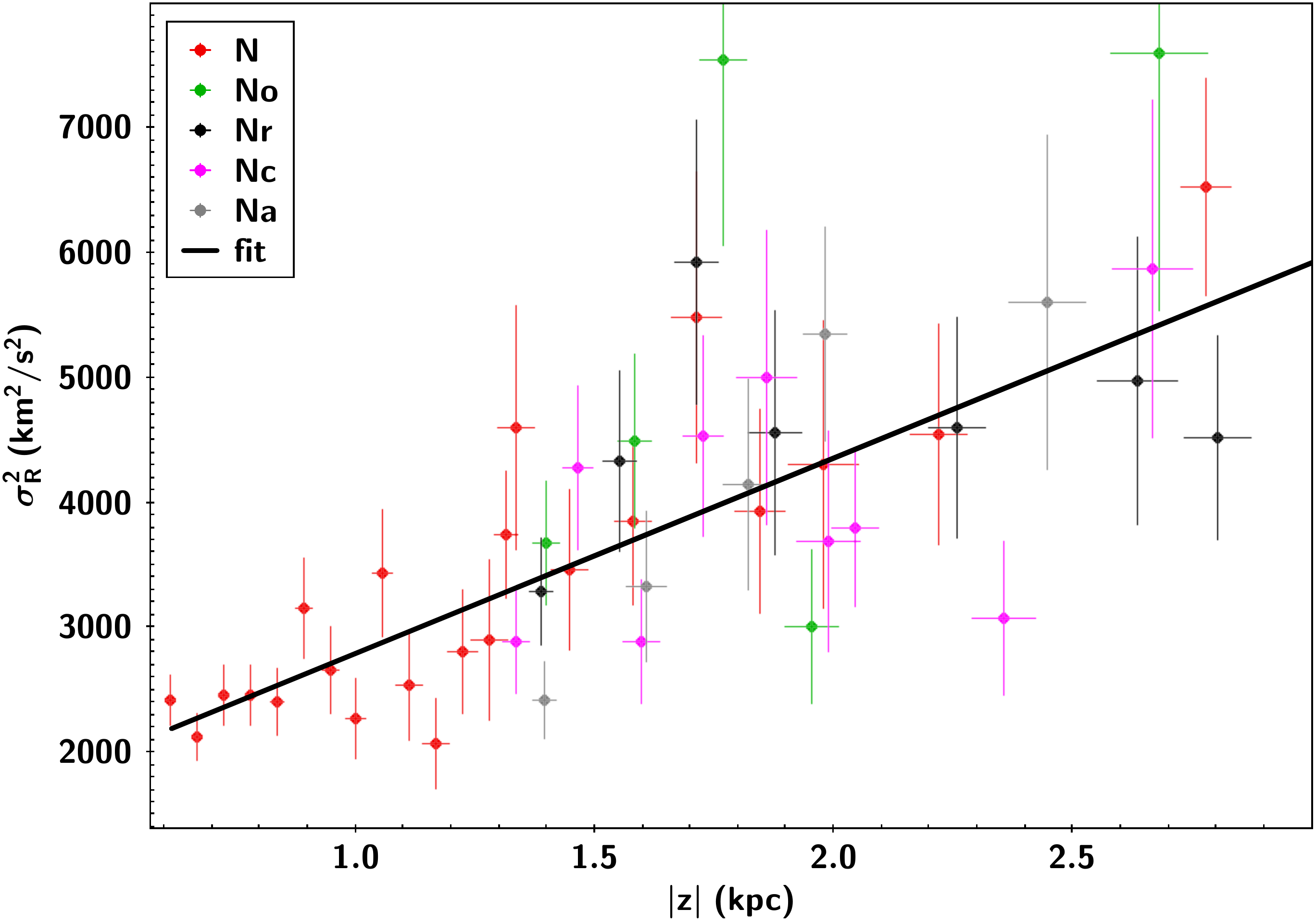}
            \includegraphics[width=\hsize]{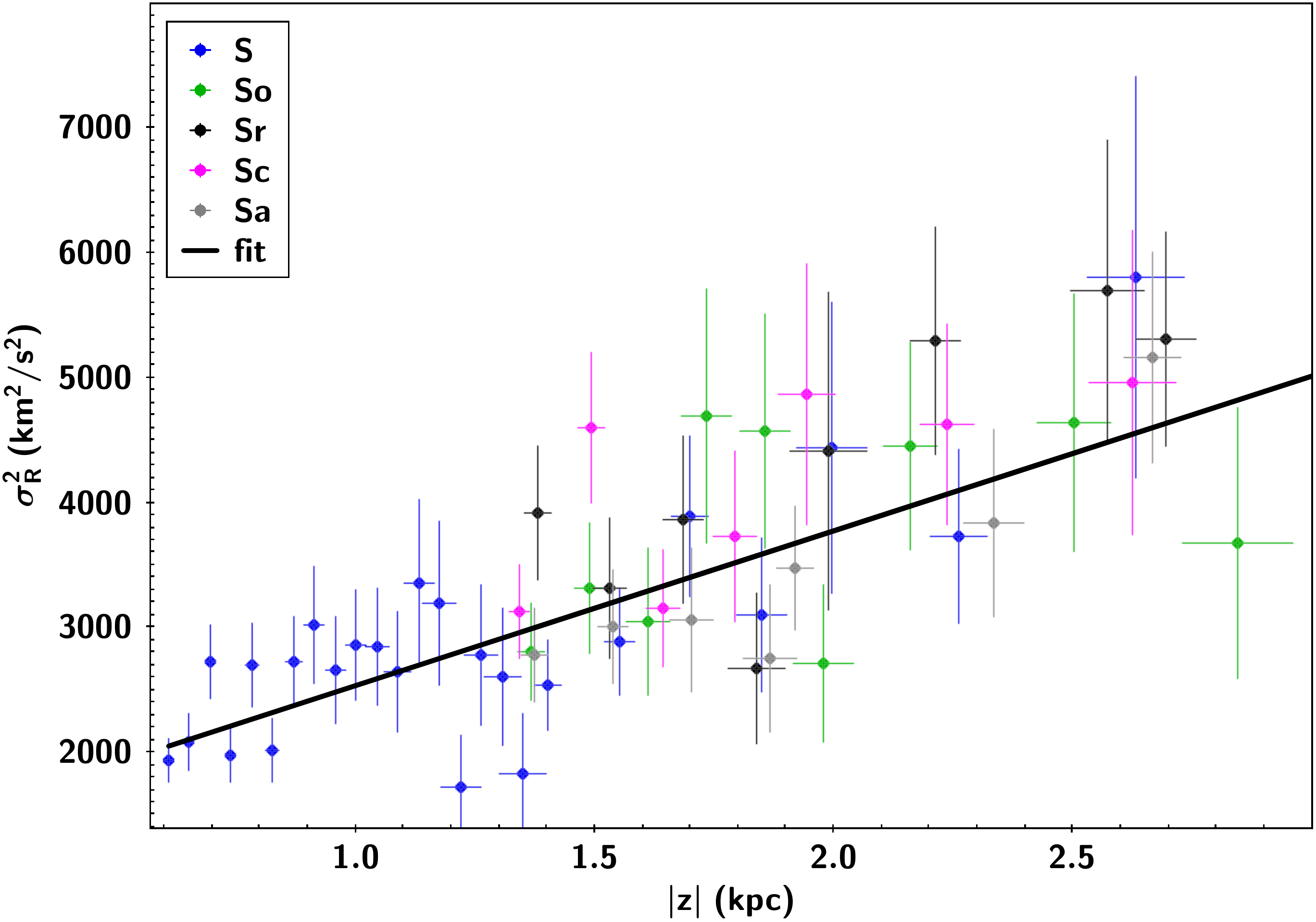}
            }
      \caption{ Vertical squared velocity dispersion on radial direction profiles binned in $|z|$, for the five sub-samples in the north (upper panel) and for the five in the south (bottom panel). The colour code per sector is the same as in Figure~\ref{subSample}. Black lines represent the best fit following Equation~\ref{eqVR} and the parameters given in Table~\ref{tableFitVR}.}
         \label{fitVr}
   \end{figure}

\begin{figure}[!ht]
   \centering
            {\includegraphics[trim = {0cm 1.7cm 0cm 0cm}, clip, width=\hsize]{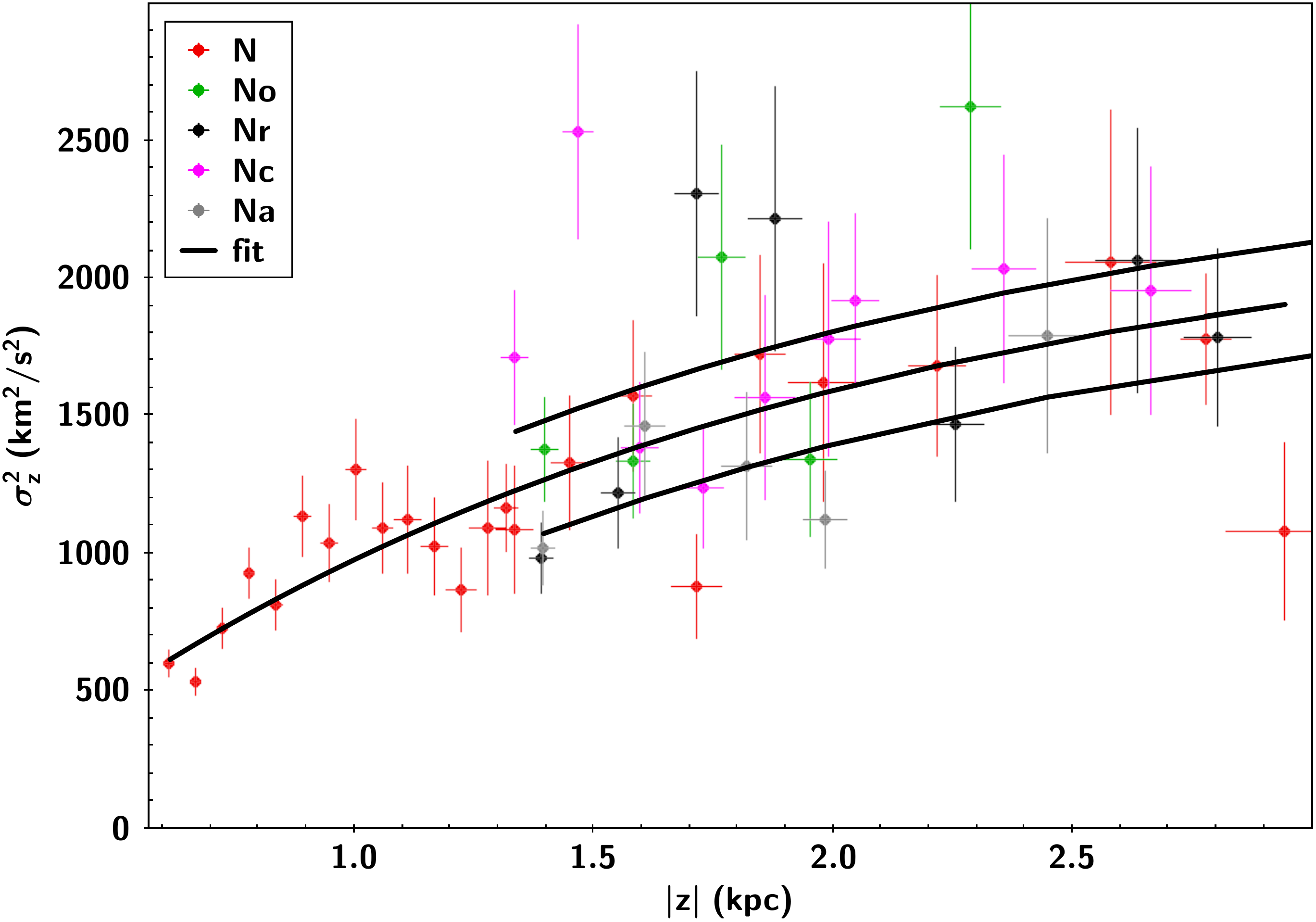}
            \includegraphics[width=\hsize]{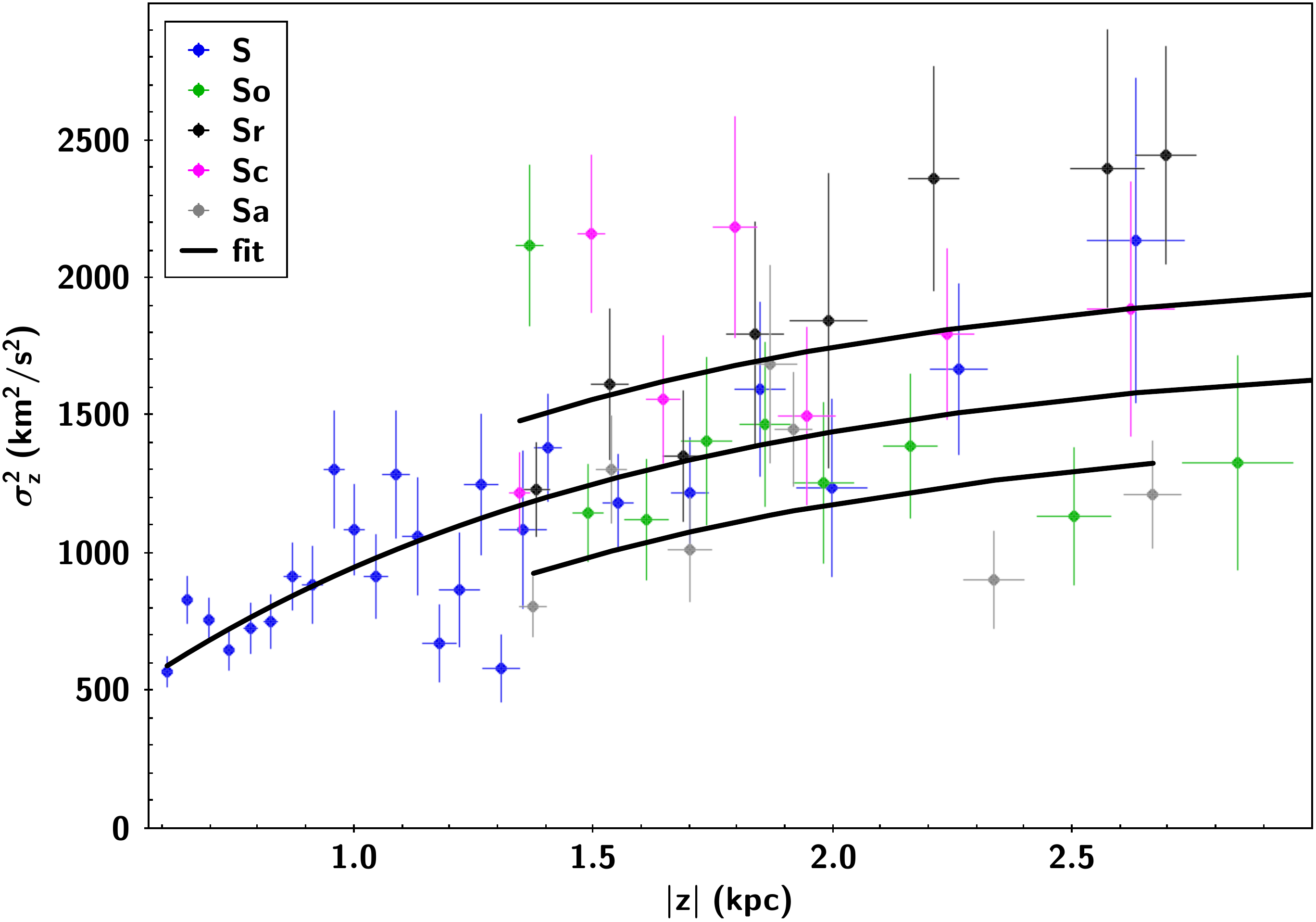}
            }
      \caption{ Vertical squared velocity dispersion on vertical direction profiles binned in $|z|$, for the five sub-samples in the north (upper panel), and in the south (bottom panel). The colour code per sector is the same as in Figure~\ref{subSample}. Black curves represent the best fit following Equation~\ref{eqVz} and the parameters given in Table~\ref{tableFitVz}, plotted at the three $R$ values (7.54, 8.34, and 9.14 kpc from the highest to the lowest).}
         \label{fitVz}
   \end{figure}

We then apply the same space partition in sectors inside the conic selection as in Section~\ref{subSampleSection}. Similarly to the way the vertical density was derived, data are binned along vertical direction keeping a constant number of 50 stars per bin in a first step. Then, the average width $\langle b \rangle$ is calculated in the four ranges in z as defined in Section~\ref{zrProfilesSection} in order to derive the density in bins of fixed width (see dots in Figures~\ref{fitVr} and~\ref{fitVz}.) 
We adopt the conventional Galactic cylindrical coordinates ($R,\phi,z$) defined in Section~\ref{sampleSelection}. Velocities are converted in this frame (as $V_R,V_{\phi},V_z$) from \gdrtwo\ proper motions and radial velocity ($\mu_{\alpha}, \mu_{\delta}, V_{los}$) taking the same solar velocity employed in \cite{Katz18}: 240 km s$^{-1}$ for the circular velocity \citep{Reid14} and $(U_\odot, V_\odot, W_\odot) =$ (11.1,12.24,7.25) km s$^{-1}$ for the Sun's peculiar velocity with respect to the local standard of rest \citep{Schonrich10}.

 To calculate the velocity profiles along z in each direction ($v_R(z)$,$v_{\phi}(z)$,$v_{z}(z)$), a $\sigma$ clipping method is applied. For each bin with $n_*$, the median and the mean velocity and the corresponding velocity dispersion ($\sigma_j$) are calculated. Values lying beyond 4$\sigma$ from the median are rejected. Then the mean, the median, and the dispersion are re-calculated. The sequence is repeated until no values are rejected. We carefully applied this process in our study to eventually eliminate strong outlier values coming from extreme velocity stars, but in the end, less than 0.5\% of stars were excluded. 

Velocities per bin, on each axis j, are then calculated as:
\begin{equation}
\langle v_j \rangle = \frac{1}{n_*-n_{ex}} \sum^{n_*-n_{ex}}_{i=1} v_j(i)
,\end{equation}
where $n_{ex}$ is the number of stars excluded by the $\sigma$ clipping. Their corresponding velocity dispersions are derived with the following equation:
\begin{equation}
\sigma_{j}^2 = \frac{1}{n_*-n_{ex}} \sum^{n_*-n_{ex}}_{i=1}\big(v_j(i) - \langle v_j \rangle \big)^2
.\end{equation}
Figures~\ref{fitVr} and~\ref{fitVz} show the distribution of the squared velocity dispersions $\sigma_R^2$ and $\sigma_z^2$ in $R$ and $z$ directions, respectively, with respect to the Galactic height. The colour code is the same as in Figure~\ref{subSample}.
Error bars are representing the classical one $\sigma$ standard deviation, on 20\% of the distance along z ($\epsilon_{d}=0.2|z|/\sqrt{n_*}$) and on the squared dispersion ($\epsilon_{\sigma_R^2}=\sigma_R^2\sqrt{2/(n_*-1)}$ and $\epsilon_{\sigma_z^2}=\sigma_z^2\sqrt{2/(n_*-1)}$).

Once vertical dispersion profiles are built, their variations have to be characterised. A similar MCMC machinery as described in Section~\ref{subSectScalehl} is used to approximate the different profiles. In the same way that we searched for the best possible simple function to fit density profiles using the Akaike information criterion values, we now test several functions to simultaneously fit the dispersion profiles of each hemisphere based on combinations of exponential and linear functions for $|z|$ between 0.6 and 3.5 kpc. To express the squared vertical dispersion in $R$, we ultimately use a simple linear equation:
\begin{equation}\label{eqVR}
\sigma_R^2(z) = a_R |z| + b_R
,\end{equation}
where a$_R$ is the slope in km$^2$ s$^{-2}$ kpc$^{-1}$ and b$_R$ is the normalisation in km$^2$ s$^{-2}$. The best-fitting parameters are reported in Table~\ref{tableFitVR}. 
 \begin{table}[!ht]
\caption{\label{tableFitVR} Best-fitting parameters of the vertical profile for the radial velocity dispersion squared.}
\centering
\begin{tabular}{lcc}
\hline\hline
 $\sigma_R^2$  &  $a_R$ ($\times 10^3$)      &  $b_R$ ($\times 10^3$) \\
             &  km$^{2}$ s$^{-2}$ kpc$^{-1}$ &  km$^2$ s$^{-2}$      \\
\hline
North       &  1.564$\pm0.010$               &  1.230$\pm0.017$       \\
South       &  1.239$\pm0.009$               &  1.297$\pm0.016$       \\
\hline
\end{tabular}
\tablefoot{ Parameters obtained from Equation~\ref{eqVR} for the northern and southern sub-samples. The parameter $a_R$ is the slope of the gradient, $b_R$ is the normalisation factor.}
\end{table}

We notice there is no need to introduce a radial dependency since the likelihood is not significantly improved when taking into account the radial information. But we may have to concede that the number of bins is more limited than for the density (as the sample is 25\% smaller). Consequently, a potential radial gradient would be detected only if it is obvious.
On top of this, the scale length of the velocity dispersion is expected to be large and it is, therefore, not unsurprising that we do not detect it. We note, however, that incorporating the uncertainty associated with a possible radial gradient as a nuisance parameter would have led to larger error estimates.

A vertical gradient exists in $\sigma_R^2(z)$, which is stronger in the north. Moreover, in this hemisphere, the dispersion around the fit is much larger than in the south with strong bumps and wiggles from one bin to another. It suggests once again that the Galaxy is not in a stationary state and is far from being north-south symmetric.

To fit the squared vertical velocity dispersion in $z$, the best choice turns out to be the equation:
\begin{equation}\label{eqVz}
\sigma_z^2(R,z) = a_z \exp\big({-R/h_{\sigma_{za}^2}\big)} + b_z \exp\big({-|z|/h_{\sigma_{zb}^2}\big)}
,\end{equation}
where $a_z$ and $b_{z}$ are normalisation factors in km$^2$ s$^{-2}$. Parameters $h_{\sigma_{za}^2}$ and $h_{\sigma_{zb}^2}$ are respectively the scale length and the scale height of $\sigma_z^2$ in kpc. We note that since the $b_{z}$ normalisation factors are typically negative, $h_{\sigma_{zb}^2}$ should not be considered along the lines of 'classical' scale heights. Indeed, the vertical velocity dispersion increases with height. Unlike variations of $\sigma_R^2$, there is a need for a radial gradient to best fit, $\sigma_z^2$ , in the different sectors. This is represented by the black lines in Figure~\ref{fitVz} for the three different radii considered in this study. The parameters of maximum likelihood are reported in Table~\ref{tableFitVz} along with their uncertainties. Following the same trend as the radial dispersion, the gradient of the vertical dispersion is smoother for the south. It accounts for the fact the Galaxy might be locally more disturbed in the north. We note, however, that in the direction of Galactic rotation, the southern velocity dispersions display an interesting bump at large heights, both in terms of radial and vertical velocity dispersions, which is probably related to the density bump noted in the previous section.
 \begin{table}[!ht]
\caption{ \label{tableFitVz} Best-fitting parameters of the profile for the vertical velocity dispersion squared.
} 
\centering
\begin{tabular}{lcccc}
\hline\hline
 $\sigma_z^2$ & $a_z$ ($\times 10^3$) & $h_{\sigma_{za}^2}$  &  $b_z$ ($\times 10^3$)  &  $h_{\sigma_{zb}^2}$ \\
              &  km$^{2}$ s$^{-2}$    &  kpc                 &  km$^{2}$ s$^{-2}$      &  kpc     \\
\hline
North       &  5.81$^{+ 0.23}_{- 0.18}$  &  8.93$^{+ 0.42}_{- 0.35}$   &  -2.47$^{+ 0.02}_{-0.03}$  &  1.57$^{+ 0.10}_{- 0.06}$ \\
South       &  9.45$^{+ 0.41}_{- 0.44}$       &  4.96$^{+ 0.16}_{- 0.12}$  &  -2.07$^{+ 0.03}_{- 0.03}$ &  1.06$^{+ 0.04}_{- 0.04}$    \\

\hline
\end{tabular}
\tablefoot{ Parameters obtained from Equation~\ref{eqVz} with the northern and southern sub-samples. The parameters $a_z$ and $b_z$ are the normalisation factors, $h_{\sigma_{za}^2}$ and $h_{\sigma_{zb}^2}$ are the scale length and scale height.}
\end{table}

\section{Vertical force and dark matter density using Jeans equations}\label{sectionForces}

Here, we investigate how making an equilibrium assumption separately for data in the northern and the southern Galactic hemispheres affects the estimates of the force field and gravitational potential with the Jeans equations. We mainly use the notation of \cite{SanchezSalcedo16} and concentrate on the local ('central') sector of the previous sections. The old method proposed by Oort and Kapteyn \citep{Kapteyn20,Oort32} assumed the decorrelation between vertical and horizontal motions. Here, with samples that are distant from the Galactic plane, we have to take into account horizontal and vertical couplings. We nevertheless simplify the problem as most past studies did, that is, by assuming an axisymmetric Galaxy and considering that the inclination of the ellipsoid is spherically aligned. The goal is to explore the different dynamical and dark matter density estimates that one can get in the north and south when ignoring disequilibrium. 

\subsection{Vertical forces}\label{sectionVerticalForces}
We define two global scale parameters. They can be directly calculated from the density profile derived in Section~\ref{sectionDensProf} (see Equation~\ref{eq2expSp}):
\begin{equation}\label{eqglobalHz}
    \begin{split}
    H_z(R,z) &= -\bigg( \frac{\partial \ln(\nu(R,z))}{\partial z} \bigg)^{-1} \\
        &= \frac{\nu(R,z)}{\frac{\alpha_{\nu}}{h_{za}} \exp{ \left( - \frac{|z|}{h_{za}}  \right)} + \frac{\beta_{\nu}}{h_{zb}} \exp{ \left( - \frac{|z|}{h_{zb}} -\frac{R}{h_{R}} \right)}}
    \end{split}
,\end{equation}
\begin{equation}\label{eqglobalHR}
    H_R(R,z) = -\bigg( \frac{\partial \ln(\nu(R,z))}{\partial R} \bigg)^{-1}
             = \frac{\nu(R,z)}{\frac{\beta_{\nu}}{h_{R}} \exp{ \left( - \frac{|z|}{h_{zb}} -\frac{R}{h_{R}} \right)}}
.\end{equation}
Vertical forces $K_z$ are calculated with the following equation:
\begin{equation}
    \begin{split}
        K_z(R,z) = \frac{-\sigma_z^2(R,z)}{H_z(R,z)} + &
        \frac{\partial \sigma_z^2(R,z)}{\partial z} + \\
        & \sigma_{Rz}^2(R,z) \bigg(\frac{1}{R}-\frac{1}{H_R(R,z)} \bigg) +
        \frac{\partial \sigma_{Rz}^2(R,z)}{\partial R}
    \end{split}\label{eqFz}
,\end{equation}
The velocity dispersion covariance term is given by:
\begin{equation}\label{eqVRZ}
\sigma_{Rz}^2(R,z) = \frac{R z (\sigma_R^2(R,z) - \sigma_z^2(R,z))}{R^2-z^2}.
\end{equation}
Based on Equations \ref{eqVR}, \ref{eqVz}, and \ref{eqVRZ}, we get:
\begin{equation}
    \begin{split}
        \frac{\partial \sigma_{Rz}^2(R,z)}{\partial R} = \bigg( \frac{z}{R^2-z^2} \bigg) \Bigg[ \bigg( \frac{R^2+z^2}{R^2-z^2} \bigg) \Big( \sigma_z^2(R,z) - \sigma_R^2(R,z) \Big) + \\
        \bigg( \frac{a_z R}{h_{\sigma_{za}^2}} \exp \big({-R/h_{\sigma_{za}^2}}\big) \bigg) \Bigg]
    \end{split}
.\end{equation}

To latter derive the vertical forces, we independently draw a solution in the last 20\% of the three Markov chains obtained from the fits of $\nu(R,z)$, $\sigma_R^2(z)$ and $\sigma_z^2(R,z)$. It lets us with a combination of 11 parameters ($\alpha_{\nu}$, $h_{za}$, $\beta_{\nu}$, $h_{zb}$, $h_{R}$, $a_{R}$, $b_{R}$, $a_{z}$, $h_{\sigma_{za}^2}$, $b_{z}$, $h_{\sigma_{zb}^2}$) reflecting a possible parametrization of the data. 

\subsection{Rotation curve corrections}
A correction term describing the radial forces has to be taken into account to accurately derive the surface mass density at large distance from the plane from the vertical force. 

We calculate this radial force correction term from the rotation curve as:
\begin{equation}\label{eqFR}
C_R(R,z) = \frac{2}{R} \int_0^z{V_c(R,z) \frac{\partial V_c(R,z)}{\partial R} {\rm d}z},
\end{equation}
such that the dynamical surface mass density ($\Sigma_{dyn} = \Sigma_* + \Sigma_{DM}$), equal to the stellar plus the dark matter mass surface density, is 
\begin{equation}\label{eqSigma}
    2\pi G \Sigma_{dyn}(R,z) = -K_z(R,z) + C_R(R,z).
\end{equation}
Here, $V_c(R,z)$ is the circular velocity related to the radial force $K_R$, through $V_c^2=-R K_R$. In the Galactic plane, this velocity has been set up to $V_c(R_0,0)$ = 240$\pm$8 km.s$^{-1}$ (See Section~\ref{sectionVelocity}). In order to encompass the majority of recent values from the literature (for example 229.0 $\pm$ 0.2 km.s$^{-1}$ from \citealt{Eilers19}\footnote{The main recent literature values are $236\pm3.8$ km s$^{-1}$ from \cite{Mroz19} based on the analysis of Cepheids, that of \cite{McGaugh18}, $233.3\pm 1.4$ km s$^{-1}$ based on a new analysis of the gas rotation and the determination of the distance to the Galactic centre \citep{Gravity18} or the analysis of \cite{Kawata19}, $236\pm3$ km s$^{-1}$, and \cite{McMillan17}, $232.8 \pm 3.0$ km s$^{-1}$.}) we extend the standard deviation to 1.5 $\sigma$, meaning $V_c(R_0,0)$ = 240$\pm$12 km.s$^{-1}$. 

As a consequence, we have to do the same for the solar radius by taking R$_{\odot}$ = 8.34$\pm$0.24 kpc which allows us to include in the uncertainty values from \cite{Gravity19}.

The variation of $V_c$ with |z| is constrained by the Besançon Galactic Model (BGM) potential \citep{Bienayme87}. We enforce the circular velocity to vary at the same rate as it evolves in the BGM. It approximately corresponds at the solar radius to a linear decrease such that $V_c(R_0,z) \approx V_c(R_0,0) -9.5 \times z_{(kpc)}$. 

For the radial gradient of the circular velocity, we chose to take an average value from the literature in the plane at the solar radius \citep{Kawata19,Mroz19,Eilers19}, which is constant with radius and height such that $\partial{V_c(R,z)} / \partial{R} = \partial{V_c(R_0,0)} / \partial{R} = -2.18\pm 1.82$ km.s$^{-1}$.kpc$^{-1}$.

\begin{figure}[!ht]
   \centering
            \includegraphics[width=\hsize]{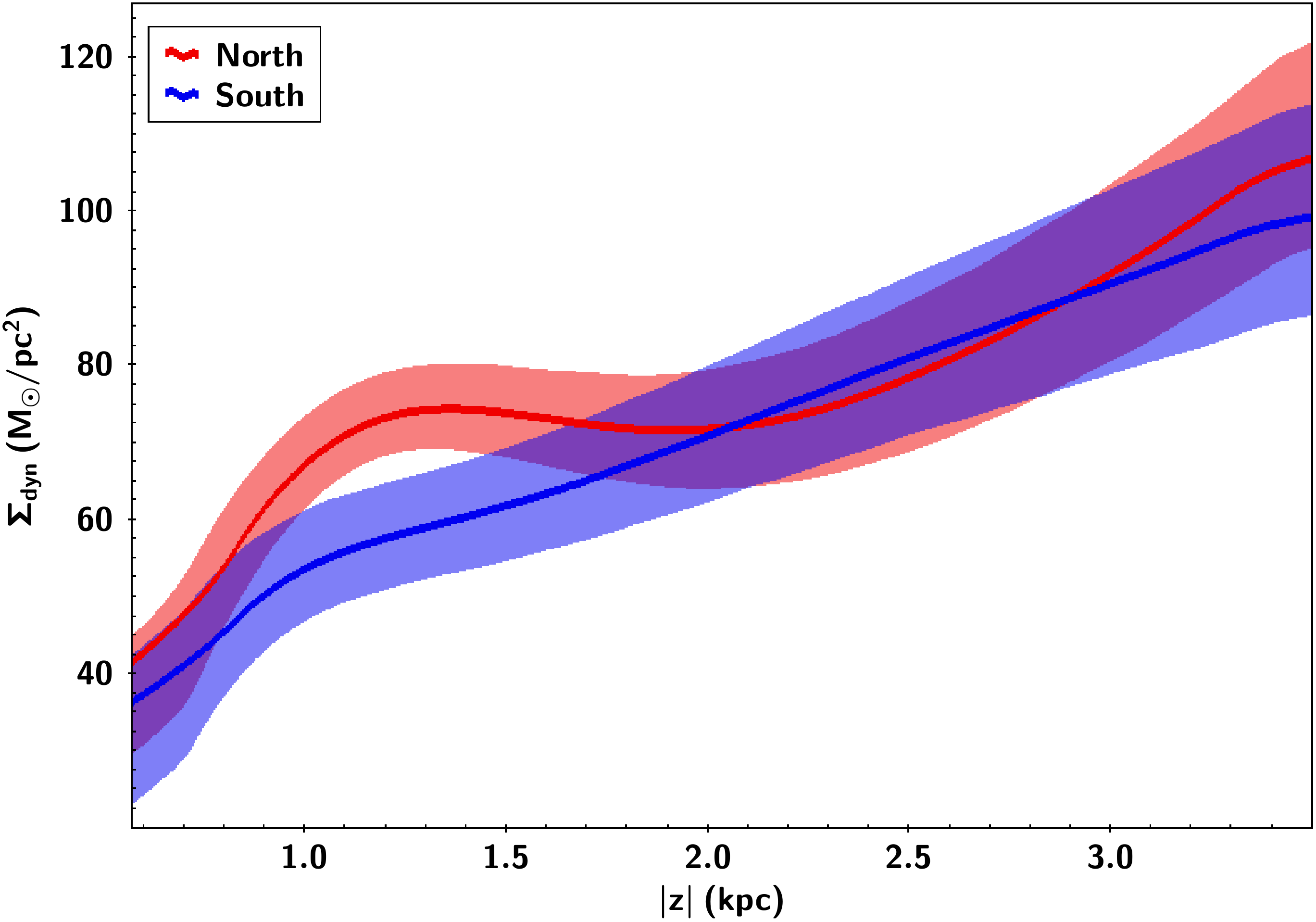}
      \caption{ Surface mass density ($\Sigma_{dyn}$) in $M_{\odot}/pc^2$ at the Solar position (Equation~\ref{eqSigma}) along the direction perpendicular to the Galactic plane ($|z|$ in kpc) for the north (red) and the south (blue). There are 10000 superimposed curves per hemisphere, which arise from the same number of draws in parameter uncertainties. The shallow red and blue shapes encompass one $\sigma$ standard deviation of solutions. The two solid lines in red and blue (approximately in the middle of each shape) represent median curves.}
         \label{figkz}
   \end{figure}

Following the dispersions of the quantities described above, we drew a set of three parameters ($R_0$, $V_c(R_0,0)$, $\partial{V_c(R_0,0)} / \partial{R}$). With the addition of the 11 parameters obtained earlier (see Section~\ref{sectionVerticalForces}), it is possible to derive both vertical (Equation~\ref{eqFz}) and radial force correction (Equation~\ref{eqFR}) with respect to galactic radius and height. We then obtain the dynamical surface mass density $\Sigma_{dyn} = \Sigma_* + \Sigma_{DM}$, from Equation~\ref{eqSigma}.
By reiterating this drawing sequence, dispersions of the forces calculation and mass surface density are natural outcomes. Figure~\ref{figkz} is depicting the evolution of the $\Sigma_{dyn}$ with respect to the height at the Solar radius for the north and south hemispheres. The width of the two curves represents the one $\sigma$ standard deviation. It stems from 10000 draws in the Markov chains for the adjusted parameters and in the standard deviation for fixed quantities.

At first glance, the north and south curves appear to be following the same trend since the two values for $\Sigma_{dyn}$ are increasing with $z$ on a similar order of magnitude. Nevertheless, the derived southern mass surface density is clearly much more regular, expressively reflecting the fact that disequilibrium are stronger in the north. Interestingly, we can immediately note that this southern surface density is smaller at low $z$, which, despite neglecting disequilibrium, does seem to reflect the slight stellar over-density in the north noted in Section~\ref{zrProfilesSection}. Then the clear slope break around 1.2 kpc for both north and south hemispheres indicates the transition between an environment largely dominated by baryons to an environment with a lower baryonic density in the thick disc, where dark matter becomes progressively dominant. Between $\sim$ 1.2 and 2.2 kpc, $\Sigma_{dyn}$ presents a kind of oscillation for the north which, despite averaging trends with simple functions, still reflects the strong northern disequilibrium noted in previous sections. Bearing in mind that unresolved systematic errors discussed in the previous sections would inevitably increase the error bars, we can nevertheless note that with our formal errors, the two deduced dynamical surface densities do not overlap at one $\sigma$ between 1 and 1.5~kpc.

\citet{Haines19} investigated, using simulations, how departures from equilibrium could affect the determination of the surface density of the disc. Their conclusion was that the derived surface densities from a perturbed disc can give similar orders of magnitude in the north and south separately while still overestimating it by a factor that can be as large as 1.5 close to the plane. Therefore those surface densities should be considered as upper limits.
\subsection{Dark matter density}\label{sectionDMdensity}
We assume here the dark matter content is dominating the baryonic part of the Galaxy at a large height from the plane, above two kpc. Hence, it gives ${\rm d} \Sigma_{dyn}/{\rm d}z = {\rm d} \Sigma_{DM}/ {\rm d}z$ for $|z| > 2$~kpc. Then it allows us to derive the dark matter mass density at these heights :
\begin{equation}
    \rho_{DM}(R_0) \approx \frac{1}{2}\Bigg<\frac{{\rm d} \Sigma_{dyn}}{{\rm d}z}\Bigg>_{[2kpc<|z|<3.5kpc]}
.\end{equation}
The density so determined is only slightly smaller (few percents at 2 kpc) than the one in the plane ($\rho_{DM}(R_0)$) under the assumption of a spherical halo.
Obviously, for a flattened oblate DM halo, the local dark matter density in the plane should then be even larger. Hence, the value derived here is a lower limit for the plane dark matter density \citep[e.g.][]{Piffl14}. 

Above 2 kpc, the variation of $\Sigma_{dyn}$ is well-approximated with a linear variation, especially in the south (see Figure~\ref{figkz}). Thus, for each of the 10000 solutions, we derive the average gradient of $\Sigma_{dyn}$ for each hemisphere. It leads us to obtain the dark matter mass densities with their associate uncertainties:
 \begin{equation}
    \begin{split}
         \rho_{DM}(R_0)_{\textrm{North}} & = 0.0134 \pm 0.0024 \, \textrm{M}_{\odot}.\textrm{pc}^{-3} \\
         &= 0.5087 \pm 0.0909  \, \textrm{GeV}.\textrm{cm}^{-3}
    \end{split}
,\end{equation}
\begin{equation}
    \begin{split}
        \rho_{DM}(R_0)_{\textrm{South}} & = 0.0098 \pm 0.0023 \, \textrm{M}_{\odot}.\textrm{pc}^{-3} \\
        &= 0.3736 \pm 0.0871 \, \textrm{GeV}.\textrm{cm}^{-3}
    \end{split}
.\end{equation}

The choice of 2~kpc as the lower limit is somewhat arbitrary, so it is interesting to check what we obtain for 1.5~kpc and 2.5~kpc, respectively. Interestingly, in the south, the density is fully unaffected, 0.0098 $\pm$ 0.0015$\textrm{M}_{\odot}.\textrm{pc}^{-3}$
and 0.0098 $\pm$ 0.0043 $\textrm{M}_{\odot}.\textrm{pc}^{-3}$, respectively. In the northern hemisphere, however, we get different values: above 1.5 kpc, we have 0.0099 $\pm$ 0.0017 $\textrm{M}_{\odot}.\textrm{pc}^{-3}$ and above 2.5 kpc, we have 0.0161 $\pm$ 0.0044 $\textrm{M}_{\odot}.\textrm{pc}^{-3}$.

Based on our finding that the northern hemisphere is much more perturbed than the southern one, the wobbling dynamical surface density yields dark matter densities that are much more varying, and a priori less reliable than in the south. With these caveats in mind, we can also proceed to carry out a direct modelling of the stellar phase-space distribution function, as a matter of comparing it with the results obtained from the Jeans analysis.

\section{An alternative modelling of the vertical force and dark matter density}\label{sectionAltern}

With the caveats of the previous sections in mind, it is instructive to proceed, as a means of comparing, to a global classical determination of $K_z$ and of the local dark matter density by modelling stellar density distributions and velocity dispersions using isothermal components, and slightly redefining the sample selection geometry. This allows us to check the robustness of the previous results under equilibrium assumptions, meaning that at least the orders of magnitudes of the obtained values are similar, and to quantify more precisely the role of the local value of the circular velocity and its gradient in the systematic uncertainties. 

We adopted, in particular, the method developed by \cite{Bienayme14}, which uses pseudo-isothermal stellar phase-space distribution function components dependent on three integrals of motion. This extends the validity of the classical method of \citet{Oort32} well beyond 1 kpc above the Galactic plane.

For the star counts, we adopt the fits obtained for the `central' sample of stars towards the Galactic poles and, thus, we simply use the observed density distributions shown in Figure \ref{fit2ExpSp} (central north and south) reproduced in Figure \ref{fig:RhoSigma}. For the vertical velocity dispersion distributions towards the poles, instead of a cylindrical sector volume, we use a conical volume directed towards the Galactic poles with a $20\times 20$ degree aperture (extending to 2.8 kpc). We compute the velocity dispersions in 200 pc $z$-wide bins (see Figure~\ref{fig:RhoSigma2}). We note that in the study of the $K_z$ of \cite{Bienayme14} as well as in that of \cite{Zhang13}, the stellar samples are subdivided into several sub-samples of different metallicities in order to reinforce the constraints on the $K_z$ measurement. As metallicities are not available for all stars in our study, we did not subdivide the  samples. The velocity dispersions $\sigma_z$ are determined per 200 pc height interval with $\sigma$ clipping.

\subsection{Fitting a self-consistent model}

For the determination of the vertical force $K_z$, we use the Galactic mass distribution model of \cite{Bienayme14}. This model incorporates components that are Stäckel potentials \citep[e.g.][]{Famaey03} including two Kuzmin-Kutuzov spheroids: one of them, more flattened, represents the galactic disc whose thickness is set at 300 pc, which is typical for the old galactic stellar disc. The other component is the dark halo. The adjusted parameters are the masses of these two components and the flattening of the dark halo. A third component with a highly concentrated mass distribution adds an additional degree of freedom to the baryonic mass distribution of the disc and thus allows us to model the contribution of the disc to the $K_z$ in a more general and realistic way.  

These three-components models allow to represent the gravitational potential of the Galaxy \citep{Famaey03}. The density and velocity dispersions of our northern and southern samples  are modelled using distribution functions that are generalisations of the \cite{Shu69} distributions \citep[see][]{Bienayme15}. They are defined using three integrals of motion $E$, $L_z$ , and $I_3$ , which are precise with regard to Stäckel potentials. The associated distribution functions allow us to then adjust the gravitational potential with an excellent level of accuracy (see \cite{Bienayme14} for a detailed description).

\begin{figure*}[!ht]
  \centerline{
   \includegraphics[height=5cm,width=7cm]{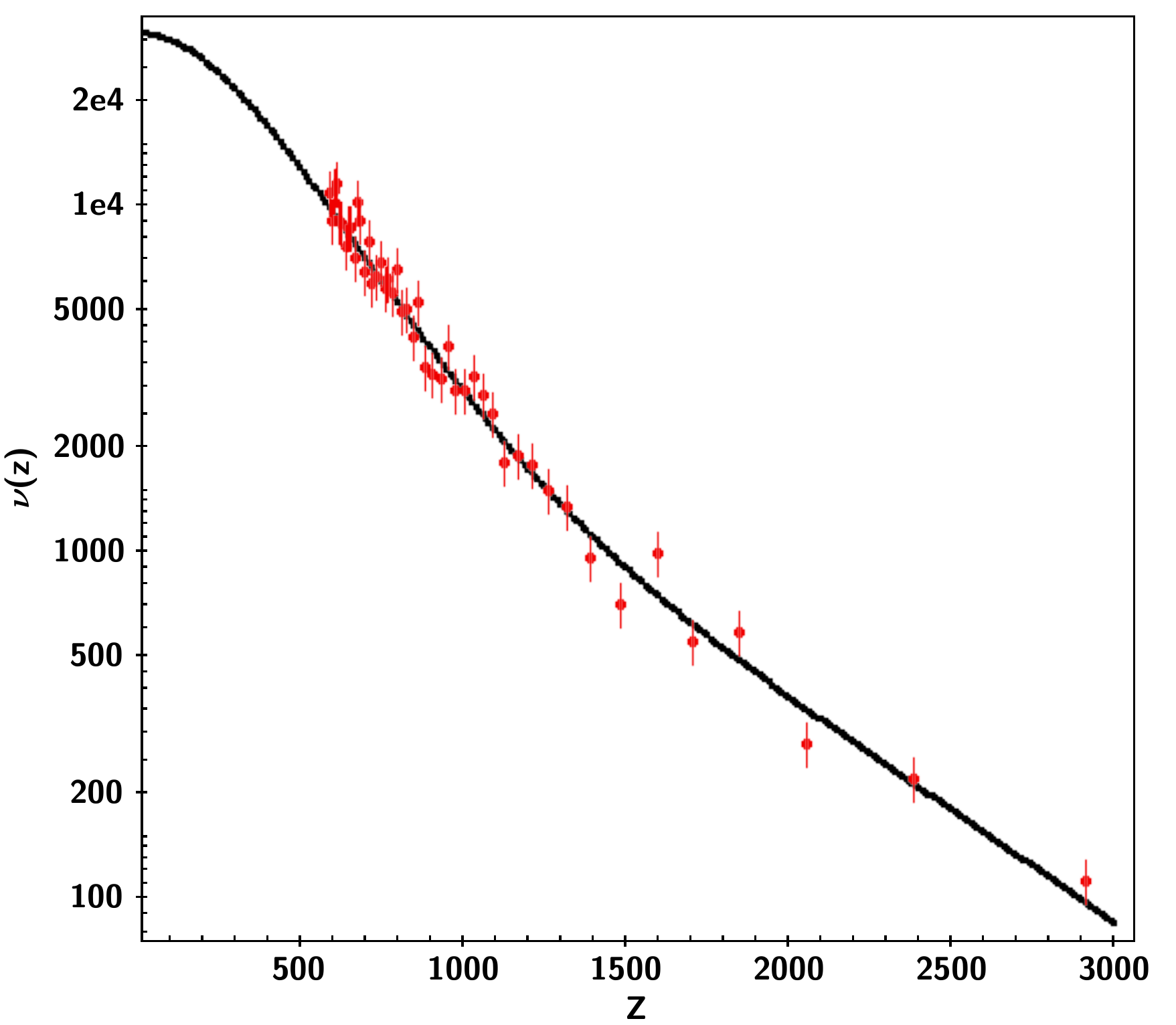},
   \includegraphics[height=5cm,width=7cm]{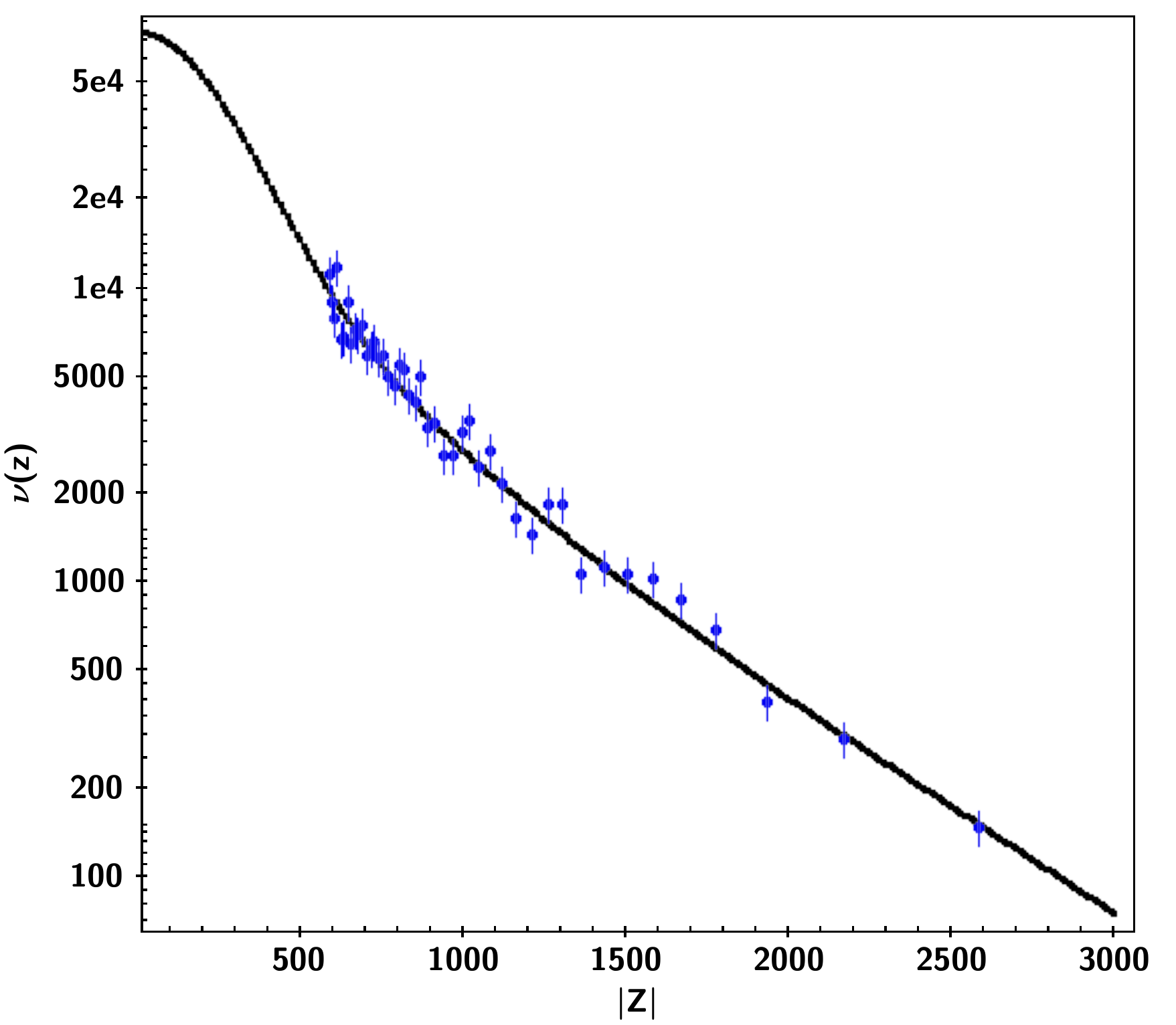},
   }   
  \caption{Vertical density profiles of red clump stars towards the north (left) and the  south (right) Galactic poles. Star counts and (continuous lines) best-fit model.
}
\label{fig:RhoSigma}
\end{figure*}

\begin{figure*}[!ht]
  \centerline{
   \includegraphics[height=5cm,width=7cm]{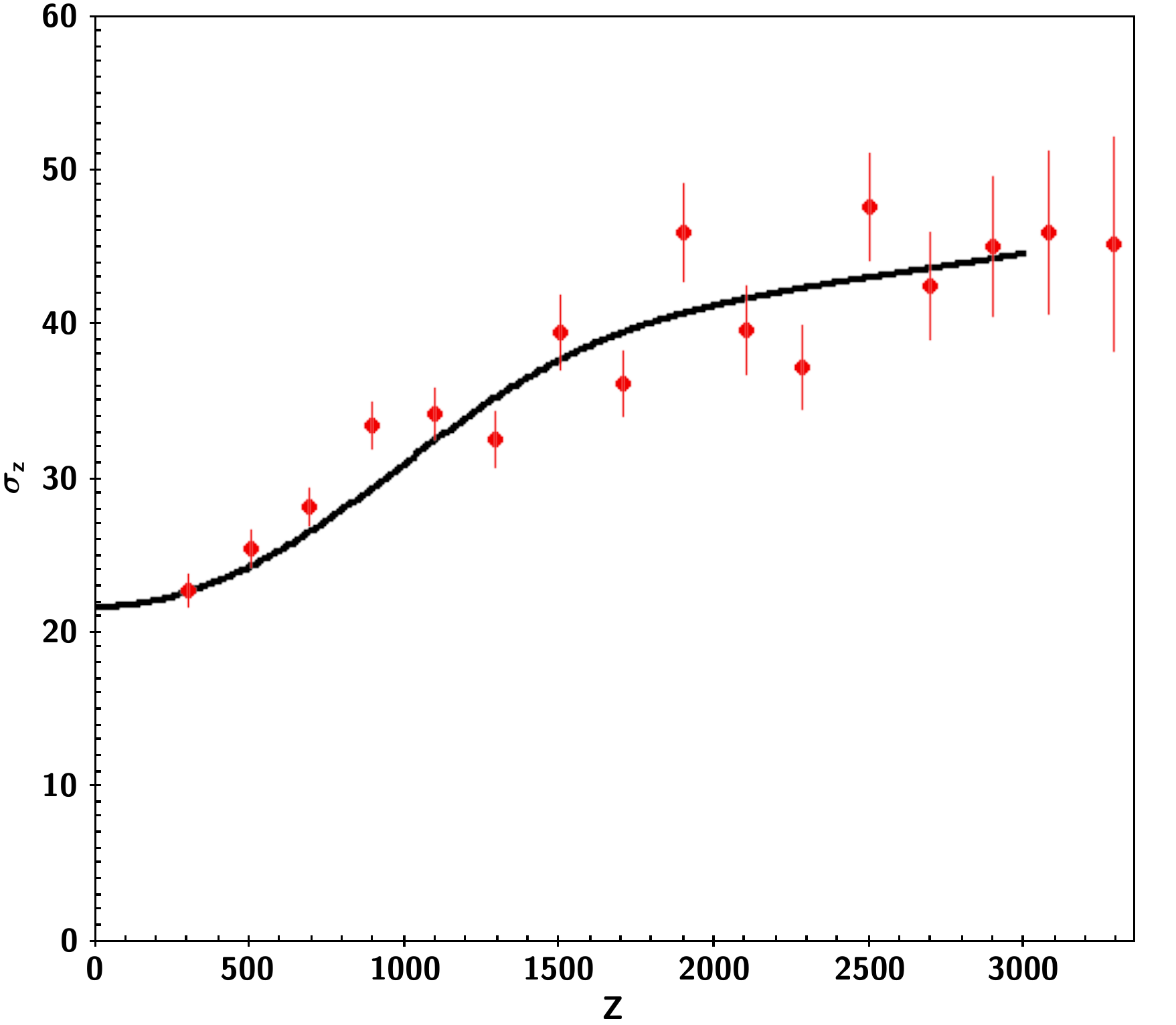},
   \includegraphics[height=5cm,width=7cm]{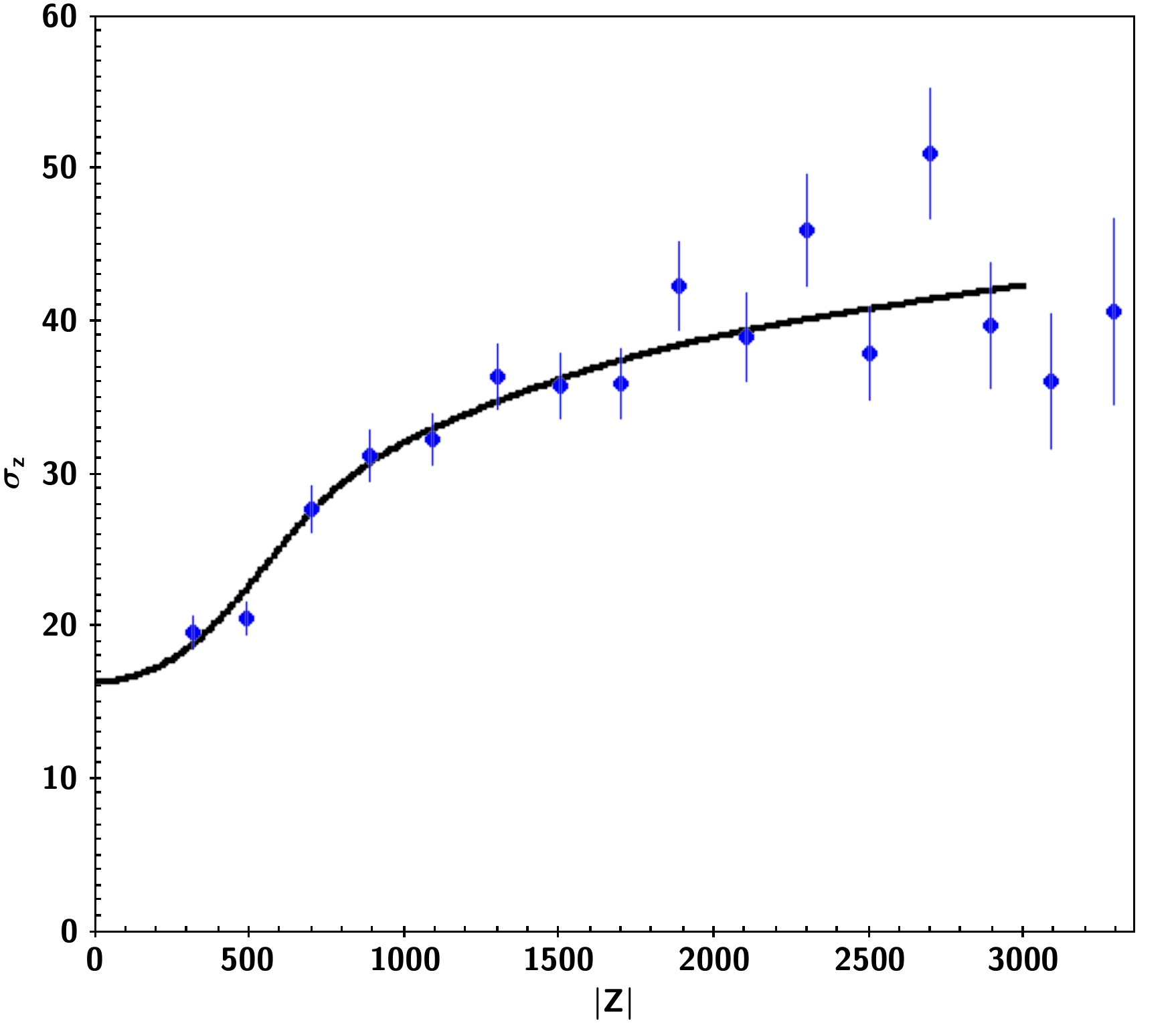}
   }   
  \caption{Vertical velocity dispersion  of red clump stars towards the north (left) and the  south (right) Galactic poles. Measured dispersions  and the (continuous line) best-fit model.
}
\label{fig:RhoSigma2}
\end{figure*}

We perform a  least-squares fit, separately in the north and south directions, of the vertical density and velocity dispersion distributions
(see  Figures \ref{fig:RhoSigma}-\ref{fig:RhoSigma2}).
The model parameters are three masses and two thicknesses for the Galaxy's gravitational potential, as well as (for each of the three distribution functions) the local stellar density in the Galactic plane at the position of the Sun and the local vertical velocity dispersion.

The best fits show significantly different mass distributions between north and south that can reproduce the observed $\nu(z)$ and $\sigma_z(z)$ in the distance intervals considered (400 pc to 3 kpc). Distinct solutions are found and the parameters obtained for the mass distribution of the different components are highly correlated. The dark matter density of the dark halo is not  constrained and a wide range of values of $\rho_{DM}$ seems to be compatible with the observed spatial and kinematic distributions of the red clump giants. However these different models differ from each other by their circular rotation curves, producing different values of the circular rotation speed $V_c(R_0)$ at the Solar Galactic radius as well as by different gradients of the rotation curve at the position of the Sun $\partial{V_c(R_0,0)} / \partial{R}$. These models all have  an almost spherical dark halo.

\begin{figure*}[!ht]
  \centerline{
   \includegraphics[height=5cm,width=7cm]{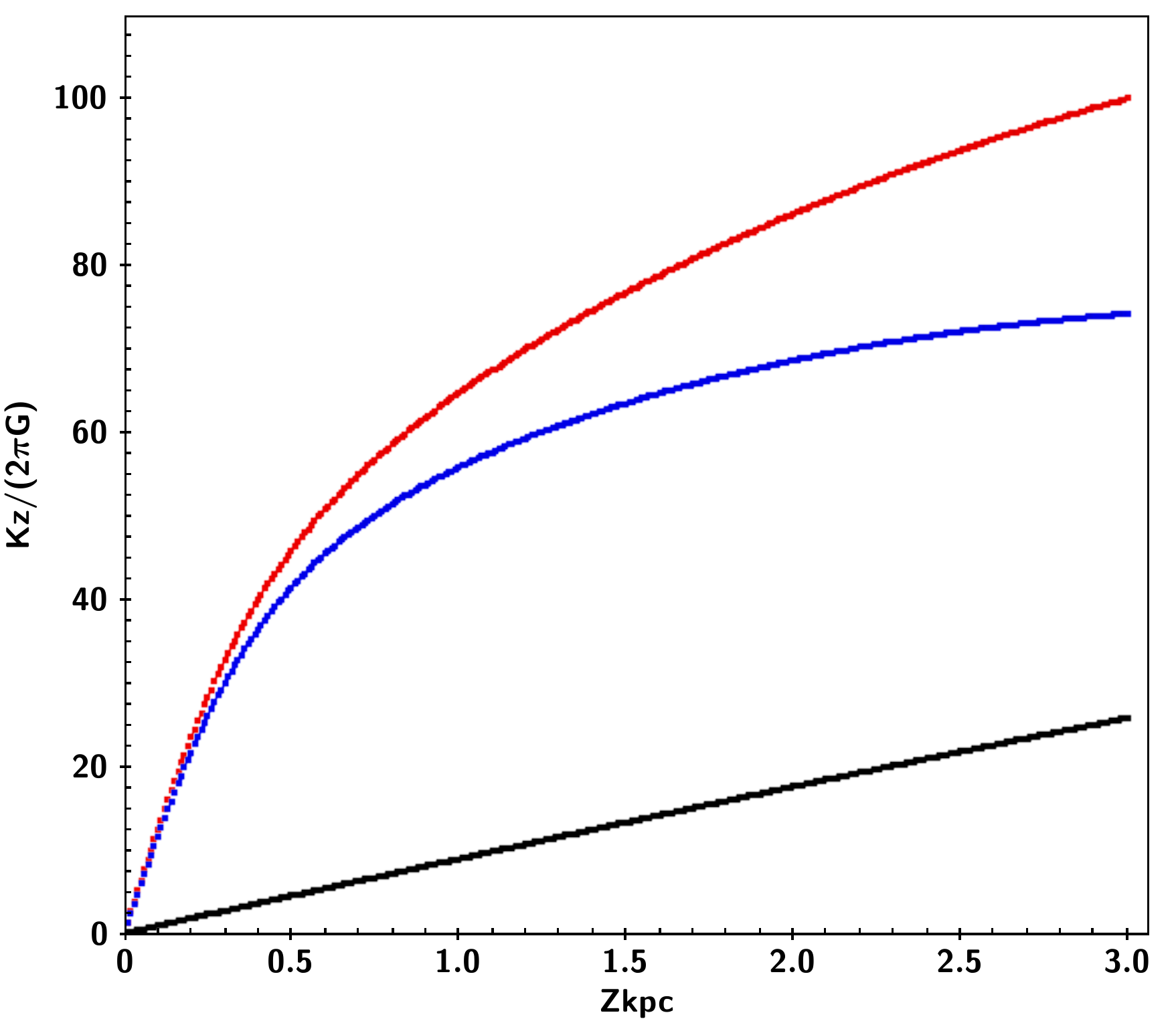},
   \includegraphics[height=5cm,width=7cm]{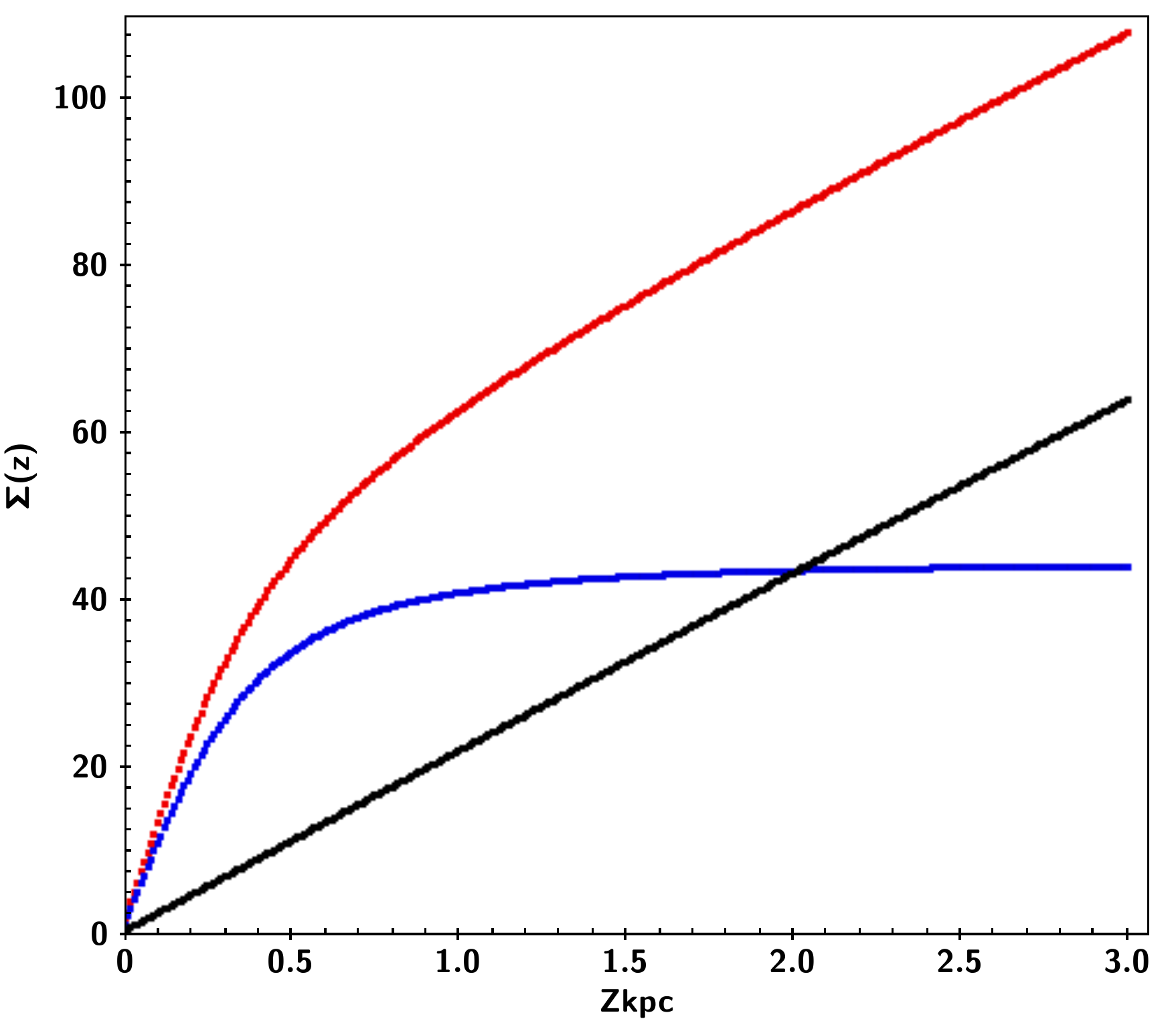}
   }   
  \caption{ Left: Vertical force (in $M_\odot pc^{-2}$ units) perpendicular to the galactic plane versus the vertical distance. Red shows the total vertical force. Blue shows the disc contribution to the vertical force. Black shows the dark matter contribution to the vertical force.
Right: surface density in $M_\odot pc^{-2}$ integrated from $-z$ to $z$. Red shows the  total surface density. Blue shows the disc contribution. Black shows the dark matter contribution to the integrated surface density. The differences between the left and right panels can be understood from Equations~\ref{eqFR} and \ref{eqSigma} for each component. The rotation curve of the disc component is decreasing and that of the halo is increasing, leaving the total rotation curve almost flat (albeit slightly decreasing). Therefore, the surface density of the disc is smaller than its corresponding vertical force (blue curves) and the opposite is true for the halo (black curves). The total surface density is slightly smaller than the corresponding vertical force up to $z \sim 2$~kpc (red curves), but the trend is then reversed due to the details of the mass distribution at larger heights.
}
\label{fig:RhoSigma3}
\end{figure*}

\subsection{Adding rotation curve constraints}

To distinguish among our different models that equally fit star counts and velocity dispersion, one can add two supplementary constraints: the circular rotation speed at $R_0$ and the slope of the rotation curve at $R_0$.

For instance, fixing those values to $V_c(R_0)=233$~km.s$^{-1}$ \citep{McGaugh18} and $\partial{V_c(R_0,0)} / \partial{R}=-1.7$~ km.s$^{-1}$.kpc$^{-1}$ \citep{Eilers19}, we find $\rho_{DM}=0.0108\pm0.004$ M$_{\odot}$ pc$^{-3}$ for the combined sample. Figure \ref{fig:RhoSigma3} (left) shows the vertical force for the north and south taken together, and also the contributions of the dark halo component and the baryonic disc to the $K_z$, while Figure \ref{fig:RhoSigma3} (right) shows the corresponding integrated surface density $\Sigma(z)$. The formal error on the estimate of $\rho_{DM}$ is on the order of a few percent. We note that the reduced $\chi^2$ of the best fit is, again, larger than 1 (namely, 1.7). However, we note that with this method, the recovered value of the local dark matter density is identical in the north and south. The formal error on this value is much smaller than the systematic error coming from the different possible values of the two local rotation curve parameters. 

A systematic correction can be applied to the local dark matter density based on the adopted value of these two parameters: 
\begin{align*}
    \rho_{\rm DM}(R_0)_{M_\odot pc^{-3}} &= 0.0108\,+\,0.00065 
    \times (V_c|_{km/s}-233)\,  \\
    & +\, 0.0019 \times [(\partial{V_c}/\partial R)|_{km/s/kpc}+1.7] . 
\end{align*} 
This means that, in adopting the same values as in the rest of the paper, namely, $V_c(R_0)=240$~km.s$^{-1}$ and $\partial{V_c(R_0,0)} / \partial{R}=-2.18$~ km.s$^{-1}$.kpc$^{-1}$, we find $\rho_{\rm DM}(R_0)_{M_\odot pc^{-3}} = 0.0144$, which is slightly larger than the results from the Jeans analysis.

The formalism used in this section would also benefit in the future from a full MCMC treatment to further explore the correlation between rotation curve parameters and associated uncertainties and the local dark matter density. Including the stellar samples away from the Sun in a single simultaneous model fit could also prove extremely interesting, bearing in mind the caveats about non-stationarity. This will be the subject of future works.

\section{Discussion and conclusion}\label{sectionCCL}

In this paper, we analyse in detail the kinematics and dynamics above the Galactic plane of a homogeneous and 98\% complete sample of 43 589 red clump giants with $G \leq 13.3$ from \gdrtwo. The spatial distribution of these stars is limited to the Galactic poles, in a cone of 45 degrees. This was done for completeness purposes, which were validated against 2MASS catalogue up to at least |z| = 3.5 kpc.

We have sliced our sample in ten sectors, five in the northern hemisphere and five in the south, separated by 0.8~kpc, covering both the inner disc and outer disc regions, as well as the regions in the direction of Galactic rotation and counter-rotation. We then analyse in detail the north-south asymmetries reflecting the non-stationarity of the Galactic disc. In particular, a global azimuthal wobble has been noted with a northern density peak at the Solar azimuth, preceded and followed by southern density peaks in the directions of rotation and counter-rotation. A similar wobble was found in the radial direction, with a global northern domination at large heights in the inner and central sectors, followed by a southern domination in the anti-centre sector. These indicate clearly the wave-like nature of the general north-south asymmetries, with a slightly shorter wavelength in the azimuthal direction. The more localised wobbles and bumps in the vertical density can be classified into three categories, with: (i) some bumps and wiggles jointly happening in the north and south; (ii) some being clearly north-south antisymmetric, meaning that they represent anti-correlated oscillation of the density, present in all sectors between 2 and 3 kpc, but vastly more so in the radial directions (centre and anti-centre); and (iii) some bumps happening only in one hemisphere, such as a striking southern bump at $z=-2.1$~kpc in the direction of rotation. Notwithstanding, one of our main conclusions is that in our Galactic suburbs, the northern hemisphere is generally more perturbed than the southern one. It will be interesting to understand how this can inform our knowledge of the interaction of the Galactic disc with external perturbers, such as the Sagittarius dwarf \citep[e.g.][]{Laporte19}. 

Averaging over the bumps and wiggles, we derive the scale-length and a double scale-height of our sample separately in the north and south. Although, in the absence of chemical information, our goal was not to make a detailed thin-thick disc decomposition, these two scale-heights roughly correspond to the contribution of the geometrical thin and thick discs. We note that the 'thick disc' scale height is significantly larger in the north ($\sim 827$~pc) than in the south ($\sim 638$~pc). Due to our selection function, the radial scale-length is essentially constrained by the thick disc and it is significantly shorter in the north ($\sim 2.2$~kpc) than in the south ($\sim 3.1$~kpc). It will be particularly interesting to conduct in the future a detailed chemical analysis of our sample of red clump giants (e.g. with WEAVE and 4MOST) to better understand these spatial structures of the thick disc in the northern and southern Galactic hemispheres.

Restricting the sample to {\it Gaia} RVS stars, which leaves us with 33 005 stars, we analyse the velocity dispersion gradients in the northern and southern hemispheres and, again, we find a much more perturbed northern Galactic hemisphere at heights above $z \sim 1$~kpc. However, we also noted an interesting bump at large heights in the southern velocity dispersions, restricted to the sector in the direction of Galactic rotation. 

After this analysis of the north-south asymmetries, we investigated how making an equilibrium assumption separately for data in the northern and southern Galactic hemispheres affects the estimates of the dynamical matter density and dark matter density. We note that many systematics could enlarge our error estimates, meaning that the formal errors quoted in this study are to be taken as lower limits. For this task, we concentrated on the local sample ('central' sector) with most of the data. Given the  findings above, this assumption of equilibrium is likely to hold better in the south than in the north.  We find that a Jeans analysis delivers relatively similar vertical forces and integrated dynamical surface densities at large heights above the plane in both hemispheres. At these heights, the densities of stars and gas are very low and the surface density is largely dominated by dark matter, which allows us to estimate, separately in the north and in the south, the local dark mater density derived under equilibrium assumptions. We find a local dark matter density (estimated above 2~kpc) $\rho_{\rm DM} \sim 0.013 \, {\rm M}_\odot/{\rm pc}^3$ ($ \sim 0.509 \, {\rm GeV/cm}^3$) in the perturbed northern hemisphere and $\rho_{\rm DM} \sim 0.010 \, {\rm M}_\odot/{\rm pc}^3$ ($ \sim 0.374 \, {\rm GeV/cm}^3$) in the much less perturbed south. 

Finally, with all these caveats in mind, we obtained a value of $\rho_{\rm DM} \sim 0.011 - 0.014 \, {\rm M}_\odot/{\rm pc}^3$ (depending on the local value of the circular velocity and its gradient) through a global phase-space distribution function fit to the data under global equilibrium assumptions. While in global agreement, we note that with the exact same rotation curve constraints, this distribution function analysis yields a slightly larger value of the local dark matter density ($0.014 \, {\rm M}_\odot/{\rm pc}^3$) than the Jeans analysis. We conclude that at this stage, and once the rotation curve is fully fixed, only future developments of non-equilibrium methods would allow for a full evaluation of potential systematics due to non-stationary effects and, thus, allowing for   more solid estimates of the dynamical matter density in the Galactic disc to be obtained.

\begin{acknowledgements}
The authors would like to thank the International Space Science Institute, Bern, Switzerland, for providing financial support and meeting facilities. This work was supported by the Programme National Cosmology et Galaxies (PNCG) of CNRS/INSU with INP and IN2P3, co-funded by CEA and CNES. JBS acknowledges financial support of the CNES. BF acknowledges funding from the Agence Nationale de la Recherche (ANR project ANR-18-CE31-0006 and ANR-19-CE31-0017) and from the European Research Council (ERC) under the European Union's Horizon 2020 research and innovation programme (grant agreement No. 834148). This work has made use of data from the European Space Agency (ESA) mission {\it Gaia} (\url{https://www.cosmos.esa.int/gaia}), processed by the {\it Gaia} Data Processing and Analysis Consortium (DPAC,
\url{https://www.cosmos.esa.int/web/gaia/dpac/consortium}). Funding
for the DPAC has been provided by national institutions, in particular
the institutions participating in the {\it Gaia} Multilateral Agreement.
\end{acknowledgements}


\bibliographystyle{aa}
\bibliography{bibdensity} 

\newpage

\appendix

\section{Parameter correlations }\label{correldensityProfileAnnex}
\begin{figure*}[h]
   \resizebox{2.9cm}{!}
{\includegraphics{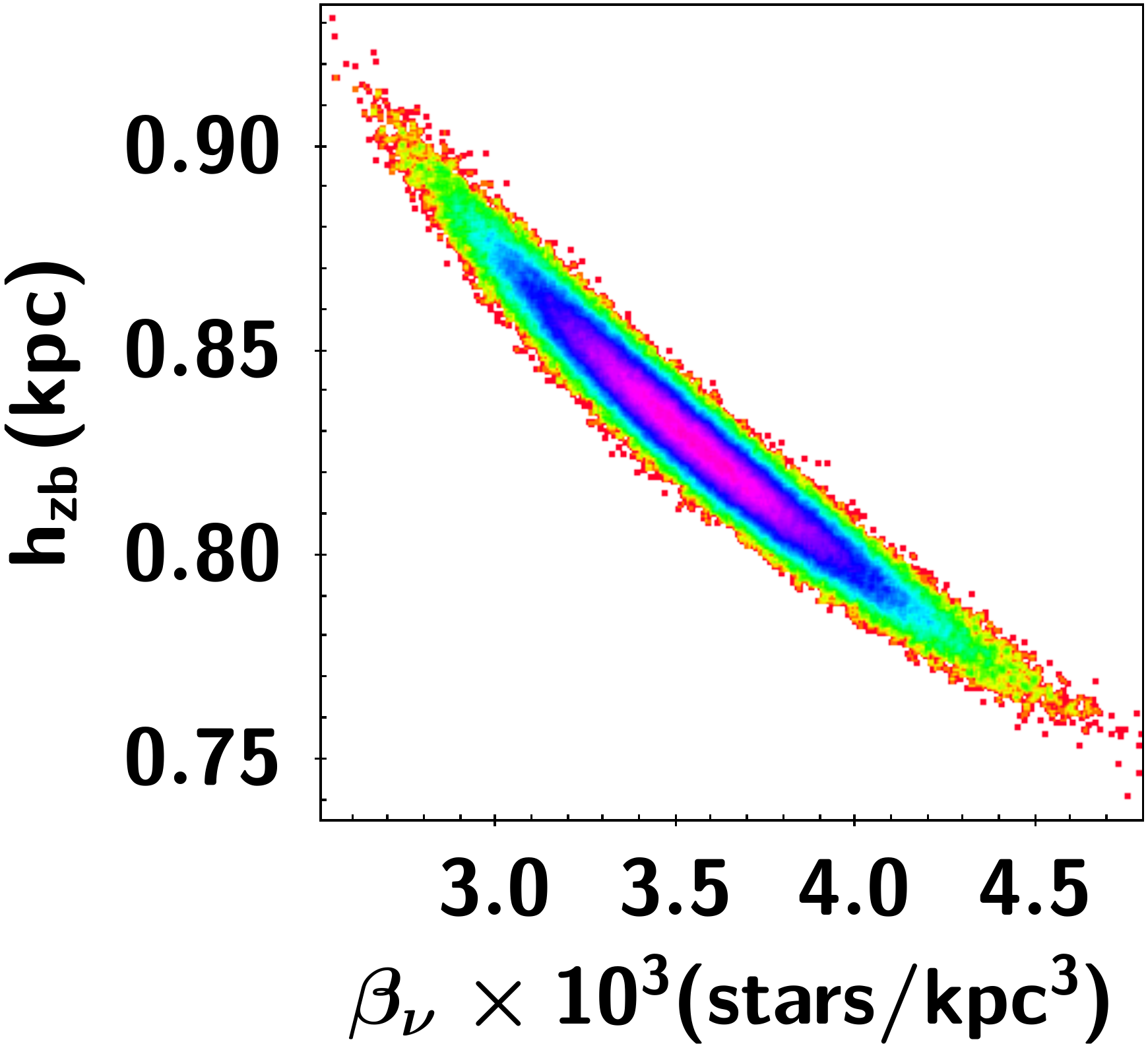}}\\
    \resizebox{5.8cm}{!}
            {\includegraphics{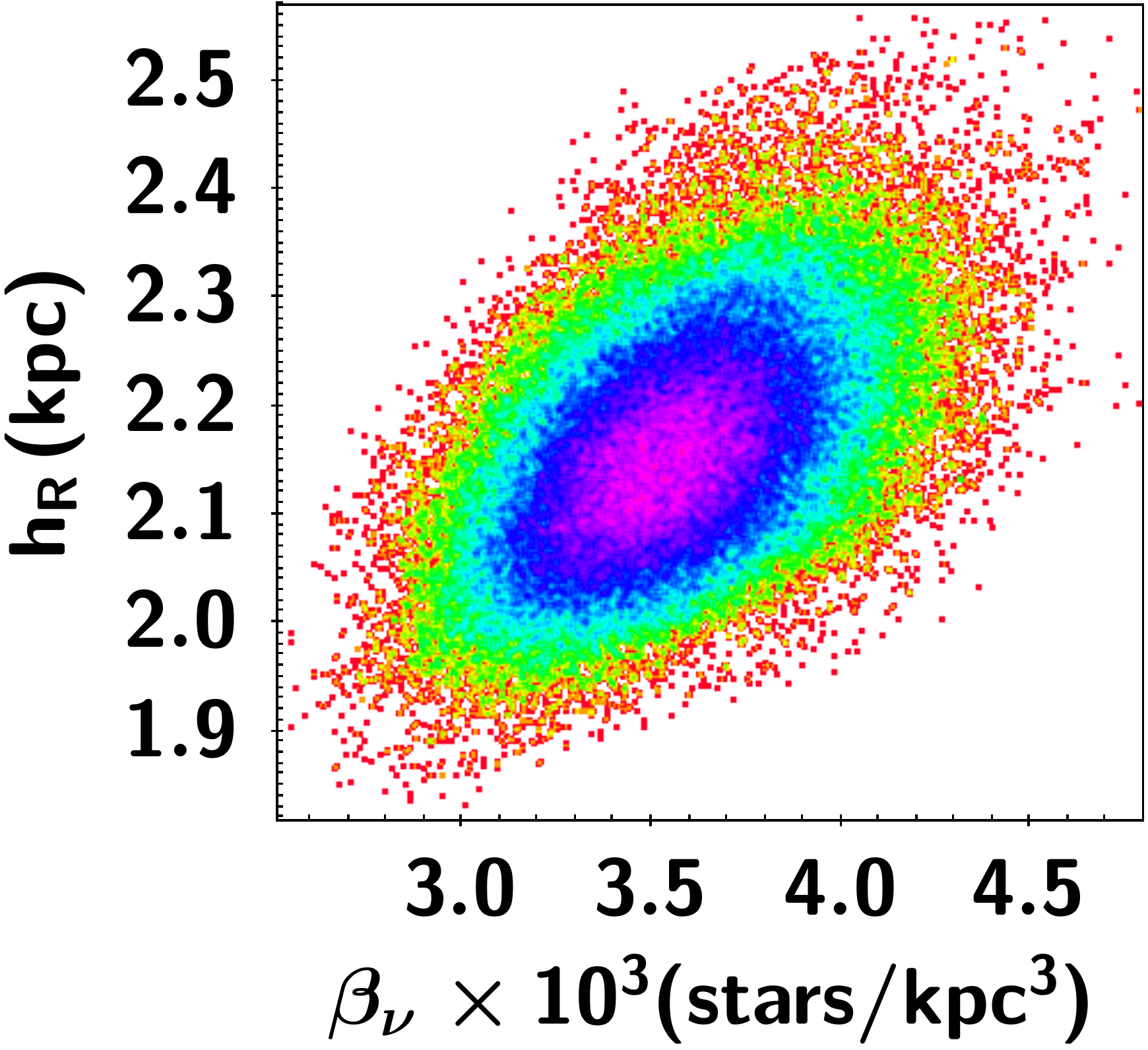}
            \includegraphics{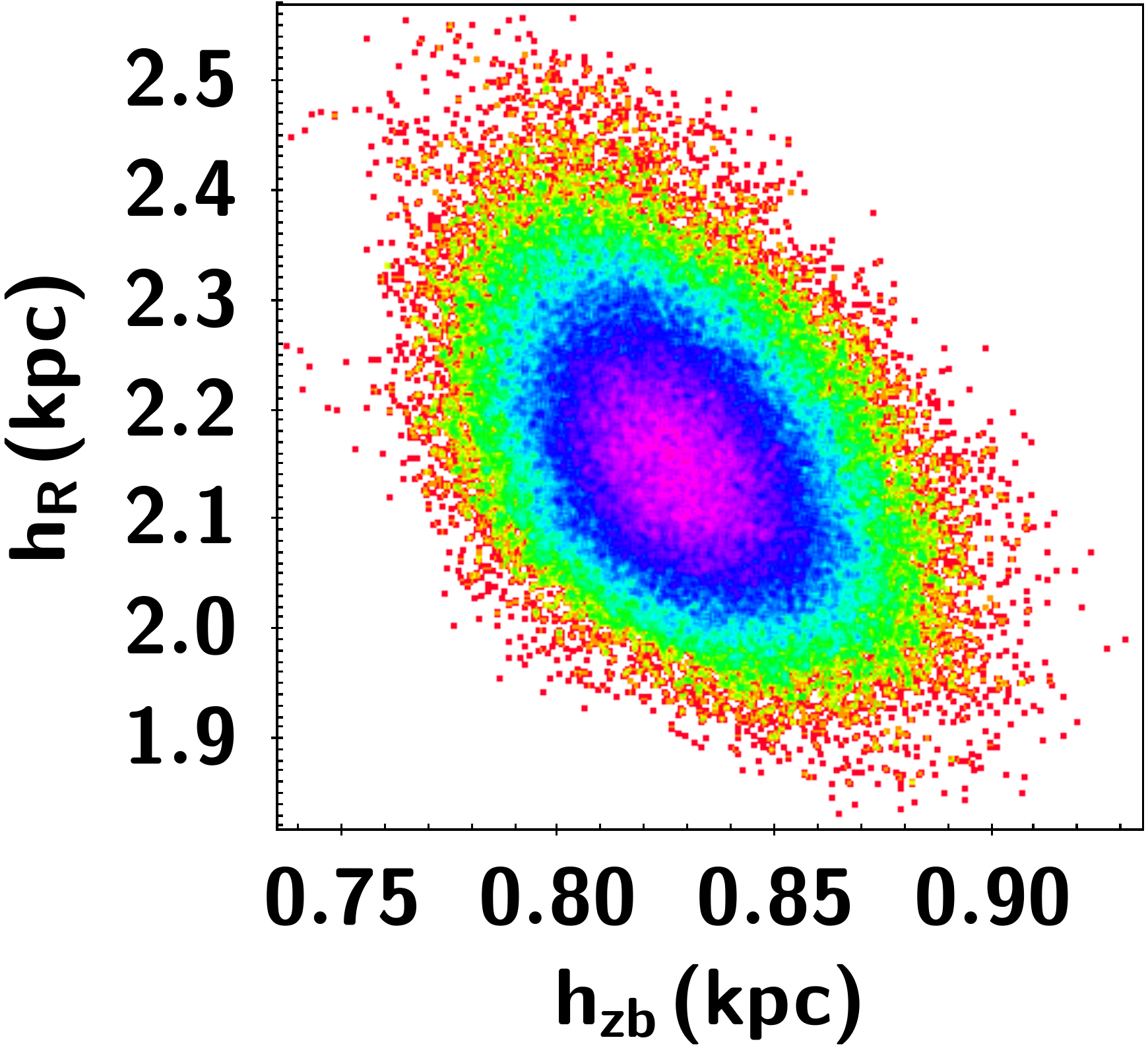}}\\
    \resizebox{8.7cm}{!}  
            {\includegraphics{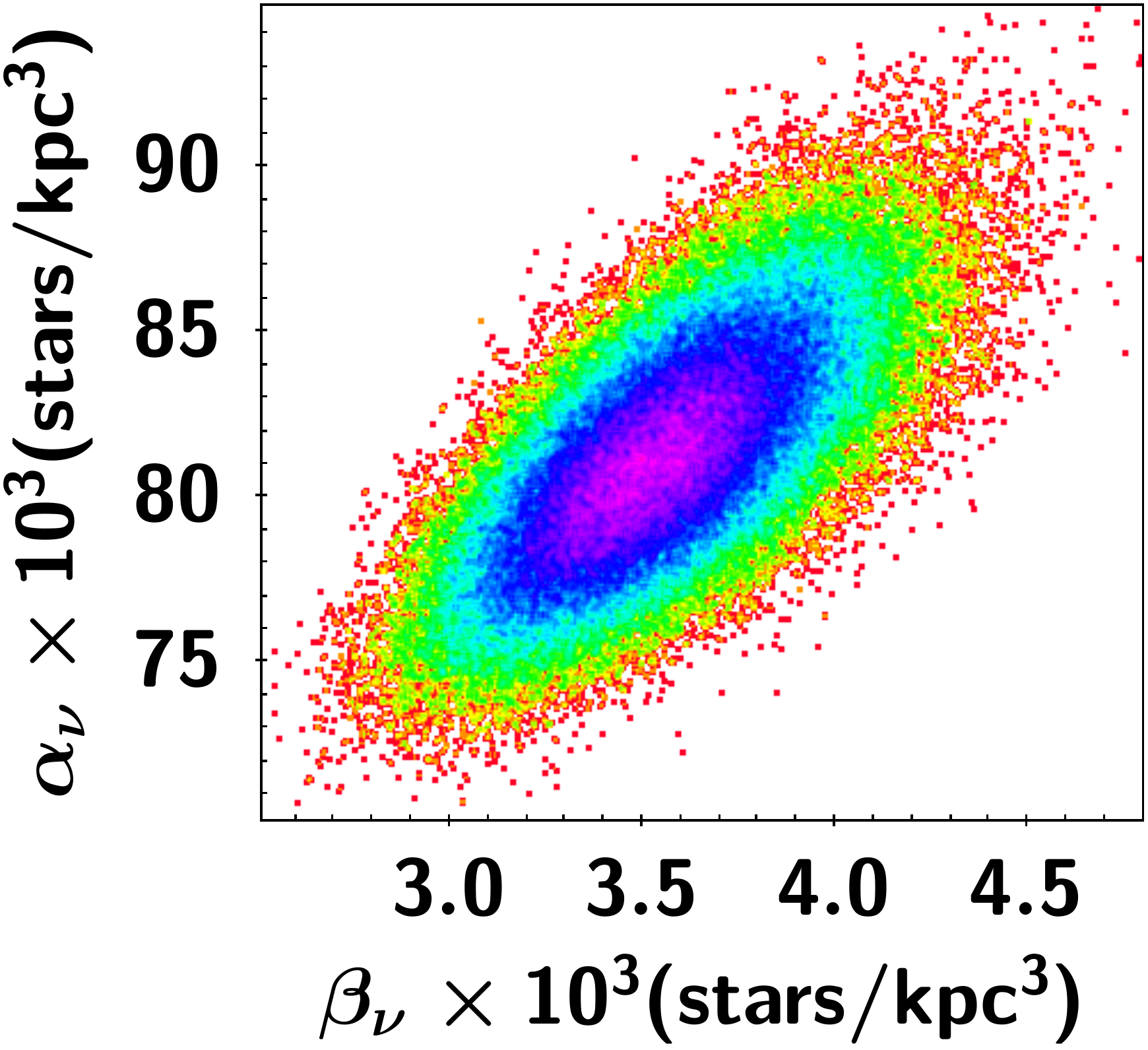}
            \includegraphics{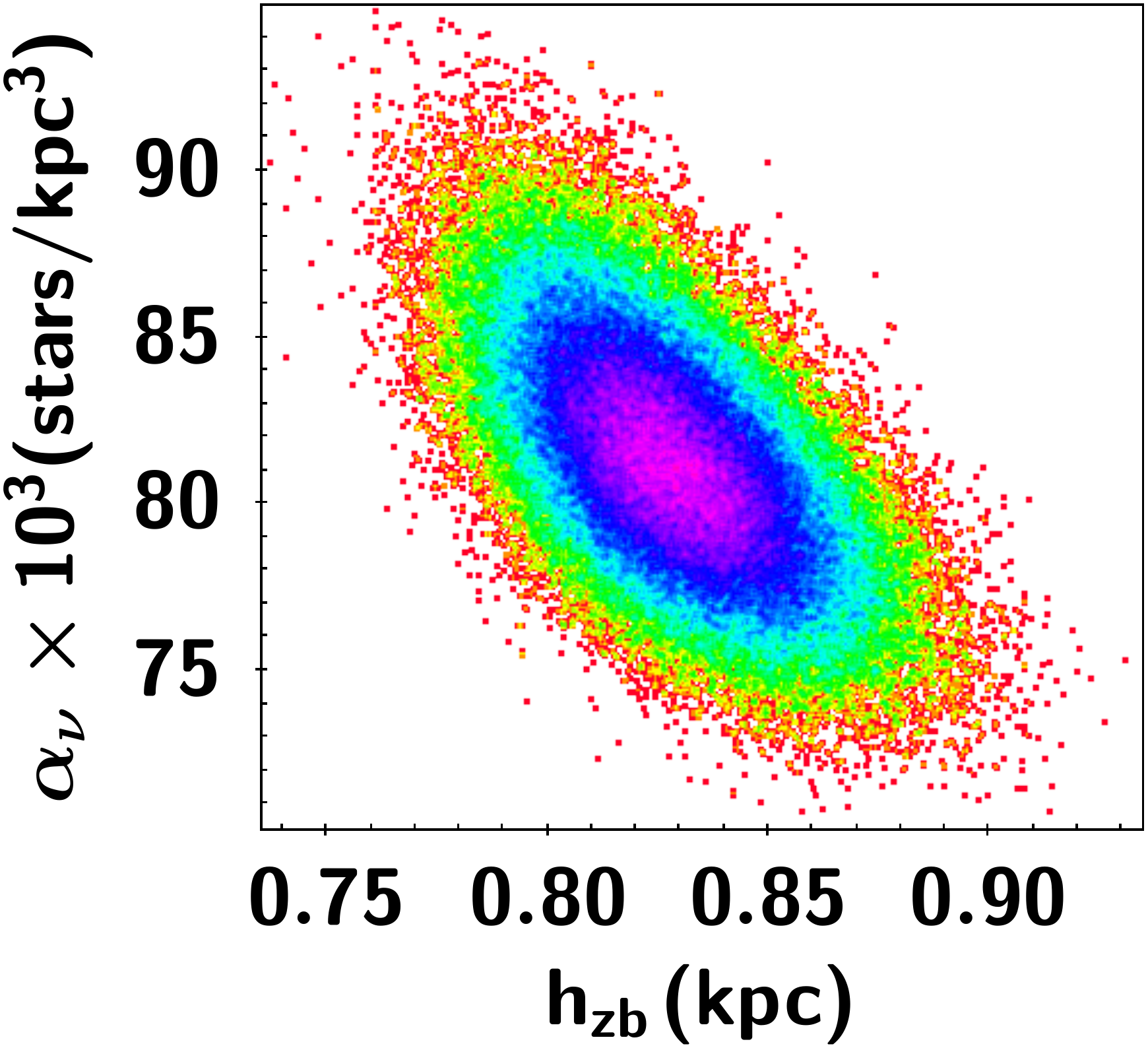}
            \includegraphics{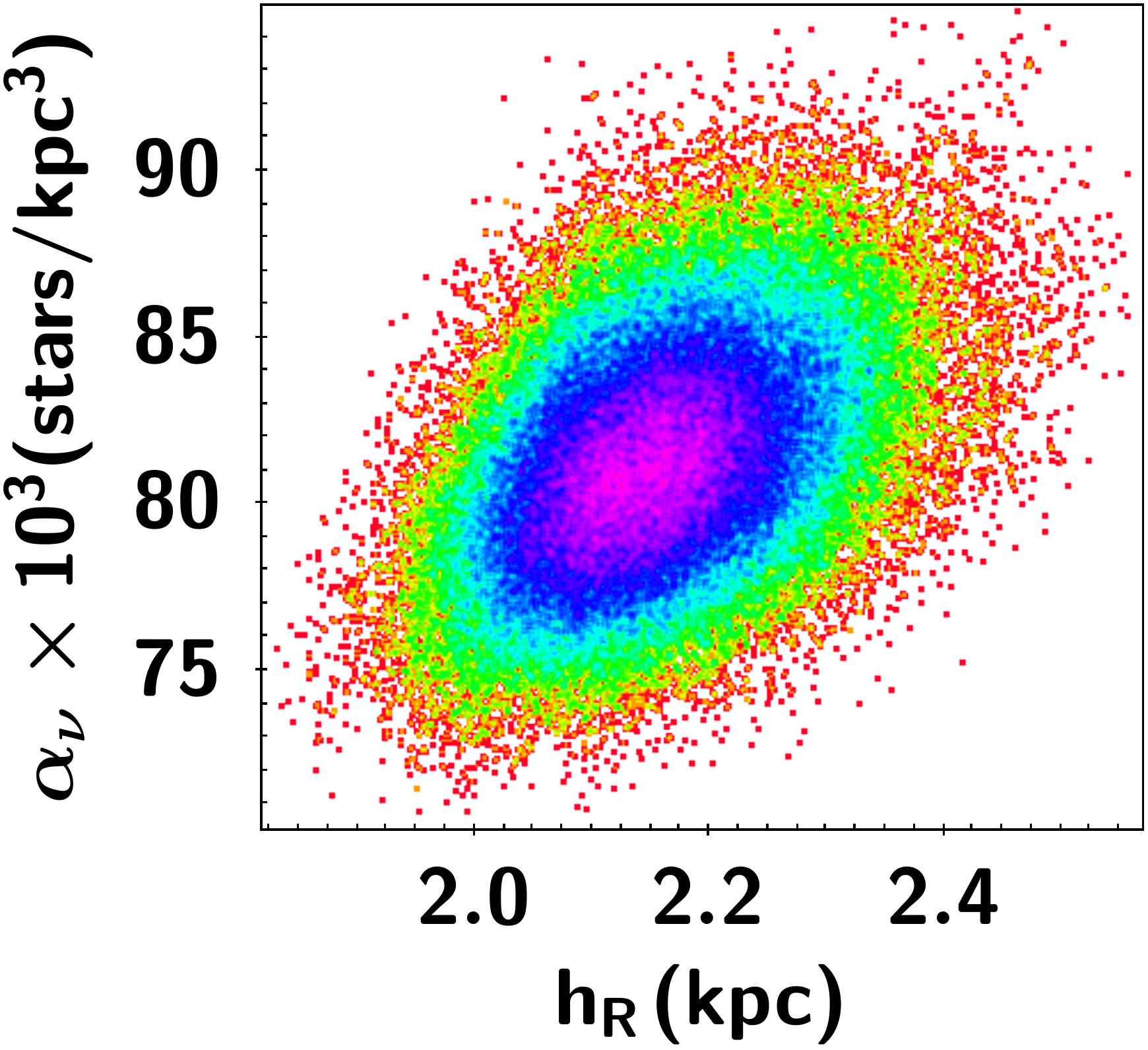}}\\
    \resizebox{11.6cm}{!}  
            {\includegraphics{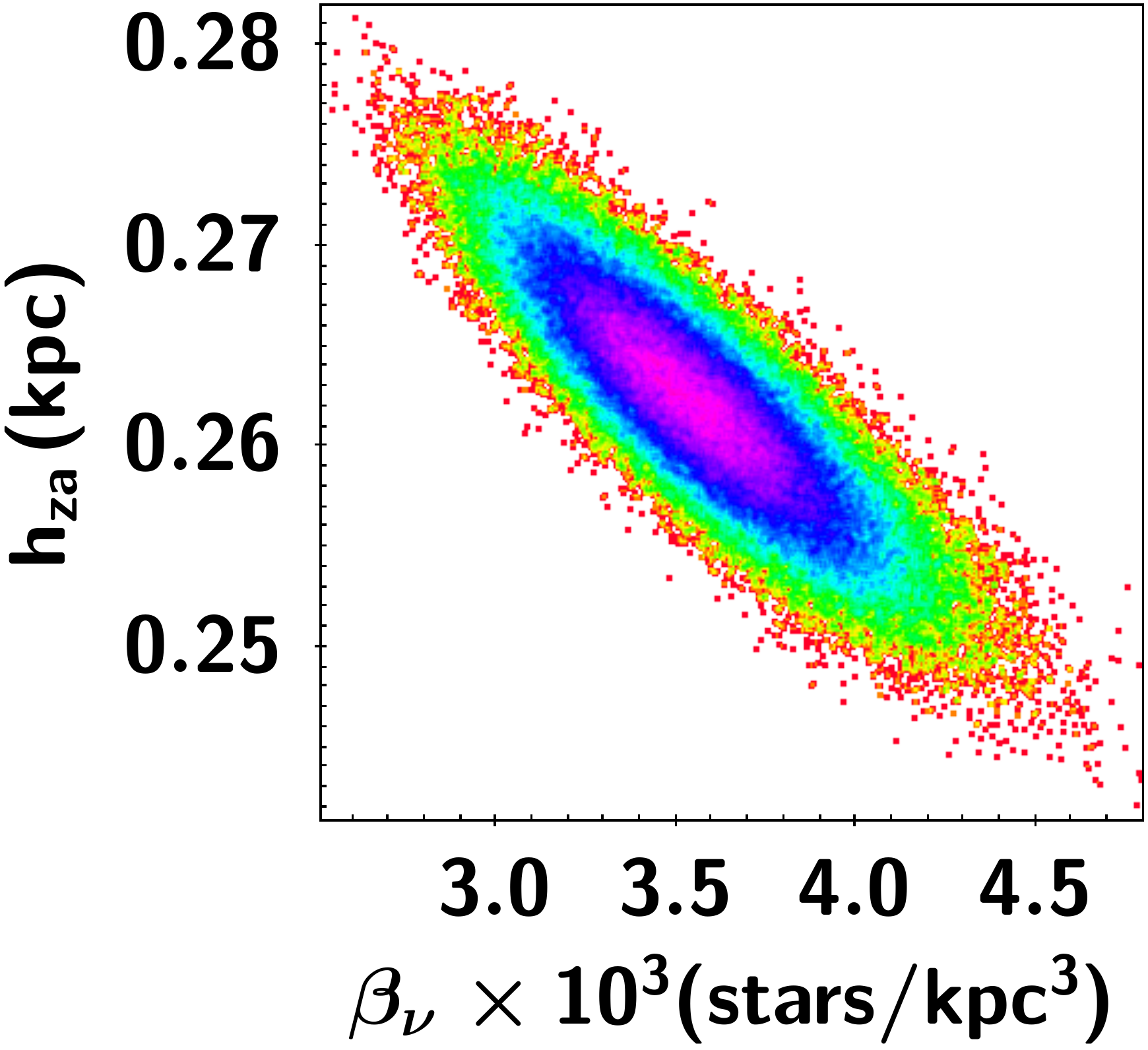}
            \includegraphics{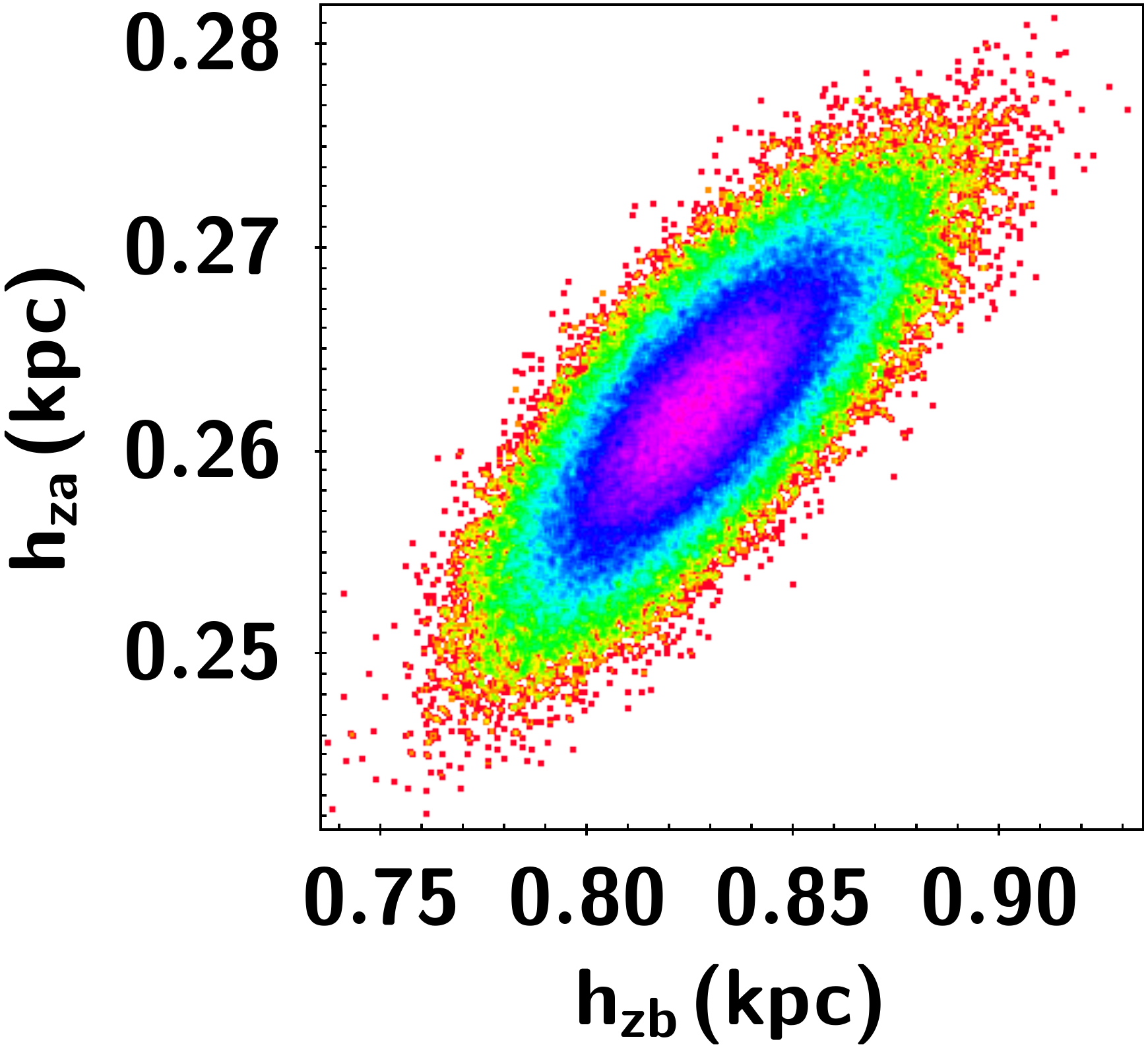}
            \includegraphics{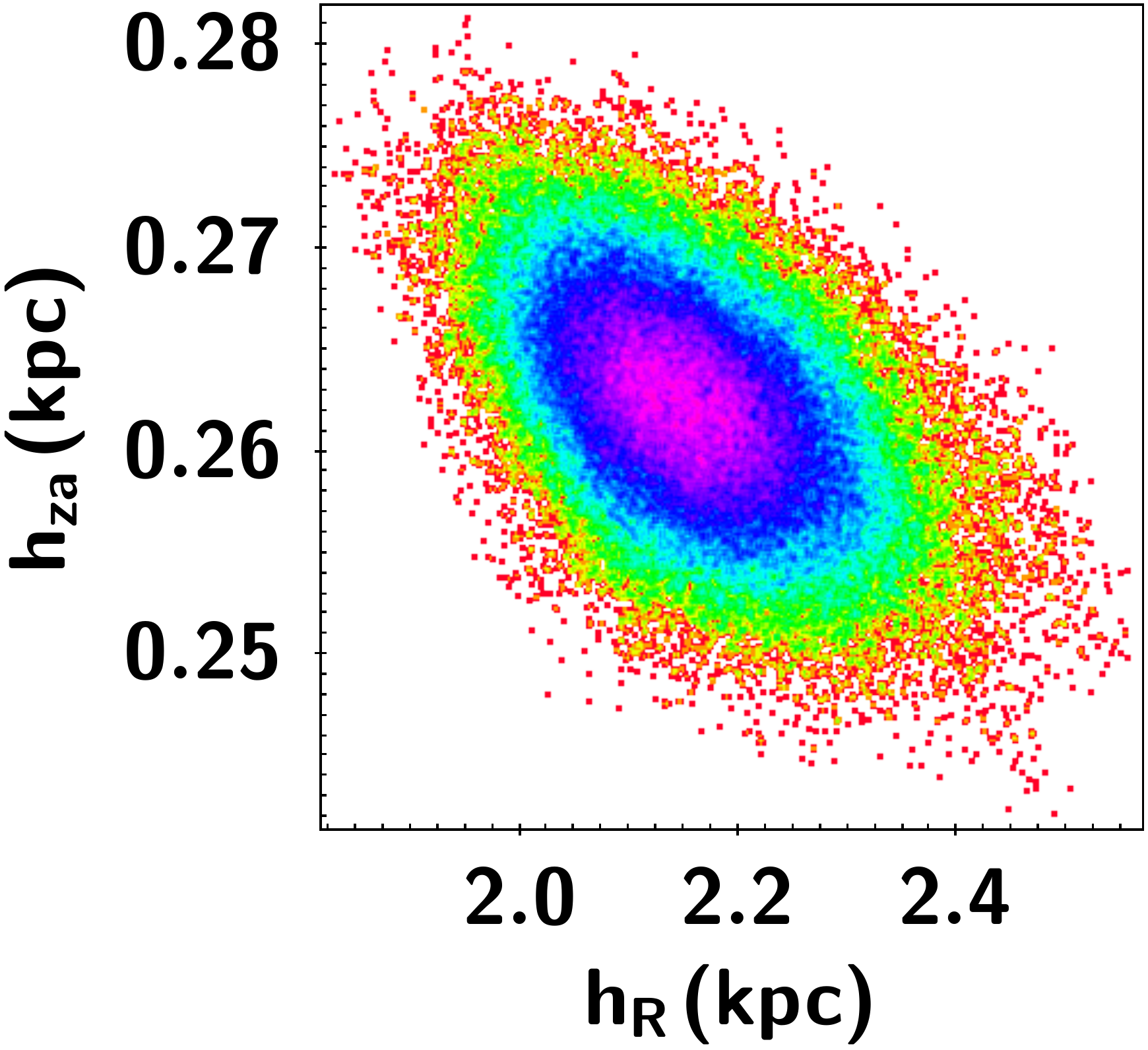}
            \includegraphics{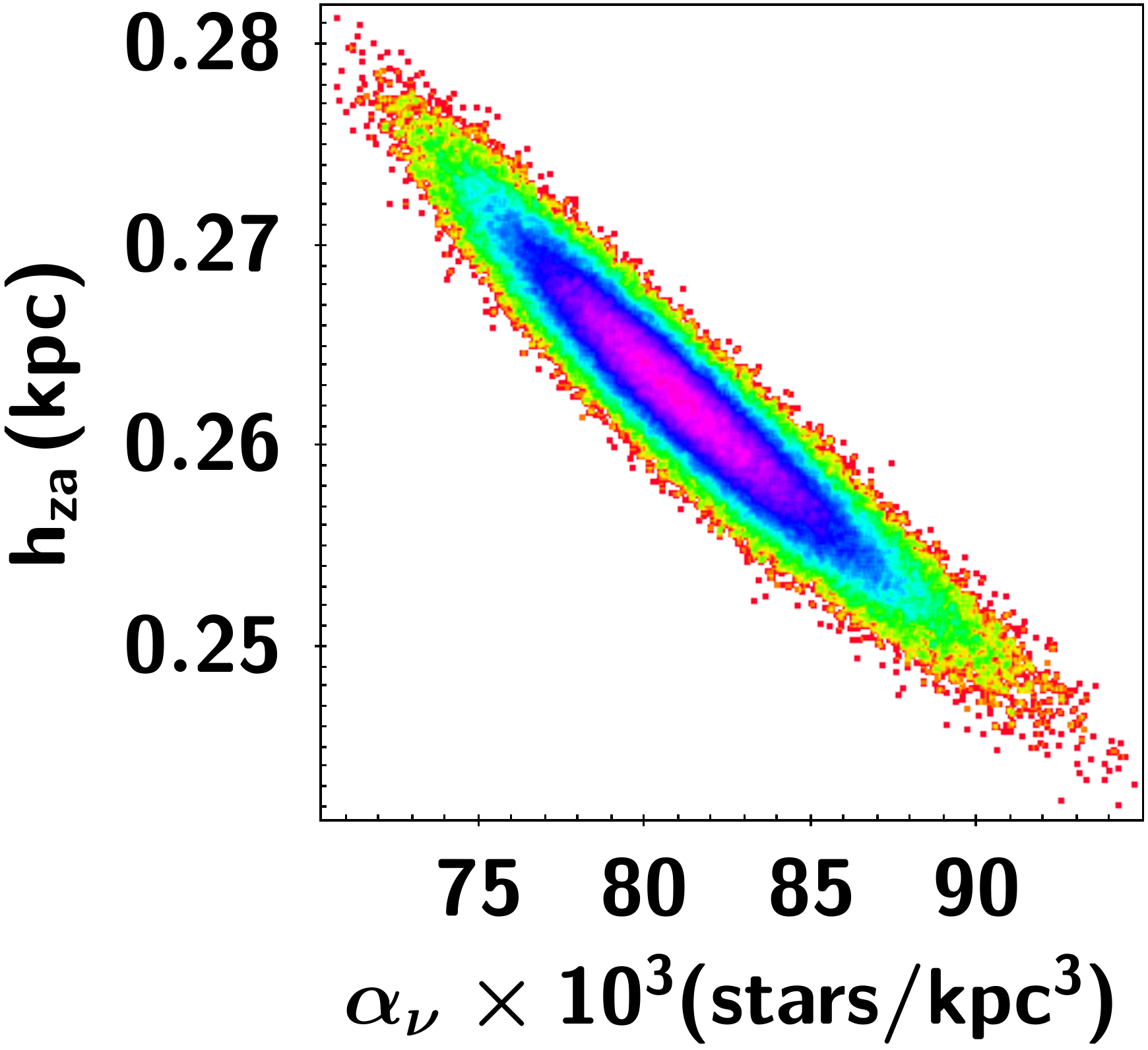}}\\
      \caption{ Two-dimensional posterior distributions (last steps of the Markov chain) of the five parameters from the adjusted Equation~\ref{eq2expSp} for the northern hemisphere, represented in a triangle correlation matrix. Colours represent the density in an arbitrary log-scale from purple (high density) to red (low density).}
               \label{northCorrelFig}
   \end{figure*}

\begin{figure*}[]
   \resizebox{2.9cm}{!}
            {\includegraphics{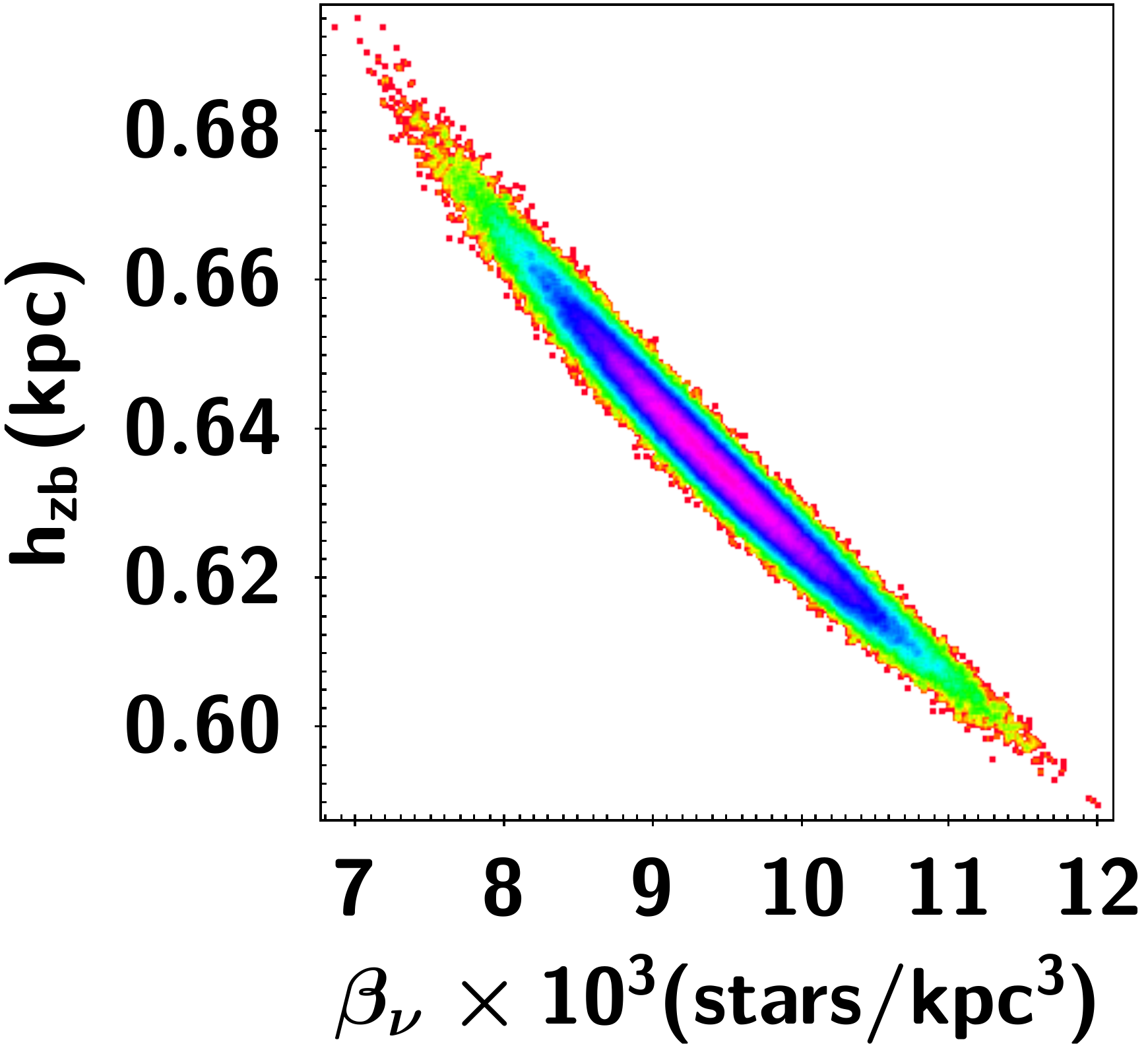}}\\
    \resizebox{5.8cm}{!}
            {\includegraphics{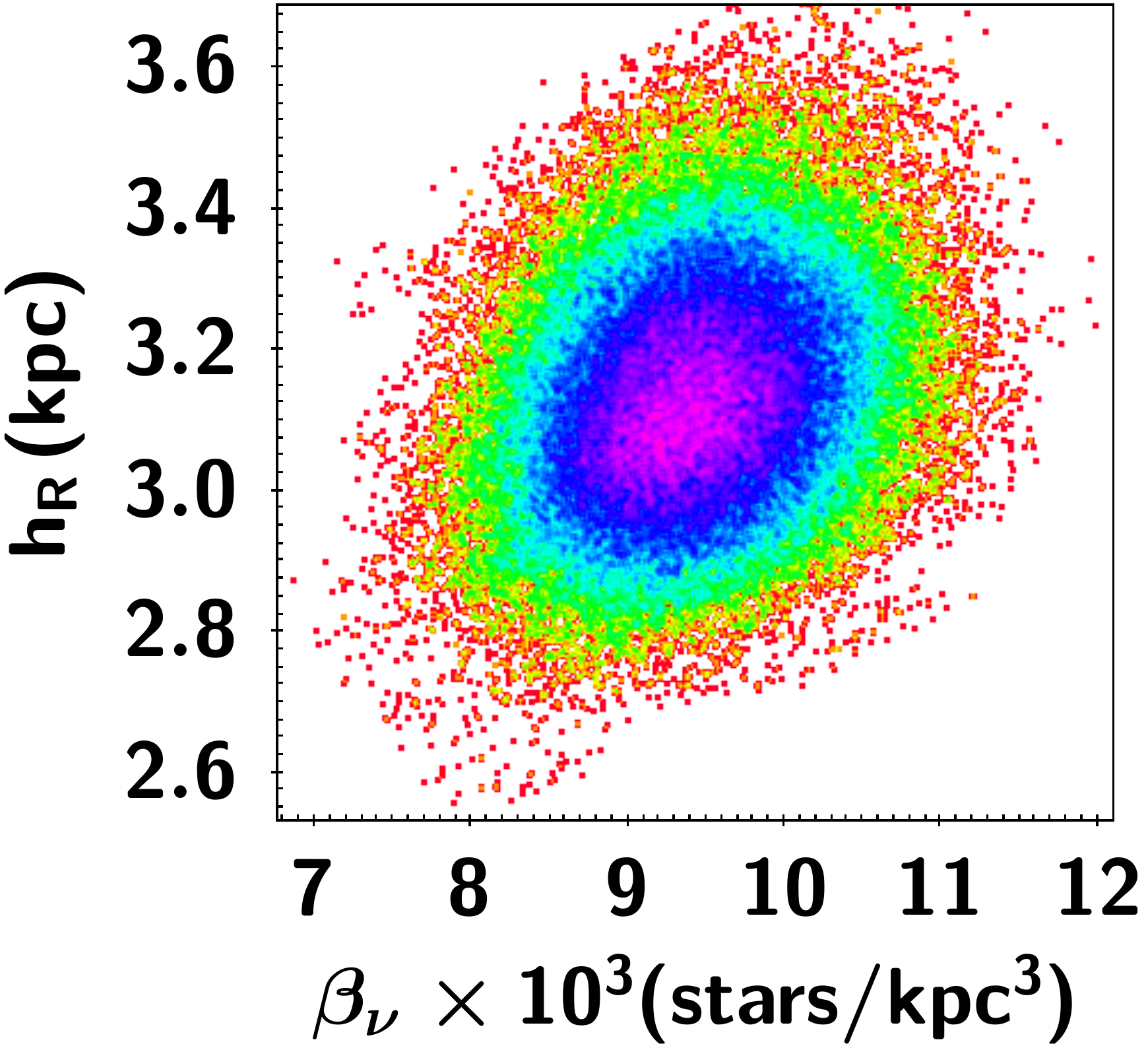}
            \includegraphics{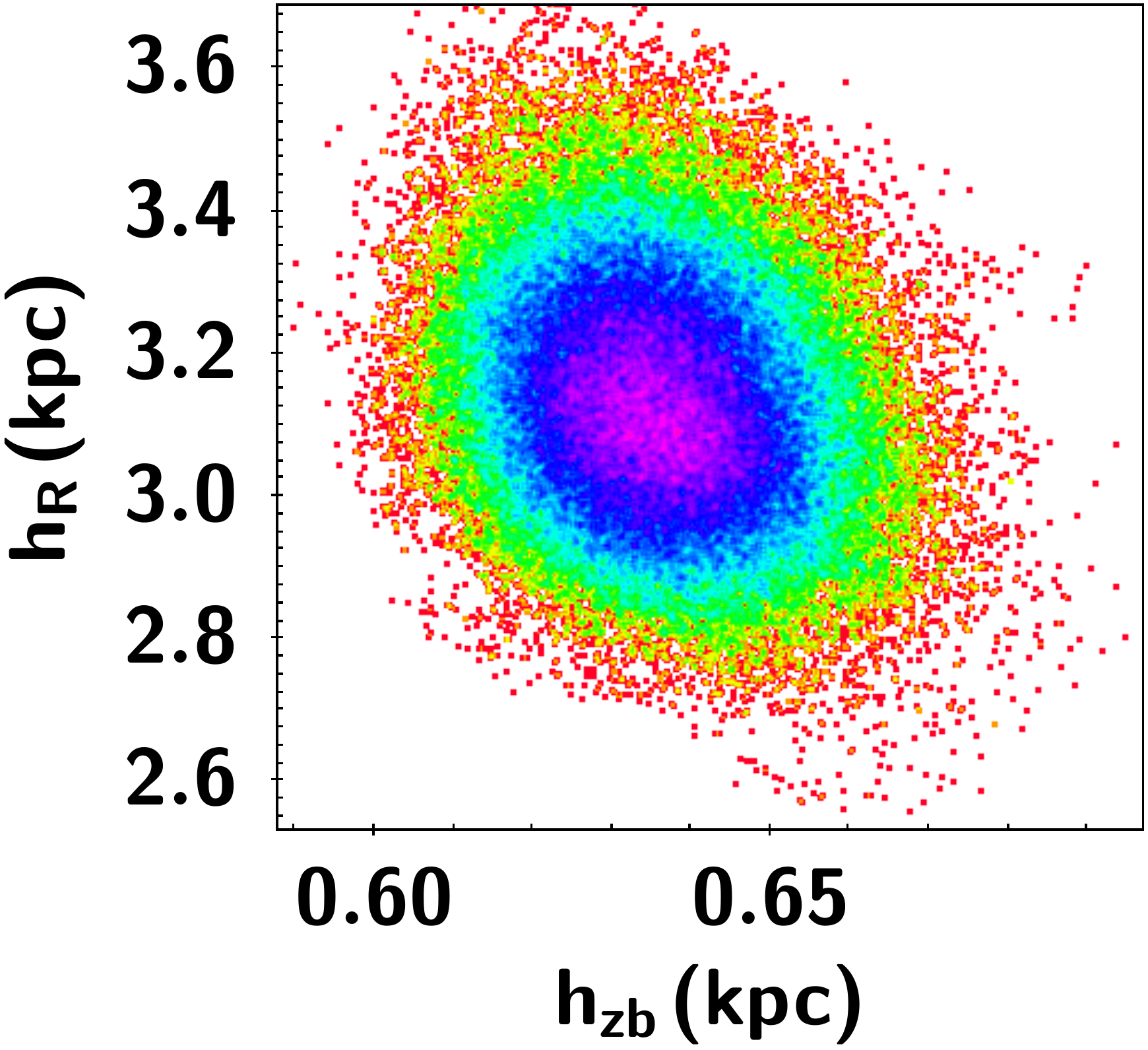}}\\
    \resizebox{8.7cm}{!}  
            {\includegraphics{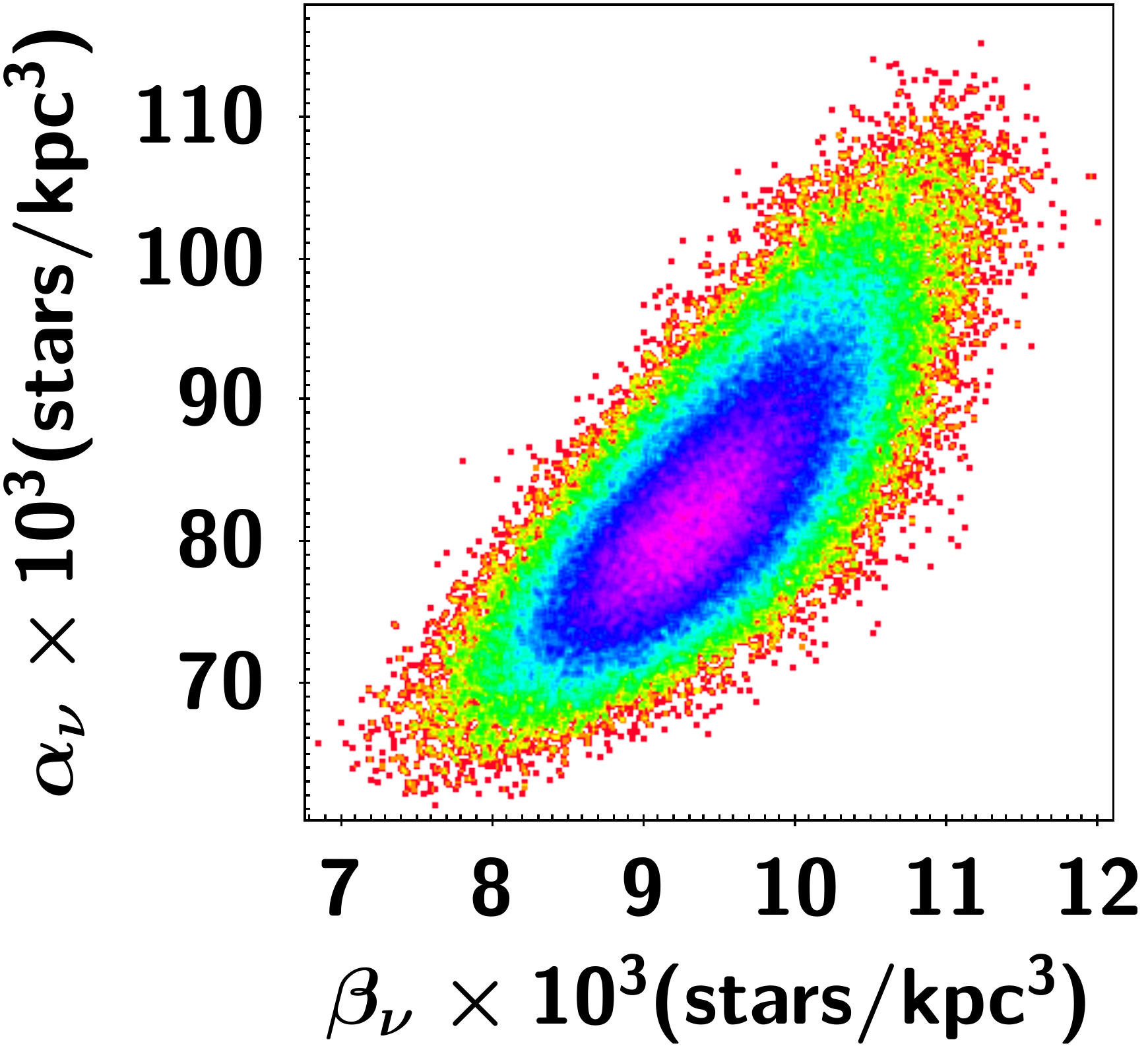}
            \includegraphics{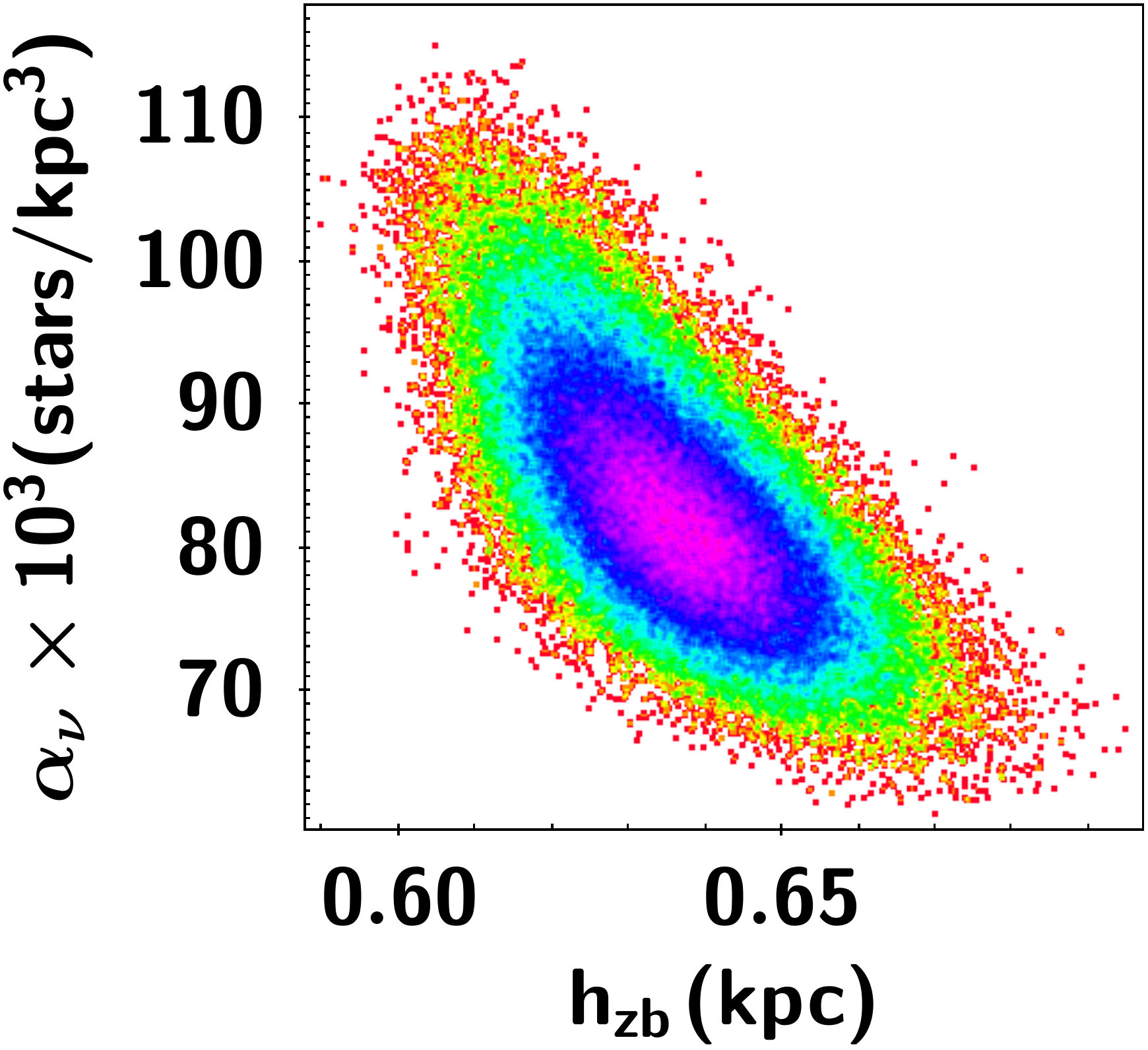}
            \includegraphics{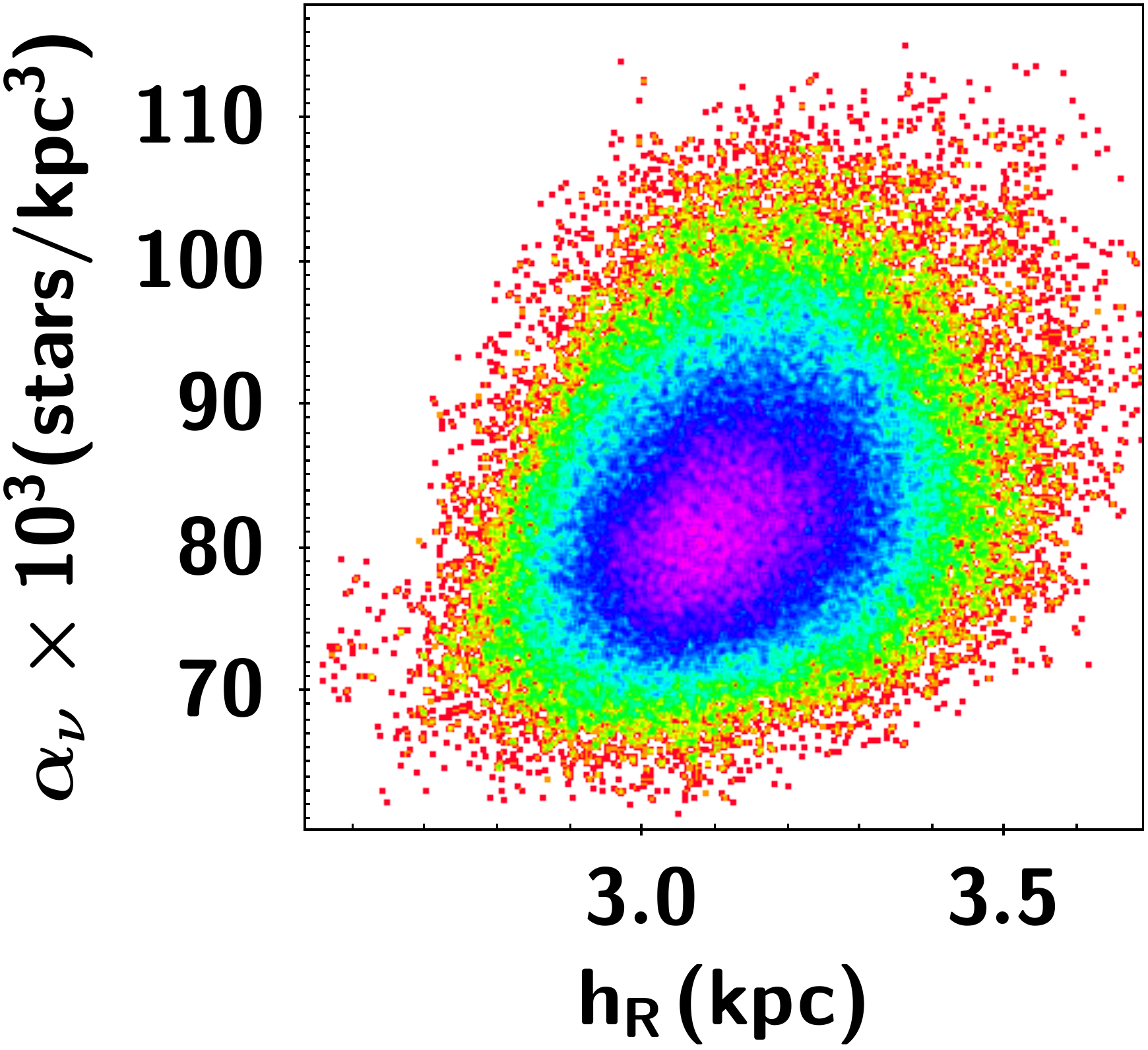}}\\
    \resizebox{11.6cm}{!}  
            {\includegraphics{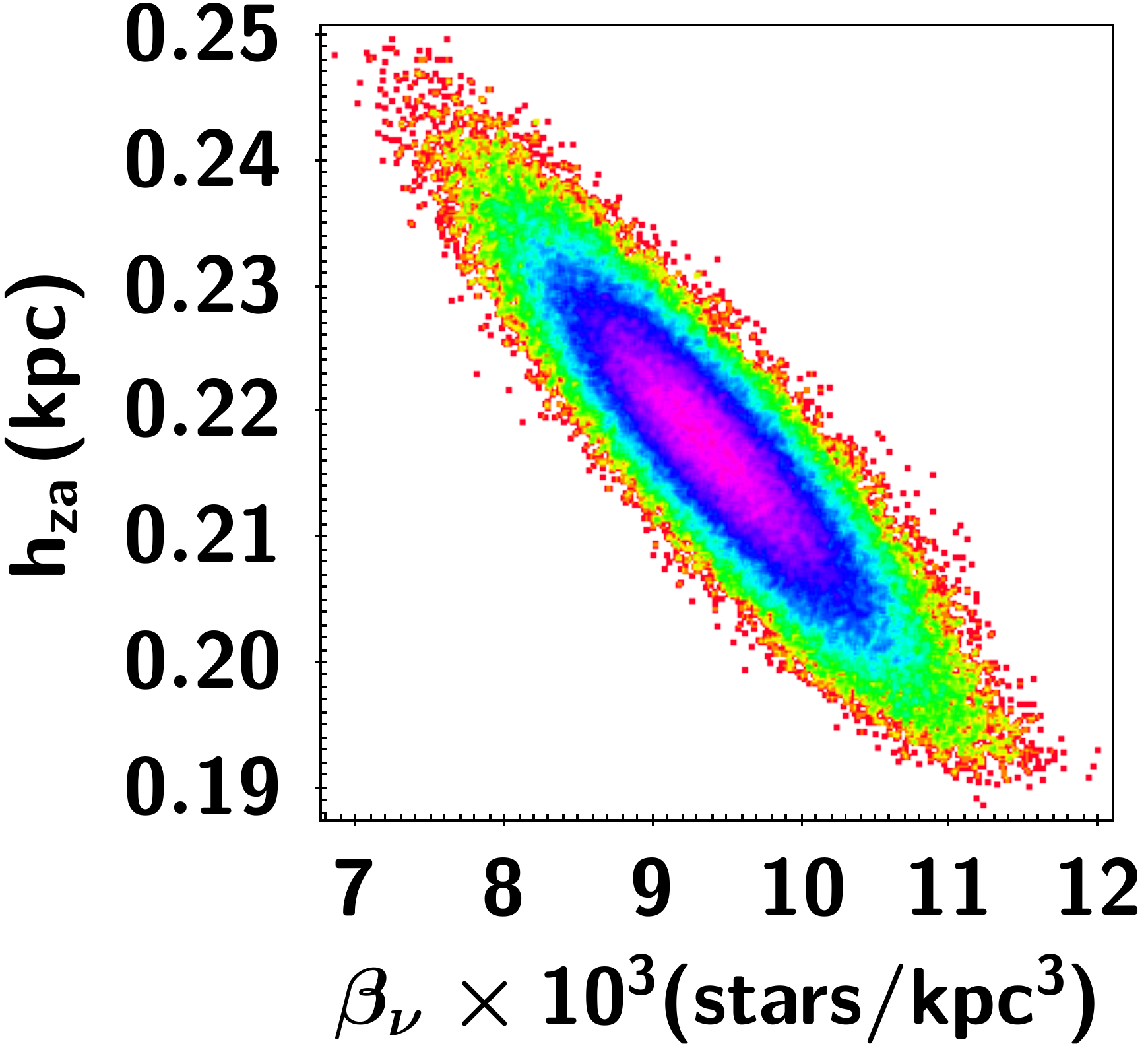}
            \includegraphics{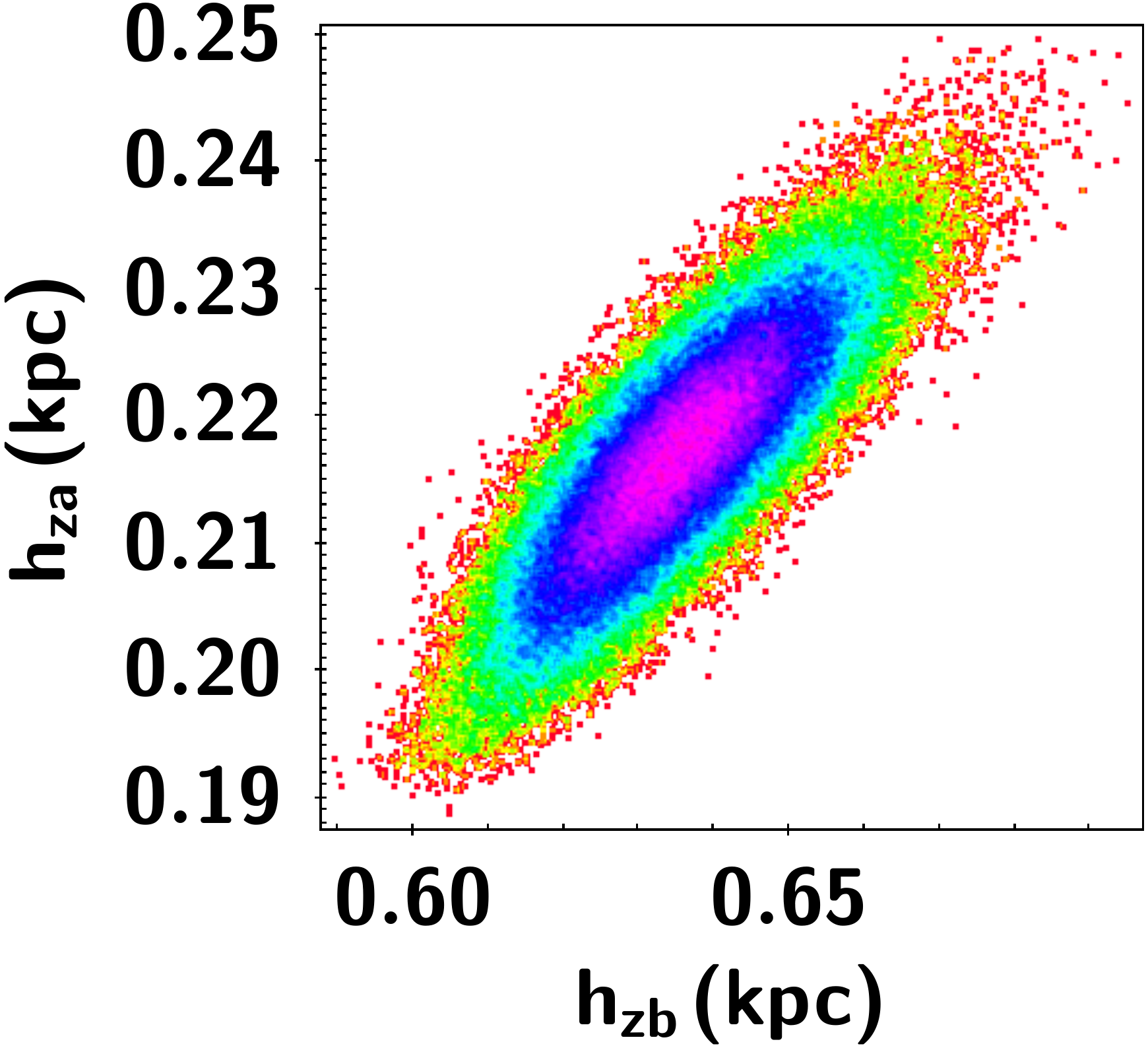}
            \includegraphics{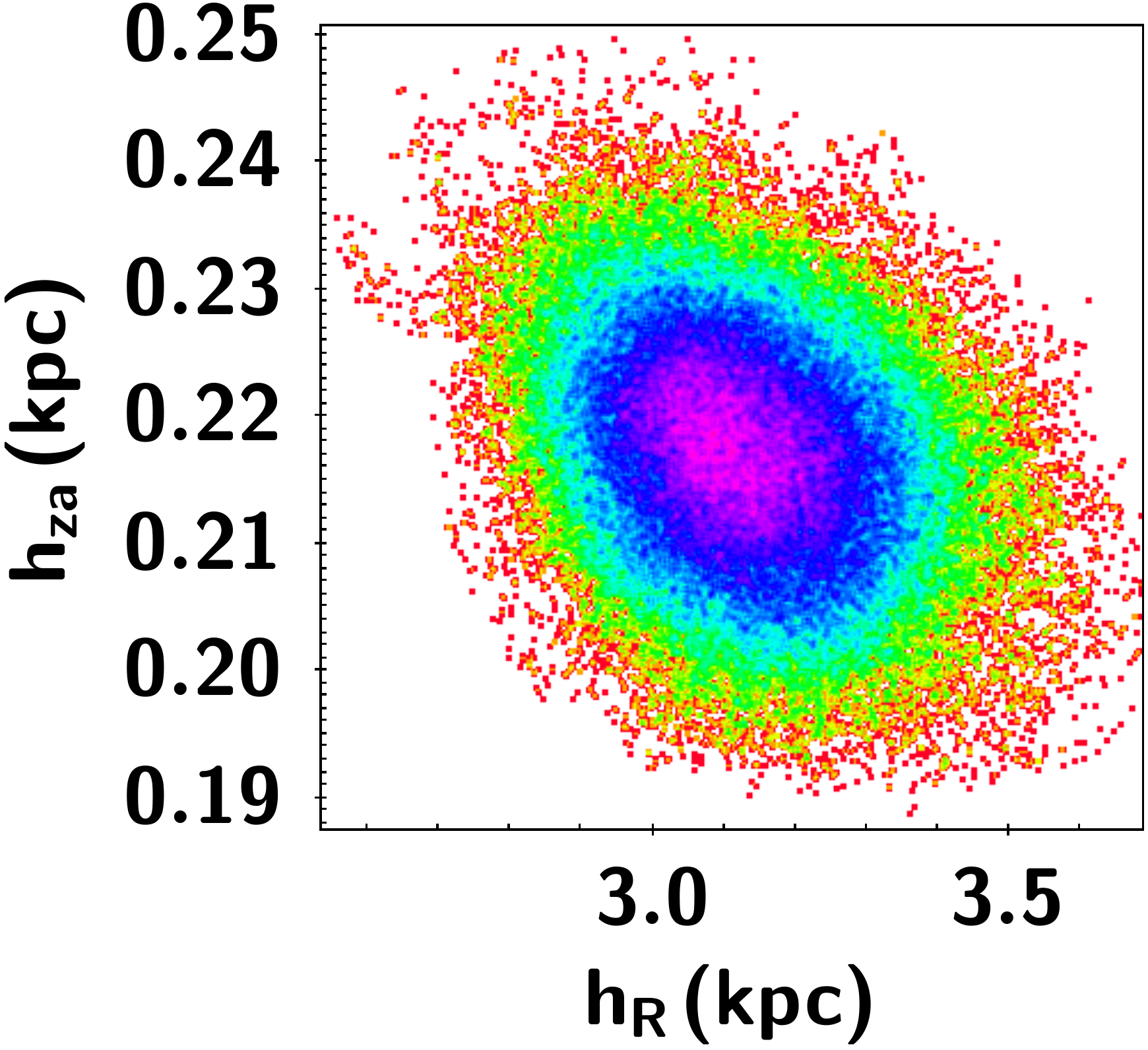}
            \includegraphics{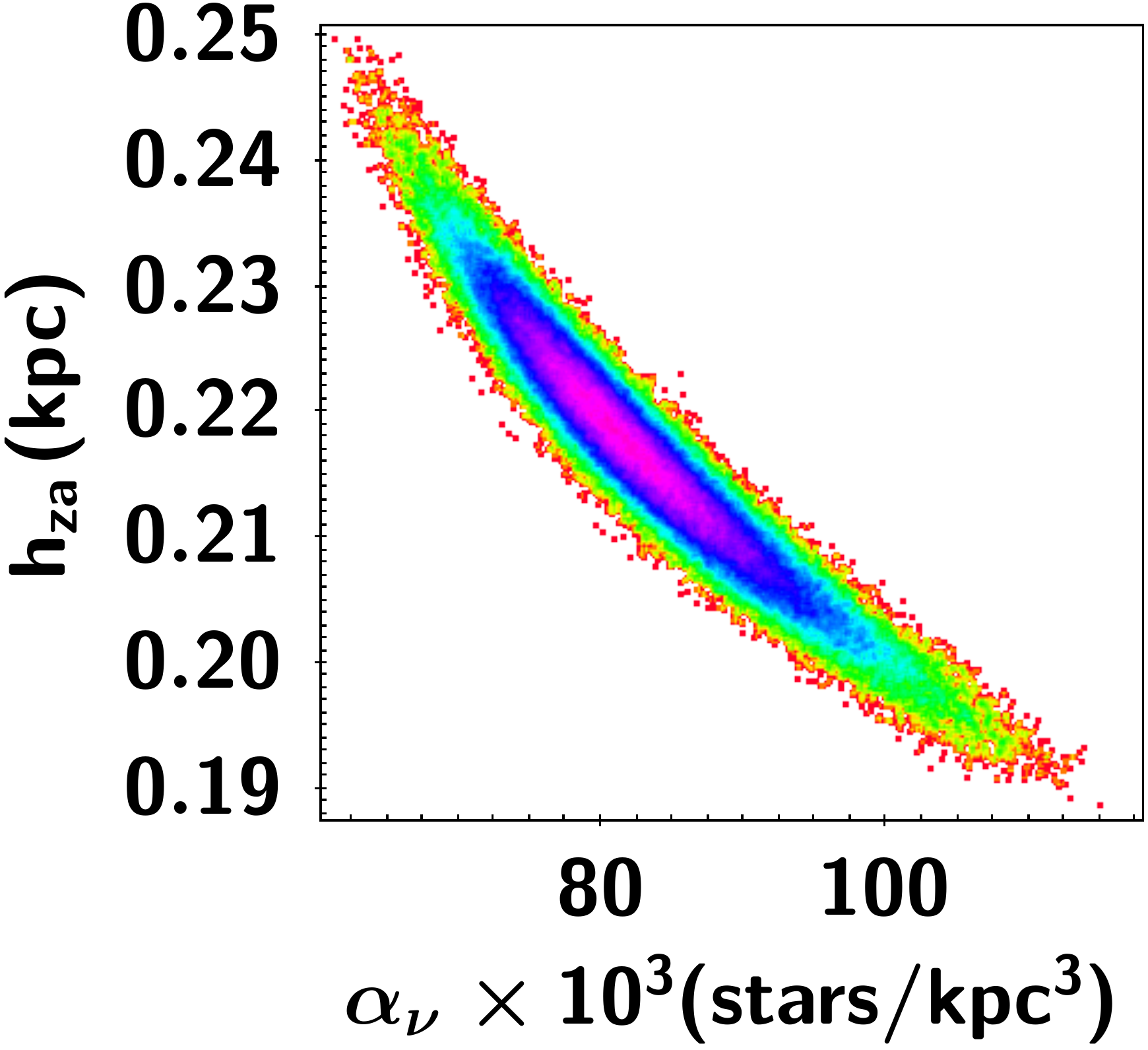}}\\
      \caption{ Same as Figure~\ref{northCorrelFig} but for the southern hemisphere. }
         \label{southCorrelFig}
   \end{figure*}

 Figures~\ref{northCorrelFig} and~\ref{southCorrelFig}  represent the triangle plots of the two-dimensional probability density distributions for the five parameters we used to fit density profiles. Most importantly, it appears clear that despite the various (anti-)correlations, chains are converging towards a defined area following the best likelihood, highlighting that there is no degeneracy.

As expected, the normalised density factor $\beta_{\nu}$ (respectively ($\alpha_{\nu}$)) is anti-correlated with its scale height $h_{zb}$ ($h_{za}$). The height of the thick (thin) disk is correlated with its density at R$_0$. In the same spirit, but to a lower level, $\beta_{\nu}$ (respectively ($\alpha_{\nu}$)) is mildly anti-correlated with its opposite scale height $h_{za}$ ($h_{zb}$) meaning the height of the thin (thick) disk relies on the density of the thick (thin) disk. The height of the thick disk $h_{zb}$ is correlated with the height of the thin disk $h_{za}$ in the same way the density of the thick disk $\beta_{\nu}$ is correlated with the density of the thin disk $\alpha_{\nu}$. Lastly, $h_R$ presents only very weak correlations with any parameters.

\end{document}